\documentclass[a4paper,11pt]{article}
\usepackage{fancyhdr}
\usepackage{lipsum}
\usepackage{tabularx}
\usepackage{color}

\usepackage{mathrsfs}
\usepackage{lipsum}
\usepackage{graphicx}
\usepackage{amsmath}
\usepackage{amssymb}
\usepackage{amsfonts}
\usepackage{amsthm}
\usepackage{bm}
\usepackage{chngcntr}
\usepackage{accents}
\usepackage[square,numbers,sort&compress]{natbib}%
\usepackage{url}
\usepackage{geometry}
\usepackage{booktabs}
\usepackage{tikz}
\usetikzlibrary{arrows.meta, decorations.pathreplacing}
\usepackage[hidelinks,bookmarks=false]{hyperref}

\numberwithin{figure}{section}
\numberwithin{table}{section}
\numberwithin{equation}{section}
\newcommand{\D}{{\mathrm{d}}}

\theoremstyle{plain}%

\newtheorem{theorem}{Theorem}[section]
\newtheorem{lemma}[theorem]{Lemma}
\newtheorem{corollary}[theorem]{Corollary}
\newtheorem{prop}[theorem]{Proposition}
\newtheorem{definition}[theorem]{Definition}
\newtheorem{remark}[theorem]{Remark}
\newtheorem{example}[theorem]{Example}
\makeatletter
\DeclareRobustCommand{\widebar}{\accentset{{\cc@style\underline{\mskip10mu}}}} %
\DeclareRobustCommand{\widecheck}[1]{\mathpalette\widecheck@{#1}}
\newcommand{\widecheck@}[2]{%
  \begingroup
  \setbox0=\hbox{$\m@th#1#2$}%
  \setbox2=\hbox{$\m@th#1\widehat{\phantom{#2}}$}%
  \setbox6=\hbox{$\m@th#1x$}%
  \dimen2=.25\ht6
  \dimen0=\ht0 \advance\dimen0 by\ht2 \advance\dimen0 by\dimen2
  \setbox4=\hbox{\rlap{\raise\dimen0\hbox{\scalebox{1}[-1]{\copy2}}}\copy0}%
  \dimen4=\ht2 \advance\dimen4 by\dimen2
  \ht4=\dimen4 \dp4=\dp0
  \box4
  \endgroup
}
\makeatother

\allowdisplaybreaks[3] 

	\title{Log-optimal portfolios for market models with regime switching\footnote{This research is fully supported by NSERC (through grant NSERC RGPIN-2025-05129-Choulli)}}
	\author{Tahir Choulli and Xunbai Yin\\
Mathematical and Statistical Sciences Dept.\\
University of Alberta, Edmonton, Canada}

\date{\today}

\begin{document}
\maketitle

\begin{abstract}
This paper considers market models under regime switching mechanism with $D$ states, and analyses the log-optimal portfolio. It is the portfolio, with finite expected utility, that maximizes the logarithmic utility from terminal wealth. We consider a family $\mathbb{X}=(X^{(i)})_{i=1,...,D}$, of $D$-semimartingale models, among which our agent can switch to-and-from following a switching regime modelled by a continuous-time Markov Chain $\alpha=(\alpha_t)_{t\geq 0}$ with $\alpha_0=1$. Then we define a general switching model $\widehat{X}$ that covers the existing cases, and  study the impact of the stochasticity of the regime switching and its intrinsic risks on the  optimal portfolio in many aspects.  On the one hand, we single out the necessary and sufficient conditions on the pair $(\mathbb{X},\alpha)$, which guarantee the existence of the optimal portfolio for $\widehat{X}$. On the other hand, we quantify the  factors that fully affect and determine the {\it increment in maximum expected logarithmic utility from terminal wealth} for both models $\widehat{X}$ and $X^{(1)}$.\end{abstract}

\section{Introduction} \label{Section:Introduction}
This paper studies the impact of random switching on two optimal portfolios that are intimately associated to the logarithmic utility, which are the num\'eraire and log-optimal portfolios. To this end, we start by defining these portfolios more precisely. Thus, we denote by $W^{\theta}$ the wealth process of the portfolio $\theta$. 
\begin{definition}\label{NP/LogOP}  Let $(X, \mathbb H, Q)$ be a market model, where $X$ is the assets' price process, $\mathbb H$ is a filtration, and $Q$ is a probability measure. Consider a fixed investment horizon $T\in(0,+\infty)$, and a portfolio $\theta^*$.
Then $\theta^*$ is called a {\it num\'eraire portfolio} for $(X, \mathbb H, Q)$ if $W^{\theta^*}>0$ and 
\begin{equation*}
{{W^{\theta}}\over{W^{\theta^*}}}\ \mbox{is an $(\mathbb H, Q)$-supermartingale for any portfolio $\theta$ with $W^{\theta}>0$}.
\end{equation*}
Portfolio  $\theta^*$ is called {\it log-optimal portfolio} for $(X, \mathbb H, Q)$ if $\theta^*\in \Theta(X,\mathbb H, Q)$ and 
\begin{eqnarray}
 u_T(X,\mathbb H, Q):=\sup_{\theta\in\Theta}E_Q\left[\ln(W^{\theta}_T)\right]= E_Q\left[\ln(W^{\theta^*}_T)\right],\label{LogInfinite}\end{eqnarray}
where $E_Q[.]$ is the expectation under $Q$, and $\Theta:=\Theta(X,\mathbb H, Q)$ is given by 
\begin{eqnarray}\label{AdmissibleSet0}
\hskip -0.6cm \Theta(X,\mathbb H, Q):=\Bigl\{\mbox{portfolio}\ \theta\ :\ W^{\theta}> 0\quad \mbox{and}\quad E_Q\left[\vert \ln(W^{\theta}_T)\vert \right]<+\infty\Bigr\}.\end{eqnarray}
\end{definition}
Since the seminal papers of Merton \cite{merton71,merton73}, utility maximization has been developed in many directions and at various levels of generality, see \cite{Karatzas&Zitkovic2003,Kramkov&Schachermayer1999} and the references therein. For logarithmic utility, G\"oll and Kallsen \cite{Goll&Kallsen2000,Goll&Kallsen2003} prove existence, describe the optimal portfolio in terms of the predictable characteristics, and establish duality under no-free-lunch-with-vanishing-risk (NFLVR hereafter) and finite maximal expected utility. This portfolio is intimately related to the num\'eraire portfolio, introduced by Long \cite{Long1990} with a martingale requirement on relative wealth and extended to the supermartingale formulation by Becherer \cite{Becherer2001}. When the num\'eraire portfolio has finite expected log utility, it is log-optimal. Without this integrability, its expected log utility may be infinite \cite{Christensen&Larsen2007,Hulley&Schweizer2010}. Karatzas and Kardaras \cite{Karatzas&Kardaras2007} characterize the num\'eraire portfolio via the predictable characteristics and connect its existence to no-unbounded-profit-with-bounded-risk (NUPBR hereafter). Choulli et al. \cite{Choullietal2015} extend this connection to general semimartingales and prove that NUPBR is equivalent to the existence of a log-optimal portfolio under equivalent probabilities arbitrarily close to the original probability in total variation. This makes NUPBR the minimal viability condition for modelling a frictionless competitive financial market \cite{Karatzas&Kardaras2007,Choullietal2015}. In this sense, the model is useful if and only if NUPBR holds. Utility maximization under NUPBR and appropriate finiteness conditions is further developed in \cite{Mostovyi2015,Chauetal2017}. Choulli and Yansori \cite{Choulli&Yansori2022c} provide a complete and explicit characterization of the log-optimal portfolio and its optimal deflator without assuming NFLVR, establish duality, and describe its direct relationship with the num\'eraire portfolio without changing the probability. The log-optimal portfolio thus provides a natural starting point for utility optimization when NFLVR fails. These results are vital for our current paper.

Regime switching models describe asset dynamics across different economic environments, such as recession and prosperity or bear and bull markets. Their use originated in econometrics and time series analysis with \cite{Quandt1958,Goldfeld&Quandt1973,Tong1983,Hamilton1989}. In financial mathematics, regime switching can be combined with jump diffusion and stochastic volatility to incorporate structural changes in market conditions. In continuous time, the switching mechanism is commonly modelled by a finite-state continuous-time Markov chain, and regime switching jump diffusion (RSJD hereafter) allows both volatility and jump activity to depend on the prevailing regime. This captures the time variation of asset dynamics across economic conditions while preserving a satisfactory level of analytical tractability. It also allows the characteristics of classical L\'evy models to vary across regimes \cite{Konikov&Madan2002,Elliott&Osakwe2006}, as illustrated by the switching L\'evy models in \cite{Dong&Vostricova2024}. Early applications address asset allocation \cite{Elliott&vanderHoek1997} and derivatives pricing \cite{Naik1993,Guo2001,Konikov&Madan2002,Norberg2003}. In particular, Naik \cite{Naik1993} incorporates price jumps caused by regime changes, while Norberg \cite{Norberg2003} considers a market driven entirely by switching jumps. For further developments in derivatives pricing, we refer to \cite{Elliott&Osakwe2006,Yuen&Yang2009,Boetal2010,Capponietal2014,Kirkby&Nguyen2020,Siu&Elliott2022}, and for stochastic optimization and optimal investment to \cite{Sotomayor&Cadenillas2009,Donnelly&Heunis2012,Zhangetal2012,Pamen2017,Dong&Vostricova2024}, and the references therein.

In this paper, we consider a family $\mathbb{X}=(X^{(i)})_{i=1,...,D}$ of semimartingale models and a switching regime modelled by a Markov chain $\alpha$. The resulting switching model $\widehat{X}$, which will be considered throughout the paper and will be defined in the forthcoming section, incorporates transition shocks that may occur jointly with the jumps of the underlying returns. For this general switching model we address the following question. 
\begin{equation}\label{RSJDQ1}
\mbox{How does regime switching affect num\'eraire portfolio or equivalently the NUPBR concept?}
\end{equation}
We characterize num\'eraire portfolio for $\widehat{X}$ and its existence (or equivalently NUPBR) in terms of the models $X^{(i)}$ and the parameters of the switching regime. As a result, we describe the set of all deflators  for $\widehat{X}$ in various manners. Then we focus on the log-optimal portfolio by answering the following question.
\begin{equation}\label{RSJDQ2}
\mbox{When does the log-optimal portfolio of $\widehat{X}$ exist, and how can it be described?}
\end{equation}
Our last main contribution lies in measuring the impact of switching on the log-optimal portfolio. This can  be achieved by analyzing deeply the {\it increment of expected logarithmic utility from terminal wealth} for both models $\widehat{X}$ and $X^{(1)}$, which is defined by
\begin{equation}\label{RSJDQ3}
\Delta_T(\widehat{X},X^{(1)}):=u_T(\widehat{X})-u_T(X^{(1)}).
\end{equation}
Precisely, we single out the main factors that perturb the quantity $\Delta_T(\widehat{X},X^{(1)})$ in a way or another. In particular, we quantify the effects of switching coefficients, transition shocks, and the impact resulting from the interaction between shocks and the jumps of the family $\mathbb{X}$. Furthermore, we address  the {\it pure switching effect}, (also called the Parrondo effect) which measures the impact of random switching after the regimes are made equivalent in expected logarithmic utility. 

The paper is organized as follows. Section~2 introduces the switching model and its predictable characteristics. Section~3 addresses the num\'eraire portfolio, deflators, and the stability of NUPBR under transition shocks. Section~4 characterizes the log-optimal portfolio and quantifies the switching impact, including the level and Parrondo effects. The appendix contains supporting results and deferred proofs.

\section{The mathematical model and its preliminaries}
Throughout the paper, we consider a filtered complete probability space $\left(\Omega,{\cal{F}}, \mathbb{F}:=(\mathcal{F}_t)_{t\geq 0},P\right)$, where the filtration $\mathbb{F}$ satisfies the usual conditions of right-continuity and completeness.%

Throughout the paper, ${\cal M}(\mathbb H,Q)$ denotes the set of all $(\mathbb H,Q)$-martingales, while ${\cal A}(\mathbb H,Q)$ (respectively ${\cal A}^+(\mathbb H,Q)$) denotes the set of all $\mathbb H$-adapted processes that are right-continuous with left limits (RCLL hereafter) and have $Q$-integrable variations (respectively nondecreasing and $Q$-integrable). If ${\cal C}(\mathbb H,Q)$ is a set of processes, then ${\cal C}_{loc}(\mathbb H,Q)$ is the set of processes, $Y$, for which there exists a sequence of $\mathbb H$-stopping times, $(T_n)_{n\geq 1}$, that increases to infinity $Q$-a.s. and $Y^{T_n}$ belongs to ${\cal C}(\mathbb H,Q)$, for each $n\geq 1$.

For a $d$-dimensional $(\mathbb H,Q)$-semimartingale $Y$, the set of $d$-dimensional $\mathbb H$-predictable processes that are $Y$-integrable in the semimartingale sense is denoted by $L(Y,\mathbb H,Q)$, and the resulting real-valued integral of $\varphi$ with respect to $Y$ is denoted by $\varphi\cdot Y$. For any $M\in {\cal M}_{loc}(\mathbb H,Q)$, we put
$$L^1_{loc}(M,\mathbb H,Q):=\{\varphi\in L(M,\mathbb H,Q)\ \big|\ \varphi\cdot M\in {\cal M}_{loc}(\mathbb H,Q)\}.$$

Throughout the paper, $^{p,\mathbb H,Q}K$ denotes the $(\mathbb H,Q)$-predictable projection of the process $K$ (when it exists), while ${\cal E}(L)$ is the Dol\'eans-Dade (stochastic) exponential given by
$$ {\cal E}_t(L)=\exp(L_t-L_0-{1\over{2}}\langle L^c\rangle_t)\prod_{0<s\leq t}(1+\Delta L_s)e^{-\Delta L_s},\quad t\geq0.$$
Let $\mathcal{O}(\mathbb H)$ and $\mathcal{P}(\mathbb H)$ denote the $\mathbb H$-optional and $\mathbb H$-predictable $\sigma$-algebras on $\Omega\times\mathbb{R}_+$, respectively, and put
$$\widetilde{\cal{O}}(\mathbb H):=\mathcal{O}(\mathbb H)\otimes\mathcal{B}(\mathbb{R}^d),\quad \widetilde{\cal{P}}(\mathbb H):=\mathcal{P}(\mathbb H)\otimes\mathcal{B}(\mathbb{R}^d).$$

We also define the set of admissible $d$-dimensional portfolios for $(Y,\mathbb H,Q)$ by
\begin{equation}\label{AdmissibleSetGeneral}
\mathcal{L}(Y,\mathbb H,Q):=\{\pi:\pi\ \mbox{is $\mathbb H$-predictable},\quad \pi^{tr}\Delta Y>-1,\quad Q\mbox{-a.s.}\}.
\end{equation}
When there is no ambiguity, we omit $(\mathbb H,Q)$, with $(\mathbb H,Q)=(\mathbb F,P)$ by default.

\subsection{The switching regime}
The switching mechanism (factor) is modelled by $\alpha:=(\alpha(t))_{t\geq0}$ such that  $\alpha(0)=1$. Throughout the paper, we assume that  $\alpha$ is a right-continuous-with-left-limits (RCLL for short hereafter) process, $\mathbb{F}$-adapted with values in $\{1,...,D\}$, and is a continuous-time Markov chain (CTMC for short hereafter) with respect to $\mathbb{F}$. Let $Q$ be  its generator (probability transition matrix) given by 
\begin{equation}\label{Qmatrix}
Q:=(q_{ij})_{1\leq i,j\leq D},\quad 	0< q_{ij}<\infty, i\neq j,\quad\mbox{and}\quad \sum_{j=1}^Dq_{ij}=0.
\end{equation}
Consider the basis vector $e_i\in\mathbb{R}^D$ where the $j^{\text{th}}$ component of $e_i$ is the Kronecker delta $\delta_{ij}$ for each $i,j\in\{1,2,...,D\}$. Then the vector-valued process $e_\alpha$ admits the semimartingale representation 
\begin{equation}
e_{\alpha(t)}:=\sum_i I_{\{\alpha(t)=i\}}e_i=e_{\alpha(0)}+\int_0^tQ^{tr}e_{\alpha(u)}\D u+ \mathfrak{M}(t), t\geq0 \label{CTMCRepresentationVector},
\end{equation}
where $\mathfrak{M}:= (\mathfrak{M}(t))_{t\geq0}:=(\mathfrak{M}_1(t),\mathfrak{M}_2(t),...,\mathfrak{M}_D(t))^{tr}_{t\geq0}$ is an $\mathbb{R}^D$-valued, square-integrable $\mathbb{F}$-martingale. For this fact, we refer the reader to \cite[Section 2]{Zhangetal2012} for more details and related results. By rewriting (\ref{CTMCRepresentationVector}) component by component, we obtain
\begin{equation}
I_{\{\alpha(t)=j\}}=I_{\{\alpha(0)=j\}}+\int_0^t\sum_i^DI_{\{\alpha(u)=i\}}q_{ij}\D u+\mathfrak{M}_{j}(t), j\in\{1,2,...,D\}. \label{Decomposition4Ialphal}
\end{equation}
Consider the counting process of transitions into state $j\in\{1,2,...,D\}$ and its compensated martingale given by 
\begin{equation}\label{MarkovJumpTotal}
\begin{split}
&J_{j}(t):=\sum_{0<u\leq t}I_{\{\alpha(u-)\neq j,\alpha(u)=j\}},\quad \widetilde{J}_{j}(t):=J_{j}(t)-\int_0^tq_{j}(u)\D u,\quad\mbox{where}\quad q_{j}:=q_{\alpha_{-}j}I_{\{\alpha_{-}\neq j\}},\\
&J(t):=(J_{1}(t),J_{2}(t),...,J_{D}(t))^{tr}, \ \ \widetilde{J}(t):=(\widetilde{J}_{1}(t),\widetilde{J}_{2}(t),...,\widetilde{J}_{D}(t))^{tr}, \ \ q(t):=(q_{1}(t),q_{2}(t),...,q_{D}(t))^{tr}. 
\end{split}
\end{equation}
\begin{remark}
We can also extend the switching regime $\alpha$ to be an $\mathbb{F}$-conditional CTMC as in \cite[Chapter 11.3]{Bielecki&Rutkowski2002}. In this case, $q$ is also understood as the intensity process of the jump $J$, which can be assumed to be predictable by \cite[I.3.13]{Jacod&Shiryaev2003}. The only change is that now $P(\alpha(t)=i)$ has no explicit expression for $i\in\{1,2,...,D\}$.
\end{remark}

\subsection{Our switching model and its parametrization}
Our mathematical model starts with $D$ given market models represented by their return processes denoted by $\mathbb{X}:=(X^{(i)})_{i=1,...,D}$. Here each $X^{(i)}$ is a $d$-dimensional semimartingale, mathematically and statistically parametrized by its predictable characteristics and the random measure of its jumps $\mu^{(i)}$ and the random measure compensator of the latter $\nu^{(i)}$, which are given by 
\begin{equation}\label{Characteristics4X(i)}
\begin{split}
&\mbox{predictable characteristics}\quad (b^{(i)},c^{(i)},F^{(i)},A^{(i)}):=(b^{X^{(i)}},c^{X^{(i)}},F^{X^{(i)}},A^{X^{(i)}}),\\
&\mbox{ and its jump measures}\quad \mu^{(i)}:=\mu^{X^{(i)}},\quad \mbox{and}\quad \nu^{(i)}:=\nu^{X^{(i)}},\quad \nu^{(i)}(\D t,\D x)=F^{(i)}_t(\D x)\D A^{(i)}(t).
\end{split}
\end{equation}
For more details about these characteristics and related properties and results we refer the reader to \cite{Jacod&Shiryaev2003,HeWangYan1992}. Thus, the canonical decomposition for each $X^{(i)}$ is given by 
\begin{equation}\label{Decompositin4X(l)}
X^{(i)}=(X^{(i)})^c+h\star(\mu^{(i)}-\nu^{(i)})+b^{(i)}\cdot{A}^{(i)}+(x-h)\star\mu^{(i)}.
\end{equation}
Without loss of generality (it is enough to take $A=\sum_{i=1}^D{A}^{(i)}+t$ and rewrite the characteristics accordingly), throughout the paper, we will assume
$$ {A}^{(i)}={A},\quad\mbox{for any $i\in\{1,...,D\}$ and}\quad dt\ll{dA}.$$
We take $A_0=0$.

From now on, for each $i\in\{1,2,...,D\}$, write $q_i:=\frac{\D t}{\D A}q_{\alpha_{-}i}I_{\{\alpha_{-}\neq i\}}$ for the intensity with respect to $A$, so that $\widetilde{J}_i=J_i-q_i\cdot A$. 

Hence, our switching model is given by
\begin{equation} \label{GRSJDDynamic}
\begin{split}
\widehat{X}:=&\widebar{X}+\gamma_\alpha\cdot \widetilde{J}=\underbrace{\sum_iI_{\{\alpha_{-}=i\}}\cdot X^{(i)}}_{\text{switching coefficients}}+\underbrace{\gamma_\alpha\cdot \widetilde{J}}_{\text{transition shock}},\quad
\gamma_\alpha:=\sum_i I_{\{\alpha_{-}=i\}}\gamma_i,
\end{split}
\end{equation} 
where $\gamma_i$ is an $\mathbb{R}^{d\times D}$-valued predictable process satisfying
$$\gamma_i\cdot\widetilde{J}\in\mathcal{M}_{loc},\quad i\in\{1,2,...,D\}.$$

It is clear that our switching model has two main components, namely the switching coefficient component modelled by the process $\widebar{X}$, and the transition shock represented by an It\^o integral with respect to the martingale $\widetilde{J}$.  Remark that the model $\widebar{X}:=\sum_iI_{\{\alpha_{-}=i\}}\cdot X^{(i)}$ is constructed by a return process switching between each $X^{(i)}$, like several ``multiverses" parallel to the real world. Recall that $J$ is the jump process recording the switching times with its compensated martingale $\widetilde{J}$. Motivated by the construction
of \citeauthor{Naik1993}\cite{Naik1993}, 
\citeauthor{Norberg2003}\cite{Norberg2003} and 
\citeauthor{Zhangetal2012}\cite{Zhangetal2012}, the transition shock should be added in the switching return. This transition shock term appears only in one form of $\gamma_\alpha\cdot \widetilde{J}$, for some integrand $\gamma_\alpha$, in all existing papers.\\ 

Our switching model (\ref{GRSJDDynamic}), which is parametrized by the family of quadruplets $(b^{(i)},c^{(i)},F^{(i)},\gamma_i)_{i=1,..,D}$, is the most general regime switching semimartingale market model. In fact, when $
\gamma_\alpha\equiv0$, then $\widehat{X}=\widebar{X}$. Thus, any result for the market $\widehat{X}$ obtained later automatically applies to  the market $\widebar{X}$ by letting $\gamma_\alpha\equiv0$. The model of  \citeauthor{Naik1993}(\citeyear{Naik1993}) \cite{Naik1993} ($D$=2) can be obtained from ours by putting $(b^{(i)},c^{(i)},F^{(i)},\gamma_i)=(b,c^{(i)},0,\gamma_i)$ with  $(c^{(i)},\gamma_i)$ and $b$ being constants. The model of \citeauthor{Norberg2003}(\citeyear{Norberg2003}) \cite{Norberg2003} can  be deduced from ours by considering $(b^{(i)},c^{(i)},F^{(i)},\gamma_i)=(0,0,0,\gamma_i)$ with $\gamma_i$ being constant. The model of \citeauthor{Elliott&Osakwe2006}(\citeyear{Elliott&Osakwe2006}) \cite{Elliott&Osakwe2006} can be deduced by putting $(b^{(i)},c^{(i)},F^{(i)},\gamma_i)=(b^{(i)},0,F^{(i)},0)$ with $(b^{(i)},F^{(i)})$ being constant. The model of \citeauthor{Yuen&Yang2009}(\citeyear{Yuen&Yang2009}) \cite{Yuen&Yang2009}  can be obtained by considering $(b^{(i)},c^{(i)},F^{(i)},\gamma_i)=(b^{(i)},c^{(i)},0,\gamma_i)$ with constant $(b^{(i)},c^{(i)},\gamma_i)$. The model of \citeauthor{Zhangetal2012}(\citeyear{Zhangetal2012}) \cite{Zhangetal2012} can  be obtained by considering $(b^{(i)},c^{(i)},F^{(i)},\gamma_i)=(b^{(i)},c^{(i)},F^{(i)},\gamma_i)$ with $(b^{(i)},c^{(i)},F^{(i)},\gamma_i)$ depending only on state processes, control processes and time.  The model of \citeauthor{Donnelly&Heunis2012}(\citeyear{Donnelly&Heunis2012}) \cite{Donnelly&Heunis2012} follows by putting $(b^{(i)},c^{(i)},F^{(i)},\gamma_i)=(b^{(i)},c^{(i)},0,0)$ for  restricted $(b^{(i)},c^{(i)})$ satisfying some certain measurability conditions. The model of  \citeauthor{Pamen2017}(\citeyear{Pamen2017}) \cite{Pamen2017} is obtained by $(b^{(i)},c^{(i)},F^{(i)},\gamma_i)=(b^{(i)},c^{(i)},F^{(i)},\gamma_i)$ with one-dimensional $(b^{(i)},c^{(i)},F^{(i)},\gamma_i)$ satisfying some certain measurability and integrability conditions. The model of  \citeauthor{Dong&Vostricova2024}(\citeyear{Dong&Vostricova2024}) \cite{Dong&Vostricova2024} results from ours by considering $(b^{(i)},c^{(i)},F^{(i)},\gamma_i)=(b^{(i)},c^{(i)},F^{(i)},0)$ for constant $(b^{(i)},c^{(i)},F^{(i)})$.
\begin{remark}
	To explain more about (\ref{GRSJDDynamic}), we assume that $\alpha_{-}=i$ at present. Then\\
{\rm(a)} Between switching times ($\Delta \alpha=0$ or $\Delta J =0$), the return process $\widehat{X}$ accumulates as the current regime $X^{(i)}$ with the compensation drift, i.e., $\D \widehat{X}=\D X^{(i)}-\gamma_iq\D A$.\\
{\rm(b)} At the switching time ($\Delta \alpha\neq 0$ or $\Delta J\neq 0$), say $\alpha=j\neq i$, $\widehat{X}$ changes its local characteristics and accumulation to be the same as $X^{(j)}$, and an extra shock $\gamma_i^{(j)}\Delta J_{j}$ is imposed on $\widehat{X}$, where $\gamma_i^{(j)}$ denotes the $j^{\text{th}}$ column vector of the  matrix $\gamma_i$.
\end{remark}
In the following we illustrate a practical example for our general switching model $\widehat{X}$. 
\begin{example}
\begin{equation}
\widehat{X}=\rho_\alpha\cdot t+\sigma_\alpha\cdot W+\zeta_\alpha\cdot\widetilde{N}_{\alpha}+\gamma_\alpha\cdot\widetilde{J},\label{RSJDdynamic}
\end{equation}
where $W$ is an $n_1$-dimensional Wiener process and $N_i:=(N_{i1},...,N_{in_2})^{tr}$ has independent Poisson components with constant intensities $\lambda(i):=(\lambda_1(i),...,\lambda_{n_2}(i))^{tr}>0$, all with respect to $\mathbb F:=\mathbb F^{W,N_1,...,N_D,\alpha}$.

For $i\in\{1,2,...,D\}$,
$$\widetilde{N}_i:=N_i-\lambda(i)t,\quad \widetilde{N}_\alpha:=\sum_iI_{\{\alpha_{-}=i\}}\cdot\widetilde{N}_i,$$
$$(\rho_\alpha,\sigma_\alpha,\zeta_\alpha,\gamma_\alpha):=\sum_iI_{\{\alpha_{-}=i\}}(\rho_i,\sigma_i,\zeta_i,\gamma_i),$$
with bounded measurable $(\rho_i,\sigma_i,\zeta_i,\gamma_i):[0,T]\to\mathbb R^d\times\mathbb R^{d\times n_1}\times\mathbb R^{d\times n_2}\times\mathbb R^{d\times D}$. The corresponding $X^{(i)}$ and its predictable characteristics are
\begin{equation}\label{JDdynamic}
X^{(i)}=\rho_i\cdot t+\sigma_i\cdot W+\zeta_i\cdot\widetilde{N}_i,
\end{equation}
\begin{equation}\label{JDTriplet}
\begin{split}
b^{(i)}=\rho_i-\int(x-h(x))F^{(i)}(\D x),\  c^{(i)}=\sigma_i\sigma_i^{tr},\ A^{(i)}(t)=t,\ F^{(i)}_t(\D x)=\sum_k{I}_{\{ x\neq0\}}\lambda_k(i)\delta_{\zeta_i^{(k)}(t)}(\D x),
\end{split}
\end{equation}
where $\zeta_i^{(k)}$ is the $k^{\text{th}}$ column of $\zeta_i$ and $h(x):=xI_{\{|x|\leq1\}}$.
\end{example}

\begin{remark}
{\rm(a)} We list different examples of the general switching model (\ref{GRSJDDynamic}).
\begin{itemize}
	\item Model (\ref{JDdynamic}).
	\item Switching L\'evy model: $(b^{(i)},c^{(i)},F^{(i)})$ is constant for all $i\in\{1,2,...,D\}$ and $A=t$.
	\item Markov-modulated model: $(b^{(i)},c^{(i)},F^{(i)},A)$ is deterministic for all $i\in\{1,2,...,D\}$, i.e., $X^{(i)}$ is a process with independent increments (PII) \cite[II.4.15]{Jacod&Shiryaev2003}.
	\item Switching It\^o semimartingale model: $A=t$.
\end{itemize}

The interplay between the time-inhomogeneity of $(b^{(i)},c^{(i)},F^{(i)})$ and regime switching induces an additional effect on the optimal utility, which is one of our principal findings.

{\rm(b)} We allow common jumps between $X^{(i)}$ and $\gamma_\alpha\cdot\widetilde{J}$. The non-negligible impact of common jumps on the optimal portfolio is another principal finding of this paper. These processes are assumed uncorrelated in \cite{Yin&Xi2010,Boetal2010,Capponietal2014}, or without common jumps in \cite{Zhangetal2012}.

{\rm(c)} Our general model (\ref{GRSJDDynamic}) is a stereoscopic integration of all existing switching models in mathematical finance within the semimartingale framework. Precisely,
\[\begin{split}	\text{most existing models}& \subseteq\text{PII } X^{(i)}\subseteq \text{model of (\ref{GRSJDDynamic})},
\end{split}
\] and
\[\text{all existing models } \subseteq\text{PII } X^{(i)}+ \text{models like \citeauthor{Donnelly&Heunis2012}\cite{Donnelly&Heunis2012}}\subseteq \text{model of (\ref{GRSJDDynamic})}.\]
The model of \cite{Donnelly&Heunis2012} also allows path-dependent coefficients.
\end{remark}

Next, we give the first key result derived in this paper. We compute the predictable characteristics, denoted by the quadruplet $(\widehat{b},\widehat{c},\widehat{F},\widehat{A})$, of the return semimartingale $\widehat{X}$ defined in (\ref{GRSJDDynamic}). This quadruplet is used to characterise the num\'eraire portfolio in the semimartingale market.
\begin{prop}\label{Proposition:GRSJDPredictableCharacteristics}
	Suppose that $X^{(i)}$ is special for each $i\in\{1,2,...,D\}$. Then the predictable characteristics of the market $\widehat{X}$, i.e., (\ref{GRSJDDynamic}), are given by
	\begin{equation}
	\begin{split}
	\widehat{b}=&\sum_i I_{\{\alpha_{-}=i\}}\left(b^{(i)}+\int \left(x-h(x)\right)F^{(i)}(\D x)\right) -\int (x-h(x))\widehat{F}(\D x), \\
	\widehat{c}=&\sum_i I_{\{\alpha_{-}=i\}}c^{(i)},\\
	\widehat{F}_t(\D x)=&\underbrace{ M^{P}_{\widebar{\mu}}\left[I_{\{\gamma_\alpha\Delta J=0\}}\big| \widetilde{\cal{P}}\right](t,x)}_{\text{exclude common jumps}}\underbrace{\sum_i I_{\{\alpha(t-)=i\}}F^{(i)}_t(\D x)}_{\text{jumps from } \widebar{X}} \\&+ \underbrace{M^{P}_{\mu^{(J)}}\left[I_{\{\Delta \widebar{X}=0\}}\big| \widetilde{\cal{P}}\right](t,x)}_{\text{exclude common jumps}}\underbrace{\sum_iq_i(t)\delta_{\gamma_\alpha^{(i)}(t)}(\D x)}_{\text{jumps from } \gamma_\alpha\cdot \widetilde{J}} \\ 
	&+\sum_i \underbrace{M^{P}_{\widebar{\mu}}\left[I_{\{\gamma_\alpha^{(i)}\Delta J_i\neq 0\}}\big| \widetilde{\cal{P}}\right](t,x-\gamma_\alpha^{(i)}(t))}_{\text{gather common jumps}}\underbrace{\widebar{F}^{(i)}_t(\D x)}_{\text{superimposed jumps}}, \\
	\widehat{A}=A,  	\label{GRSJDTriplet}
	\end{split}
	\end{equation}
	where 
	\begin{align*}
	\widebar{F}^{(i)}_t( \mathscr{A}):=&\int_{\mathbb{R}^d}I_\mathscr{A}(x+\gamma_\alpha^{(i)}(t))I_{\{x+\gamma_\alpha^{(i)}(t)\neq 0 \}}\widebar{F}_t(\D x)\\
	=&\int_{\mathbb{R}^d}\sum_{j}I_\mathscr{A}(x+\gamma_{j}^{(i)}(t))I_{\{x+\gamma_{j}^{(i)}(t)\neq 0 \}}I_{\{\alpha(t-)=j\}}F^{(j)}_t(\D x),\ \forall \mathscr{A}\in \mathcal{B}(\mathbb{R}^d),\\
	\widebar{\mu}(\D t ,\D x):=&\sum_s I_{\{\Delta \widebar{X}(s)\neq0\}}\delta_{(s,\Delta \widebar{X}(s))}(\D t ,\D x),\\
	\mu^{(J)}(\D t,\D x):=&\sum_s I_{\{\gamma_\alpha(s)\Delta J(s)\neq 0\}}\delta_{(s,\gamma_\alpha(s)\Delta J(s))}(\D t ,\D x),
	\end{align*}
	$\gamma_\alpha^{(i)}$ denotes the $i^{\text{th}}$ column vector of the matrix $\gamma_\alpha$ and $h$ is a canonical truncation function on $\mathbb{R}^d $ defined as $h(x):=xI_{\{|x|\leq 1\}}$.
\end{prop}
The proof of the proposition is relegated to Appendix \ref{Appendix: PredictableCharacteristics}, while herein we discuss some remarks.
\begin{remark}
{\rm(a)} By \cite[Corollary II.2.38]{Jacod&Shiryaev2003}, the truncation function $h$ could always be chosen as the identity function $x$ if $X^{(i)}$ is special. This way $\widehat{b}=\widebar{b}$ and is interpreted as the drift process. Keeping this $h$ is convenient for our derivation to directly interact with the theorems in reference papers where $h$ appears in the expression. $B^{X^{(i)}}+(x-h)\star\nu^{(i)}$ is then the drift process of $X^{(i)}$ by  \cite[Proposition II.2.29-(a)]{Jacod&Shiryaev2003}. In the following results, one may always understand $h(x)=x$ whenever convenient and this changes nothing.\\
{\rm(b)} The condition that $X^{(i)}$ is special is technically used to provide a nice expression of $\widehat{b}$. This condition is not necessary when $\gamma_i\equiv 0,i\in\{1,2,...,D\}$.\\
{\rm(c)} If $A(t)\equiv t$, then the return process is quasi-left-continuous (see \cite[II.2.9(i)]{Jacod&Shiryaev2003}) and more precisely, an It\^o semimartingale.\\
{\rm(d)} We see in Proposition \ref{Proposition:GRSJDPredictableCharacteristics} that the use of the conditional expectation w.r.t. random measure is to filter out or select the common jumps in a predictable manner. This trick seems essential in its own interest.

\end{remark}

We also provide the predictable characteristics of the specific market (\ref{RSJDdynamic}) as a consequence of Proposition \ref{Proposition:GRSJDPredictableCharacteristics}.
\begin{corollary}\label{Corollary:RSJDPredictableCharacteristics}
	Suppose that $\widehat{X}$ follows the model (\ref{RSJDdynamic}). Then the predictable characteristics\\ $(\widehat{b},\widehat{c},\widehat{F},\widehat{A})$ take the form of
	\begin{equation}\label{RSJDTriplet}
	\begin{split}
	&\widehat{b}=\rho_\alpha-\int (x-h(x))\widehat{F}(\D x),\quad \widehat{c}=\sigma_\alpha\sigma_\alpha^{tr},\quad \widehat{A}(t)=t, \\
&\widehat{F}_t(\D x)=\sum_k\widebar{\lambda}_k(t,\alpha(t-))\delta_{\zeta_\alpha^{(k)}(t)}(\D x)
	+\sum_i\widebar{q}_i(t) \delta_{\gamma_\alpha^{(i)}(t)}(\D x) \\
	&+\sum_k\sum_i\frac{\D \langle\widetilde{N}_{\alpha k},\widetilde{J}_i\rangle(t)}{\D t}\delta_{(\zeta_\alpha^{(k)}(t)+\gamma_\alpha^{(i)}(t))}(\D x)  ,\\
	\end{split}  
	\end{equation}
	where $\zeta_\alpha^{(k)}(\gamma_\alpha^{(i)},\text{respectively})$ denotes the $k^{\text{th}}  (i^{\text{th}},\text{respectively})$ column vector of the matrix $\zeta_\alpha(\gamma_\alpha,\text{respectively})$, $h$ is a canonical truncation function on $\mathbb{R}^d $ defined as $h(x):=xI_{\{|x|\leq 1\}}$ and
	\begin{align*}
	\widebar{\lambda}_k(t,\alpha(t-)):=\lambda_k(\alpha_{-})-\sum_i\frac{\D \langle\widetilde{N}_{\alpha k},\widetilde{J}_i\rangle(t)}{\D t},\quad
	\widebar{q}_i:=q_i-\sum_k\frac{\D \langle\widetilde{N}_{\alpha k},\widetilde{J}_i\rangle(t)}{\D t}.
	\end{align*}
\end{corollary}

\section{Num\'eraire portfolio  and deflators}
This section addresses the num\'eraire portfolio for the switching model proposed in (\ref{GRSJDDynamic}). Precisely, we discuss its existence, its characterisation, and a closely related concept named deflator. In fact, thanks to \cite{Karatzas&Kardaras2007,Choullietal2015,Choulli&Yansori2022c}, the existence of num\'eraire portfolio for any model is equivalent to the existence of deflators for the model. Thus, we start this section by defining num\'eraire portfolio and deflator in the following. 
\begin{definition}\label{Definition: DeflatorNumeraireLogoptimal} Let $Y$ be an arbitrary $d$-semimartingale model and $Q$ be a probability. \\
{\rm(a)} The num\'eraire portfolio for the model $(Y,Q)$, when it exists, is $\widetilde{\pi}\in{L}(Y,Q)\cap{\cal{L}}(Y,Q)$ such that ${\cal{E}}(\pi\cdot Y) \slash {\cal{E}}({\widetilde{\pi}}\cdot{Y})$ is a $Q$-supermartingale for any $\pi\in{L}(Y,Q)\cap{\cal{L}}(Y,Q)$. \\
{\rm(b)} A  deflator for the model $(Y,Q)$ is any positive process $Z$ such that $Z{\cal{E}}({\pi}\cdot{Y})$ is a supermartingale for any portfolio $\pi\in{L}(Y,Q)\cap{\cal{L}}(Y,Q)$. The set of all deflators for the model $Y$ will be denoted by $\mathcal{D}(Y,Q)$.\\
{\rm(c)} A local martingale deflator for the model $(Y,Q)$ is any positive process $Z$ with $Z_0=1$ such that $Z$ and $Z(\varphi\cdot Y)$ are $Q$-local martingales for some predictable scalar process $\varphi$ with $0<\varphi\leq1$. The set of all local martingale deflators for the model $Y$ will be denoted by $\mathcal{Z}_{loc}(Y,Q)$.\\
When $Q=P$, we simply omit the probability in the notation  and terminologies for the sake of simplification. \end{definition}
Throughout the paper, equality and uniqueness of portfolios for a given market are understood up to null investments.
In order to describe the set of deflators and the num\'eraire portfolio for the general switching model (\ref{GRSJDDynamic}), we will use the following lemma that simplifies tremendously the task.
\begin{lemma}\label{Theorem: SwitchingDeflator} Let $X$ be a $d$-dimensional semimartingale. Then $Z\in\mathcal{D}(X)$ if and only if for any $i\in\{1,2,...,D\}$, there exists $Z^{(i)}\in\mathcal{D}(I_{\{\alpha_{-}=i\}}\cdot X)$ satisfying the following
\begin{equation*}
 Z=\prod_{i=1}^D Z^{(i)}\quad\mbox{and}\quad I_{\{\alpha_{-}\not=i\}}\cdot Z^{(i)}=0.\end{equation*}
\end{lemma}
Lemma \ref{Theorem: SwitchingDeflator} brings an insight that the structure of a switching market can always be studied regime by regime. This rule is pivotal everywhere, and provides a quick guide in analysis of switching market by decomposing it into each regime. This motivates us to consider the division as below.  \begin{equation*}
\widehat{X}=\sum_{i=1}^DI_{\{\alpha_{-}=i\}}\cdot\left(X^{(i)}+\gamma_i\cdot\widetilde{J}\right),
\end{equation*}

By similar computations to (\ref{GRSJDTriplet}), we have
\begin{corollary}\label{Corollary: TGJDPredictableCharacteristics} 
	Suppose that $X^{(i)}$ is special for each $i\in\{1,2,...,D\}$. Then the predictable characteristics of $\widehat{X}^{(i)}:=I_{\{\alpha_{-}=i\}}\cdot (X^{(i)}+\gamma_i\cdot\widetilde{J})$, denoted by $(\widehat{b}^{(i)},\widehat{c}^{(i)},\widehat{F}^{(i)})$, are given by 
	\begin{equation}\label{TGJDTriplet} 
	\begin{split}
	\widehat{b}^{(i)}=&I_{\{\alpha_{-}=i\}}\left(b^{(i)}+\int (x-h(x))F^{(i)}(\D x)\right) -\int (x-h(x))\widehat{F}^{(i)}(\D x),\quad \widehat{c}^{(i)}=I_{\{\alpha_{-}=i\}} c^{(i)},  \quad \widehat{A}^{(i)}=A, \\
	\widehat{F}^{(i)}_t(\D x)=&I_{\{\alpha_{-}=i\}}\bigg( M^{P}_{\mu^{(i)}}\left[I_{\{\gamma_i\Delta J=0\}}\big| \widetilde{\cal{P}}\right](t,x)F^{(i)}_t(\D x)+ M^{P}_{\mu^{(Ji)}}\left[I_{\{\Delta X^{(i)}=0\}}\big| \widetilde{\cal{P}}\right](t,x)\sum_{j}q_{j}(t)\delta_{\gamma_i^{(j)}(t)}(\D x) \\ 
	&+\sum_{j} M^{P}_{\mu^{(i)}}\left[I_{\{\gamma_i^{(j)}\Delta J_{j}\neq 0\}}\big| \widetilde{\cal{P}}\right](t,x-\gamma_i^{(j)}(t))F^{(ij)}_t(\D x)\bigg), \\ 
	\end{split}
	\end{equation}
	where 
	\begin{equation*}
	\begin{split}
	F^{(ij)}_t( \mathscr{A}):=&\int_{\mathbb{R}^d}I_\mathscr{A}(x+\gamma_i^{(j)}(t))I_{\{x+\gamma_i^{(j)}(t)\neq 0 \}}F^{(i)}_t(\D x),\\
	\mu^{(ij)}([0,t]\times\mathscr{A}):=&\int_{[0,t]\times\mathbb{R}^d}I_\mathscr{A}(x+\gamma_i^{(j)}(s))I_{\{x+\gamma_i^{(j)}(s)\neq 0\}}\mu^{(i)}(\D s,\D x),\\
	\mu^{(Ji)}(\D t,\D x):=&\sum_s I_{\{\gamma_i(s)\Delta J(s)\neq 0\}}\delta_{(s,\gamma_i(s)\Delta J(s))}(\D t ,\D x),
	\end{split}
	\end{equation*}
	and as before $\gamma_i^{(j)}$ denotes the $j^{\text{th}}$ column of the matrix $\gamma_i$.
	The common-jump part
	$$\sum_{0<s\leq\cdot}I_{\{\Delta X_s^{(i)}\neq 0,\ \gamma_i^{(j)}(s)\Delta J_j(s)\neq 0\}}\left(\Delta X_s^{(i)}+\gamma_i^{(j)}(s)\Delta J_j(s)\right)$$
	has jump measure $I_{\{\gamma_i^{(j)}\Delta J_j\neq 0\}}.\mu^{(ij)}$.
\end{corollary}
\begin{proof}
	The condition that $X^{(i)}$ is special implies that $I_{\{\alpha_{-}=i\}}\cdot X^{(i)}$ is also special by applying \cite[I.3.5 and I.4.34]{Jacod&Shiryaev2003} to the canonical decomposition of $X^{(i)}$. And then the result is derived very similarly to Proposition \ref{Proposition:GRSJDPredictableCharacteristics}, see Appendix \ref{Appendix: PredictableCharacteristics} for full details. This ends the proof of the corollary. 
\end{proof}

Upon studying the num\'eraire portfolio, another technical lemma is given below.
\begin{lemma}\label{Lemma: SwitchingConditionalExpectation}
	Suppose that there are $D$ random measures $\mu^{i},i\in\{1,2,...,D\}$ with their compensator measures $\nu^{i},i\in\{1,2,...,D\}$.
	Define a switching random measure $\widebar\mu$ and its compensator $\widebar\nu$ \begin{equation*}
	\widebar{\mu}(dt,dx):=\sum_i I_{\{\alpha_{t-}=i\}}\mu^{i}(dt,dx),\quad \widebar{\nu}(dt,dx):=\sum_i I_{\{\alpha_{t-}=i\}}\nu^{i}(dt,dx).
	\end{equation*}
	 Then 
	\begin{equation*}
	M_{\widebar{\mu}}\left[U\big|\widetilde{\cal{P}} \right]=\sum_{i=1}^D I_{\{\alpha_{-}=i\}}M_{\mu^{i}}\left[U\big|\widetilde{\cal{P}} \right],\ P\otimes \widebar{\nu}-a.e.,
	\end{equation*}
	for $i\in\{1,2,...,D\}$ and any nonnegative $U\in\widetilde{\cal{O}}$. 
\end{lemma}
The proof of the lemma is straightforward and will be omitted herein. We now provide one of our main theorems of this section. 

\begin{theorem} \label{Theorem:GRSJDNumerairePortfolio}
	The following assertions are equivalent.\\
{\rm(a)} The num\'eraire portfolio for $\widehat{X}$,  denoted by $\widetilde{\pi}_\alpha$, exists.\\
{\rm(b)} For each state $i\in\{1,2,...,D\}$, the num\'eraire portfolio for $I_{\{\alpha_{-}=i\}}\cdot(X^{(i)}+\gamma_i\cdot  \widetilde{J})$, denoted by $\widetilde{\pi}_i$, exists. \\
Furthermore, when one of these assertions holds, then the portfolios $\widetilde{\pi}_\alpha$ and $(\widetilde{\pi}_i)_{i=1,...,D}$ satisfy 
\begin{equation}\label{Relationship}
\widetilde{\pi}_\alpha=\sum_{i=1}^D{I}_{\{\alpha_{-}=i\}}\widetilde{\pi}_i.\end{equation}
\end{theorem}
\begin{proof}
....
\end{proof}
This theorem clearly simplifies our study by focusing on each model $I_{\{\alpha_{-}=i\}}\cdot(X^{(i)}+\gamma_i\cdot  \widetilde{J})$. Thus for the rest, for the sake of simplifying the notation, we consider a $d$-dimensional semimartingale $X$ with its random measure of its jumps $\mu$ and the compensator of it $\nu$ and the predictable characteristics $(b,c,F,A)$.

For each $i\in\{1,2,...,D\}$, $\widetilde{\pi}_i\in\mathcal{L}(X^{(i)}+\gamma_i\cdot \widetilde{J})$
 and 	
 \begin{equation}\label{TGJDNumeraireEquation}
		\begin{split}
		&(\pi-\widetilde{\pi}_i)^{tr}\chi_i(\widetilde{\pi}_i)\leq 0,\quad \text{on}\quad\{\alpha_{-}=i\}, \quad\mbox{for any}\ \pi\in\mathcal{L}(X^{(i)}+\gamma_i\cdot \widetilde{J})
		\end{split}
		\end{equation}
		where, for any $\pi\in \mathcal{L}(X^{(i)}+\gamma_i\cdot \widetilde{J})$, the formal operator $\chi_i$ \footnote{Here we abuse the expression for neatness. The integration in $\chi_i$ may not make sense. But  $(\pi-\widetilde{\pi}_i)^{tr}\chi_i(\widetilde{\pi}_i)$ is understood as letting $(\pi-\widetilde{\pi}_i)^{tr}$ go inside the integration, which is then always well-defined. See  \cite[Remark 2.1-(b)]{Choulli&Yansori2022c} and \cite[Section 3.2]{Karatzas&Kardaras2007} for more explanations.} is given by
		\begin{equation*}
		\begin{split}
		\chi_i(\pi):=&b^{(i)}+\int (x-h(x))F^{(i)}(\D x)-c^{(i)}\pi
		-\int M^{P}_{\mu^{(i)}}\left[I_{\{\gamma_i\Delta J= 0\}}\big| \widetilde{\cal{P}}\right](x)x(1-\dfrac{1}{1+\pi^{tr}x})F^{(i)}(\D x)\\
		-&\sum_{j} q_{j} M^{P}_{\mu^{(Ji)}}\left[I_{\{\Delta X^{(i)}=0\}}\big| \widetilde{\cal{P}}\right](\gamma_i^{(j)})\gamma_i^{(j)}(1-\dfrac{1}{1+\pi^{tr}\gamma_i^{(j)}})\notag \\
		&-\sum_{j}\int \left(M^{P}_{\mu^{(i)}}\left[I_{\{\gamma_i^{(j)}\Delta J_{j}\neq 0\}}\big| \widetilde{\cal{P}}\right](x)(x+\gamma_i^{(j)})(1-\dfrac{1}{1+\pi^{tr}(x+\gamma_i^{(j)})})\right)F^{(i)}(\D x),\\ 
		\end{split}
		\end{equation*}
		and as before $\gamma_i^{(j)}$ denotes the $j^{\text{th}}$ column of the matrix $\gamma_i$.\\

Theorem \ref{Theorem:GRSJDNumerairePortfolio} is practical in two ways:\\
{\rm(a)} First it instructs how to find a num\'eraire portfolio in the general switching model with broad applicability: examining the inequality in the form of (\ref{TGJDNumeraireEquation}) with the explicit expression of the operator $\chi$ given.\\
{\rm(b)} Second, it guarantees a guideline in studying the market structure with switching: one only needs to check the transformed market $\widehat{X}^{(i)}=I_{\{\alpha_{-}=i\}}\cdot (X^{(i)}+\gamma_i\cdot \widetilde{J})$ one by one. In the sense of integrand, this is a partition of values into different regimes, which usually simplifies the analysis especially in some certain cases.

There are some special cases simplifying the expression in (\ref{TGJDNumeraireEquation}).
\begin{corollary}
	The following assertions hold.\\
{\rm{(a)}} If $[\gamma_\alpha\cdot J,\widebar{X}]\equiv0$ (i.e. no common jumps), then 
		\begin{equation*}
		M^{P}_{\mu^{(i)}}\left[I_{\{\gamma_i\Delta J=0\}}\big| \widetilde{\cal{P}}\right]\equiv 1,\quad M^{P}_{\mu^{(Ji)}}\left[I_{\{\Delta X^{(i)}=0\}}\big| \widetilde{\cal{P}}\right]\equiv 1,\quad M^{P}_{\mu^{(i)}}\left[I_{\{\gamma_i^{(j)}\Delta J_{j}\neq 0\}}\big| \widetilde{\cal{P}}\right]\equiv 0,
		\end{equation*} 
		and the operator $\chi_i$ defined in (\ref{TGJDNumeraireEquation}) becomes
		\begin{equation}\label{TGJDNumeraireEquation1}
		\begin{split}
		&\chi_i(\widetilde{\pi}_i)=b^{(i)}-c^{(i)}\widetilde{\pi}_i+\int \left(\dfrac{x}{1+\widetilde{\pi}_i^{tr}x}-h(x)\right)F^{(i)}(\D x)
		-\sum_j q_j \dfrac{\widetilde{\pi}_i^{tr}\gamma_i^{(j)}}{1+\widetilde{\pi}_i^{tr}\gamma_i^{(j)}}\gamma_i^{(j)}.
		\end{split}
		\end{equation}
		
		If $\gamma_\alpha\equiv0 $, which means $ \widehat{X}\equiv \widebar{X}$, then
		\begin{equation*}
		M^{P}_{\mu^{(i)}}\left[I_{\{\gamma_i\Delta J=0\}}\big| \widetilde{\cal{P}}\right]\equiv 1,\quad M^{P}_{\mu^{(Ji)}}\left[I_{\{\Delta X^{(i)}=0\}}\big| \widetilde{\cal{P}}\right]\equiv 0,\quad M^{P}_{\mu^{(i)}}\left[I_{\{\gamma_i^{(j)}\Delta J_{j}\neq 0\}}\big| \widetilde{\cal{P}}\right]\equiv 0.
		\end{equation*} 
		and the operator $\chi_i$ in (\ref{TGJDNumeraireEquation}) becomes
		\begin{equation}\label{TGJDNumeraireEquation2}
		\begin{split}
			&\chi_i(\widetilde{\pi}_i)=b^{(i)}-c^{(i)}\widetilde{\pi}_i+\int \left(\dfrac{x}{1+\widetilde{\pi}_i^{tr}x}- h(x)\right)F^{(i)}(\D x).
		\end{split}
		\end{equation}
		In this case, we don't need the condition that $X^{(i)}$ is special to have  (\ref{TGJDNumeraireEquation}) hold.\\
		{\rm{(b)}} If $F^{(i)}\equiv0$ for any $i\in\{1,2,...,D\}$, then the natural constraint set (\ref{NaturalConstraint}) is open. So the inequality (\ref{TGJDNumeraireEquation}) is forced to be the following equality 
		\begin{equation} \label{TGJDNumeraireEquation3}
		\begin{split}
		\chi_i(\widetilde{\pi}_i)=0,\quad  \text{on}\quad \{\alpha_{-}=i\},
		\end{split}
		\end{equation}
		and the operator $\chi_i$ in (\ref{TGJDNumeraireEquation}) becomes
		\begin{equation*}
		\begin{split}
		&\chi_i(\widetilde{\pi}_i)=b^{(i)}-c^{(i)}\widetilde{\pi}_i-\sum_j q_j \gamma_i^{(j)}\left(1-\dfrac{1}{1+\widetilde{\pi}_i^{tr}\gamma_i^{(j)}}\right).
		\end{split}
		\end{equation*}
		{\rm{(c)}}  If the return $\widehat{X}$ follows the specific form  in (\ref{RSJDdynamic}), then recall  from (\ref{JDTriplet}),
		\begin{equation*}
		\begin{split}
		b^{(i)}=&\rho_i-\int (x-h(x))F^{(i)}(\D x) ,\quad  c^{(i)}=\sigma_i\sigma_i^{tr},   \\
		F^{(i)}=&F^{(i)}_t(\D x)=\sum_k\lambda_k(i)\delta_{\zeta_i^{(k)}(t)}(\D x), \quad	A^{(i)}(t)=t. 
		\end{split}
		\end{equation*}
		As the natural constraint set (\ref{NaturalConstraint}) is open, the inequality in Proposition \ref{Proposition: Numeraire} (a) applied to
$$(\widehat{b}^{(i)},\widehat{c}^{(i)},\widehat{F}^{(i)})$$
from (\ref{TGJDTriplet}) is forced to be an equality. And it is similarly decomposed into each regime as
		\begin{equation} \label{TJDNumeraireEquation}
		\chi_i(\widetilde{\pi}_i)=0,\quad  \text{on}\quad \{\alpha_{-}=i\},
		\end{equation}
		where
		\begin{equation*}
		\begin{split}
		\chi_i(\widetilde{\pi}_i)=&\rho_i-\zeta_i\lambda(i)-\gamma_i q-\sigma_i\sigma_i^{tr}\widetilde{\pi}_i\\
		+&\sum_k\widebar{\lambda}_k(t,i)\dfrac{\zeta_i^{(k)}}{1+\widetilde{\pi}_i^{tr}\zeta_i^{(k)}}
		+\sum_{j}\widebar{q}_{j} \dfrac{\gamma_i^{(j)}}{1+\widetilde{\pi}_i^{tr}\gamma_i^{(j)}} 
		+\sum_{k}\sum_{j}\frac{\D \langle\widetilde{N}_{i k},\widetilde{J}_{j}\rangle}{\D t}\dfrac{\zeta_i^{(k)}+\gamma_i^{(j)}}{1+\widetilde{\pi}_i^{tr}(\zeta_i^{(k)}+\gamma_i^{(j)})} . 
		\end{split}
		\end{equation*}
		\end{corollary}

As is demonstrated in Theorem \ref{Theorem:GRSJDNumerairePortfolio}, the common jumps between $\widebar{X}$ and $\gamma_\alpha\cdot\widetilde{J}$ are the main factor in disturbing the structure of the market and its deflators. In fact, it is clear that when there are no common jumps we have the following. 
\begin{corollary}\label{Corollary: StableTS}
	Suppose that $[X^{(i)},\gamma_i\cdot\widetilde{J}]\equiv 0$,  for each $i\in\{1,2,...,D\}$. Then $\mathcal{D}(X^{(i)})\neq\varnothing$ for any $i\in\{1,...,D\}$ implies that $\mathcal{D}(\widehat{X})\neq\varnothing$.
\end{corollary}
The converse does not hold, as the following example shows.
\begin{example}
	$d=1,D=2,q_{12}=q_{21}=q'>0$. Consider the model (\ref{RSJDdynamic}) with constant coefficients.
	\begin{align*}
	& \rho_\alpha>0,\quad \sigma_\alpha=0,\quad\gamma_\alpha<0,\quad\zeta_\alpha=0,\quad \rho_\alpha=0 \quad \text{has no solution, while\ } \\
	&\rho_1-\gamma_1^{(2)} q'+q'\frac{\gamma_1^{(2)}}{1+\widetilde{\pi}_1\gamma_1^{(2)}}=0,\quad\text{and} \quad\rho_2-\gamma_2^{(1)} q'+q'\frac{\gamma_2^{(1)} }{1+\widetilde{\pi}_2\gamma_2^{(1)}}=0, \quad\text{have solutions\ }\\
	&\widetilde{\pi}_1=\frac{\rho_1}{(\gamma_1^{(2)}q'-\rho_1)\gamma_1^{(2)}},\quad 1+\widetilde{\pi}_1\gamma_1^{(2)}=\frac{\gamma_1^{(2)}q'}{\gamma_1^{(2)}q'-\rho_1}>0.\\
	&\widetilde{\pi}_2=\frac{\rho_2}{(\gamma_2^{(1)}q'-\rho_2)\gamma_2^{(1)}},\quad 1+\widetilde{\pi}_2\gamma_2^{(1)}=\frac{\gamma_2^{(1)}q'}{\gamma_2^{(1)}q'-\rho_2}>0.
	\end{align*}
	According to Proposition \ref{Proposition: Numeraire}-(b), the above computations show that  $\mathcal{D}(\widehat{X})\neq\varnothing$, but $\mathcal{D}(\widebar{X})=\mathcal{D}(X^{(i)})=\varnothing$ for $i\in\{1,2\}$.
\end{example}

Since $\gamma_\alpha\cdot\widetilde{J}\in\mathcal{M}_{loc}$, the finite jump activity of $J$ and the boundedness of $\gamma_\alpha$ imply that 
$$\sum_{s\leq\cdot}|\gamma_\alpha(s)\Delta J_s|\in\mathcal{A}_{loc}^+,\quad |x|\star\nu^{(J)}\in\mathcal{A}_{loc}^+.$$
Therefore, throughout the paper, we consider  the following decomposition of the model (\ref{GRSJDDynamic}), where we single out the common jumps. 
\begin{equation}\label{Decomposition4Xhat}
\begin{split}
\widehat{X}=&\widecheck{X}+\underline{J}=\sum_{i=1}^D \left(\widecheck{X}^{(i)}+\underline{J}^{(i)}\right), \\
\widecheck{X}:=&\widebar{X}+(I_{\{\Delta \widebar{X}\neq 0\}}\gamma_\alpha)\cdot J-\left((I_{\{\Delta \widebar{X}\neq 0\}}\gamma_\alpha)\cdot J\right)^p,\quad\quad\quad \widecheck{X}^{(i)}:=I_{\{\alpha_{-}=i\}}\cdot\widecheck{X},\ i\in\{1,2,...,D\},\\	
\underline{J}:=&\widehat{X}-\widecheck{X}=(\gamma_\alpha I_{\{\Delta \widebar{X}=0\}})\cdot J-\left((\gamma_\alpha I_{\{\Delta \widebar{X}=0\}})\cdot J\right)^p,\quad\quad \underline{J}^{(i)}:=I_{\{\alpha_{-}=i\}}\cdot\underline{J},\ i\in\{1,2,...,D\}.
\end{split}
\end{equation} 
The associated measures $\widecheck{F}$, $\widecheck{\nu}$ and $\widecheck{\mu}$ are the corresponding kernel, jump compensator and jump measure of $\widecheck{X}$.

\begin{theorem}\label{Theorem: SwitchingShockComparison}
Suppose that $\gamma_i$ is locally bounded for each $i\in\{1,2,...,D\}$.

{\rm (a)} Let
$$\nu^{\underline{J}}=\theta\widecheck{\nu}+\nu^{(\underline{J},\perp)},\quad\nu^{(\underline{J},\perp)}\perp\widecheck{\nu}$$
be the Lebesgue decomposition, where
$$\nu^{(\underline{J},\perp)}(\D t,\D x)=F^{(\underline{J},\perp)}_t(\D x)\D A_t.$$
Then
\begin{equation*}
\begin{split}
&\mathcal{Z}_{loc}(\widehat{X})\neq\varnothing\\
\Longleftrightarrow\quad&
\mathcal{Z}_{loc}\left(\widecheck{X}-x\star\nu^{\underline{J}}
+\left(xI_{\mbox{\rm supp}F^{(\underline{J},\perp)}}\right)\star\mu^{\underline{J}}\right)\neq\varnothing\\
\Longleftrightarrow\quad&
\exists f\in\widetilde{\cal{P}}:\quad\mathcal{Z}_{loc}\left(\widecheck{X}-x\star\nu^{\underline{J}}
+(xf)\star\nu^{(\underline{J},\perp)}\right)\neq\varnothing,
\end{split}
\end{equation*}
where
\begin{equation*}
\begin{split}
&f>0\quad P\otimes\nu^{(\underline{J},\perp)}\mbox{-a.e.},\quad f\star\nu^{(\underline{J},\perp)}_T<\infty\quad P\mbox{-a.s.}.
\end{split}
\end{equation*}

{\rm (b)} On any finite horizon $[0,T]$, for each $i\in\{1,2,...,D\}$,
$$\mathcal{Z}_{loc}(\widecheck{X}^{(i)})\neq\varnothing
\quad\Longleftrightarrow\quad
\bigcup_{r\in\mathcal{R}_i}\left[\mathcal{Z}_{loc}(X^{(i,r)})\cap\bigcap_{j\neq i}\mathcal{Z}_{loc}(\mathcal{J}^{(ij)}-r_{ij}\cdot A)\right]\neq\varnothing,$$
where
$$\mathcal{J}^{(ij)}_t:=\sum_{0<s\leq t}I_{\{\alpha_{s-}=i,\alpha_s=j,\Delta X^{(i)}_s\neq0\}},\quad j\neq i,$$
$$\mathfrak{q}_{ij}(t):=I_{\{\alpha_{t-}=i\}}\int M^{P}_{\mu^{(i)}}\left[I_{\{\alpha_t=j\}}\big|\widetilde{\cal{P}}\right](t,x)F^{(i)}_t(\D x),$$
$$X^{(i,r)}:=I_{\{\alpha_{-}=i\}}\cdot X^{(i)}+\sum_{j\neq i}\gamma_i^{(j)}\cdot\left((r_{ij}-\mathfrak{q}_{ij})\cdot A\right),$$
and $\mathcal{R}_i$ is the set of predictable $r=(r_{ij})_{j\neq i}$ satisfying
$$r_{ij}\geq0,\quad\{r_{ij}>0\}=\{\mathfrak{q}_{ij}>0\}\quad P\otimes\D A\mbox{-a.e.},\quad\sum_{j\neq i}(r_{ij}\cdot A)_T<\infty\quad P\mbox{-a.s.}.$$
For $r\in\mathcal{R}_i$ and $j\neq i$,
$$ (\mathcal{J}^{(ij)})^p=\mathfrak{q}_{ij}\cdot A,\quad0\leq\mathfrak{q}_{ij}\D A\leq I_{\{\alpha_{-}=i\}}q_{ij}\D t,\quad\Delta(r_{ij}\cdot A)=0,$$
and
\begin{equation}\label{CommonJumpIntensityDecomposition}
\widecheck{X}^{(i)}=X^{(i,r)}+\sum_{j\neq i}\gamma_i^{(j)}\cdot(\mathcal{J}^{(ij)}-r_{ij}\cdot A).
\end{equation}
\end{theorem}

\begin{remark}
The local boundedness of $\gamma_i$ is not essential. Put
$$\varphi:=\left(1+\sum_{i=1}^D\sum_{j=1}^D|\gamma_i^{(j)}|\right)^{-1}.$$
Then Theorem \ref{Theorem: SwitchingShockComparison} applies to $(\varphi\cdot X^{(i)},\varphi\gamma_i)$, $i\in\{1,2,...,D\}$, since $|\varphi\gamma_i^{(j)}|\leq1$. Moreover,
$$\mathcal{Z}_{loc}(\varphi\cdot Y)=\mathcal{Z}_{loc}(Y)$$
for every semimartingale $Y$.
\end{remark}

\begin{proof}
{\rm (a)} Take
$$ X:=\widecheck{X},\quad
\widehat{X}=X+\underline{J},\quad [X,\underline{J}]=0,$$
and write
$$\nu^{\underline{J}}=\theta\nu^X+\nu^{(\underline{J},\perp)},\quad
\underline{J}=x\star(\mu^{\underline{J}}-\nu^{\underline{J}}),\quad
\nu^{(\underline{J},\perp)}(\D t,\D x)=F^{(\underline{J},\perp)}_t(\D x)\D A_t.$$
The jumps of $\underline{J}$ occur only at the jumps of $J$, and
$$1\star\nu^{\underline{J}}_T\leq T\max_i\sum_{j\neq i}q_{ij},\quad
\nu^{\underline{J}}(\{t\},\D x)=0.$$
Since $\gamma_\alpha$ is locally bounded, $\Delta\underline{J}$ is locally bounded.

Set
$$X_1:=X-x\star\nu^{\underline{J}}+\left(xI_{\mbox{\rm supp}F^{(\underline{J},\perp)}}\right)\star\mu^{\underline{J}},\quad p:=\frac{\D(\theta\nu^X)}{\D\widehat{\nu}},\quad Z:=\mathcal{E}\left(-p\star(\widehat{\mu}-\widehat{\nu})\right).$$
Since
$$\widehat{\nu}=(1+\theta)\nu^X+\nu^{(\underline{J},\perp)},\quad 0\leq p<1,\quad p\star\widehat{\nu}=1\star(\theta\nu^X)\leq1\star\nu^{\underline{J}},$$
we have
$$0<Z_t\leq\exp\left(p\star\widehat{\nu}_t\right)\leq\exp\left(T\max_i\sum_{j\neq i}q_{ij}\right),\quad0\leq t\leq T.$$
Thus $Z$ is a uniformly integrable martingale on $[0,T]$, and $\D\mathbb{Q}:=Z_T\D P$ gives $\mathbb{Q}\sim P$.

With $h(x)=x$, Girsanov's theorem gives
$$\left(\widehat{B}^{\mathbb{Q}},\widehat{C}^{\mathbb{Q}},\widehat{\nu}^{\mathbb{Q}}\right)=\left(B^X-x\star(\theta\nu^X),C^X,\nu^X+\nu^{(\underline{J},\perp)}\right)=\left(B^{X_1,P},C^{X_1,P},\nu^{X_1,P}\right).$$
Consequently, invariance under equivalent probabilities and \cite[Theorem A.1]{Aksamitetal2017} yield
$$\mathcal{Z}_{loc}(\widehat{X})\neq\varnothing
\quad\Longleftrightarrow\quad\mathcal{Z}_{loc}(\widehat{X},\mathbb{Q})\neq\varnothing
\quad\Longleftrightarrow\quad\mathcal{Z}_{loc}(X_1)\neq\varnothing.$$

For $f\in\widetilde{\cal{P}}$ with $f>0$ and $f\star\nu^{(\underline{J},\perp)}_T<\infty$ $P$-a.s., set
$$\mu^{(\underline{J},\perp)}:=I_{\mbox{\rm supp}F^{(\underline{J},\perp)}}\cdot\mu^{\underline{J}},\quad X_2:=X-x\star\nu^{\underline{J}}+(xf)\star\nu^{(\underline{J},\perp)}.$$
It remains to prove
$$\mathcal{Z}_{loc}(X_1)\neq\varnothing\quad\Longleftrightarrow\quad\exists f\in\widetilde{\cal{P}},\quad f>0,\quad f\star\nu^{(\underline{J},\perp)}_T<\infty,\quad\mathcal{Z}_{loc}(X_2)\neq\varnothing.$$

For necessity, suppose that $\mathcal{Z}_{loc}(X_1)\neq\varnothing$. By \cite[Theorem A.1]{Aksamitetal2017} and \cite[Theorem B.1]{Choulli&Yansori2022c}, choose $\mathcal{E}(N)\in\mathcal{Z}_{loc}(X_1)$ with its Jacod component generated by $X_1$. Splitting $N$ on $\mbox{\rm supp}F^{(\underline{J},\perp)}$ gives
\begin{equation*}
\begin{split}
&N=N_1+N_2,\quad
N_2=(f-1)\star\left(\mu^{(\underline{J},\perp)}-\nu^{(\underline{J},\perp)}\right),\quad f>0,\\
&N_1,N_2\in\mathcal{M}_{loc},\quad
\Delta N_1\in\{0,\Delta N\},\quad\Delta N_1>-1,\quad\Delta N_2>-1,\\
&
[N_1,N_2]=[X,N_2]=[N_1,x\star\mu^{(\underline{J},\perp)}]=0.
\end{split}
\end{equation*}
Since $\nu^{(\underline{J},\perp)}(\{t\},\D x)=0$, the predictable-jump terms remain in $N_1$.

By the Burkholder--Davis--Gundy inequality,
$$\sqrt{(f-1)^2\star\mu^{(\underline{J},\perp)}}\in\mathcal{A}_{loc}^+.$$
Since $1\star\mu^{(\underline{J},\perp)}$ is locally bounded,
$$|f-1|\star\mu^{(\underline{J},\perp)}\leq\sqrt{1\star\mu^{(\underline{J},\perp)}}\sqrt{(f-1)^2\star\mu^{(\underline{J},\perp)}}\in\mathcal{A}_{loc}^+.$$
Taking predictable compensators (\cite[Proposition II.1.28]{Jacod&Shiryaev2003}) gives
$$|f-1|\star\nu^{(\underline{J},\perp)}\in\mathcal{A}_{loc}^+,\quad
f\star\nu^{(\underline{J},\perp)}_T
\leq|f-1|\star\nu^{(\underline{J},\perp)}_T
+1\star\nu^{(\underline{J},\perp)}_T<\infty\quad P\mbox{-a.s.}.$$
For the strictly positive bounded predictable scalar process $\varphi$ in the definition of a local martingale deflator,
\begin{equation*}
\begin{split}
\varphi\cdot X_1+[\varphi\cdot X_1,N]
&=\varphi\cdot\left(X-x\star\nu^{\underline{J}}\right)
+[\varphi\cdot X,N_1]+(\varphi xf)\star\mu^{(\underline{J},\perp)}\\
&=\varphi\cdot X_2+[\varphi\cdot X_2,N_1]
+(\varphi xf)\star\left(\mu^{(\underline{J},\perp)}-\nu^{(\underline{J},\perp)}\right).
\end{split}
\end{equation*}
Since $\Delta\underline{J}$ is locally bounded and $f\star\nu^{(\underline{J},\perp)}_T<\infty$,
$$ (\varphi xf)\star\left(\mu^{(\underline{J},\perp)}-\nu^{(\underline{J},\perp)}\right)\in\mathcal{M}_{loc},\quad\mathcal{E}(N_1)\in\mathcal{Z}_{loc}(X_2).$$

For sufficiency, let $f\in\widetilde{\cal{P}}$ satisfy $f>0$, $f\star\nu^{(\underline{J},\perp)}_T<\infty$ and $\mathcal{Z}_{loc}(X_2)\neq\varnothing$. Choose $\mathcal{E}(N_1)\in\mathcal{Z}_{loc}(X_2)$ with its Jacod component generated by $X_2$, and set
$$N_2:=(f-1)\star\left(\mu^{(\underline{J},\perp)}-\nu^{(\underline{J},\perp)}\right).$$
Since
$$|f-1|\star\nu^{(\underline{J},\perp)}_T
\leq(f+1)\star\nu^{(\underline{J},\perp)}_T<\infty,$$
and $\nu^{(\underline{J},\perp)}(\{t\},\D x)=0$, \cite[Proposition II.1.28]{Jacod&Shiryaev2003} gives $N_2\in\mathcal{M}_{loc}$ with $\Delta N_2>-1$. Also,
$$[N_1,N_2]=[X,N_2]=[N_1,x\star\mu^{(\underline{J},\perp)}]=0,\quad
\mathcal{E}(N_1+N_2)=\mathcal{E}(N_1)\mathcal{E}(N_2)>0.$$
Choose a strictly positive bounded predictable scalar process $\varphi$ such that $\varphi\cdot X_2+[\varphi\cdot X_2,N_1]\in\mathcal{M}_{loc}$. Then
$$\varphi\cdot X_1+[\varphi\cdot X_1,N_1+N_2]=\varphi\cdot X_2+[\varphi\cdot X_2,N_1]+(\varphi xf)\star\left(\mu^{(\underline{J},\perp)}-\nu^{(\underline{J},\perp)}\right)\in\mathcal{M}_{loc},$$
so $\mathcal{E}(N_1+N_2)\in\mathcal{Z}_{loc}(X_1)$.

{\rm (b)} For necessity, let $Z=\mathcal{E}(N)\in\mathcal{Z}_{loc}(\widecheck{X}^{(i)})$. By Definition \ref{Definition: DeflatorNumeraireLogoptimal}-(c) and integration by parts, there is $\varphi\in\mathcal{P}$ such that
$$N\in\mathcal{M}_{loc},\quad\Delta N>-1,\quad0<\varphi\leq1,\quad\varphi\cdot(\widecheck{X}^{(i)}+[\widecheck{X}^{(i)},N])\in\mathcal{M}_{loc}.$$

To obtain $Z\in\bigcap_{j\neq i}\mathcal{Z}_{loc}(\mathcal{J}^{(ij)}-r_{ij}\cdot A)$, we construct $r\in\mathcal{R}_i$ satisfying
$$\mathcal{J}^{(ij)}+[\mathcal{J}^{(ij)},N]-r_{ij}\cdot A\in\mathcal{M}_{loc},\quad j\neq i.$$
For $j\neq i$, the Burkholder--Davis--Gundy inequality and the local boundedness of $\mathcal{J}^{(ij)}$ give
$$0\leq\mathcal{J}^{(ij)}+[\mathcal{J}^{(ij)},N]=\sum_{0<s\leq\cdot}(1+\Delta N_s)\Delta\mathcal{J}^{(ij)}_s\leq\mathcal{J}^{(ij)}+\sqrt{\mathcal{J}^{(ij)}[N,N]}\in\mathcal{A}_{loc}^+.$$
Define
$$r_{ij}\cdot A:=(\mathcal{J}^{(ij)}+[\mathcal{J}^{(ij)},N])^p\in\mathcal{A}_{loc}^+.$$
Since $1+\Delta N>0$ and $(\mathcal{J}^{(ij)})^p=\mathfrak{q}_{ij}\cdot A$,
$$r_{ij}\D A\sim\mathfrak{q}_{ij}\D A\ll\D t,\quad(r_{ij}\cdot A)_T<\infty,\quad\Delta(r_{ij}\cdot A)=0,\quad r\in\mathcal{R}_i.$$
Consequently,
$$\mathcal{J}^{(ij)}-r_{ij}\cdot A+[\mathcal{J}^{(ij)}-r_{ij}\cdot A,N]=\mathcal{J}^{(ij)}+[\mathcal{J}^{(ij)},N]-r_{ij}\cdot A\in\mathcal{M}_{loc}.$$
As $\varphi\gamma_i^{(j)}$ is locally bounded and predictable, (\ref{CommonJumpIntensityDecomposition}) yields
$$\varphi\cdot(X^{(i,r)}+[X^{(i,r)},N])=\varphi\cdot(\widecheck{X}^{(i)}+[\widecheck{X}^{(i)},N])-\sum_{j\neq i}(\varphi\gamma_i^{(j)})\cdot(\mathcal{J}^{(ij)}+[\mathcal{J}^{(ij)},N]-r_{ij}\cdot A)\in\mathcal{M}_{loc}.$$
Integration by parts gives
$$Z\in\mathcal{Z}_{loc}(X^{(i,r)})\cap\bigcap_{j\neq i}\mathcal{Z}_{loc}(\mathcal{J}^{(ij)}-r_{ij}\cdot A).$$

For sufficiency, let $r\in\mathcal{R}_i$ and
$$Z=\mathcal{E}(N)\in\mathcal{Z}_{loc}(X^{(i,r)})\cap\bigcap_{j\neq i}\mathcal{Z}_{loc}(\mathcal{J}^{(ij)}-r_{ij}\cdot A).$$
By Definition \ref{Definition: DeflatorNumeraireLogoptimal}-(c) and integration by parts, choose $\varphi_0,\varphi_j\in\mathcal{P}$ with $0<\varphi_0,\varphi_j\leq1$ such that
$$\varphi_0\cdot(X^{(i,r)}+[X^{(i,r)},N])\in\mathcal{M}_{loc},\quad\varphi_j\cdot(\mathcal{J}^{(ij)}+[\mathcal{J}^{(ij)},N]-r_{ij}\cdot A)\in\mathcal{M}_{loc},\quad j\neq i.$$
Put
$$\varphi:=\varphi_0\prod_{j\neq i}\varphi_j,\quad0<\frac{\varphi}{\varphi_0},\frac{\varphi}{\varphi_j}\leq1.$$
Then (\ref{CommonJumpIntensityDecomposition}) and stochastic integration give
$$\varphi\cdot(\widecheck{X}^{(i)}+[\widecheck{X}^{(i)},N])=\frac{\varphi}{\varphi_0}\cdot(\varphi_0\cdot(X^{(i,r)}+[X^{(i,r)},N]))+\sum_{j\neq i}\frac{\varphi\gamma_i^{(j)}}{\varphi_j}\cdot(\varphi_j\cdot(\mathcal{J}^{(ij)}+[\mathcal{J}^{(ij)},N]-r_{ij}\cdot A))\in\mathcal{M}_{loc}.$$
Hence $Z\in\mathcal{Z}_{loc}(\widecheck{X}^{(i)})$ by integration by parts.

\end{proof}

As a result, we give a sufficient condition for the switching model (\ref{GRSJDDynamic}) to admit deflators. 
\begin{prop}
	Suppose that for each $i\in\{1,2,...,D\}$, $X^{(i)}$ is special and $\mathcal{D}(
	X^{(i)})\neq\varnothing$. If furthermore, for each $i\in\{1,2,...,D\}$,
	\begin{equation}\label{Condition: AbsoluteContinuous}
	\begin{split}
	\D F^{(i)}&\ll\left(M^{P}_{\mu^{(i)}}\left[I_{\{\gamma_i\Delta J=0\}}\big| \widetilde{\cal{P}}\right]\D F^{(i)}+\sum_{j}M^{P}_{\mu^{(ij)}}\left[I_{\{\gamma_i^{(j)}\Delta J_{j}\neq 0\}}\big| \widetilde{\cal{P}}\right]\D F^{(ij)}\right),\quad\mbox{on}\quad\{\alpha_{-}=i\},\\
	\mbox{and}&\quad\dfrac{\D F^{(i)}}{M^{P}_{\mu^{(i)}}\left[I_{\{\gamma_i\Delta J=0\}}\big| \widetilde{\cal{P}}\right]\D F^{(i)}+\sum_{j}M^{P}_{\mu^{(ij)}}\left[I_{\{\gamma_i^{(j)}\Delta J_{j}\neq 0\}}\big| \widetilde{\cal{P}}\right]\D F^{(ij)}}\leq 1\quad\mbox{on}\quad\{\alpha_{-}=i\},
	\end{split}
	\end{equation}  then  $\mathcal{D}(\widehat{X})\neq\varnothing$.
\end{prop}
\begin{proof} By Theorem \ref{Theorem: StableTSNSCondition}-(c), with $\theta^{(i)}=0$, it suffices to show that $\mathcal{D}(\widecheck{X}^{(i)})\neq\varnothing$ for each $i\in\{1,2,...,D\}$. Therefore, in the remaining proof, we fix an arbitrary $i\in\{1,2,...,D\}$. Since $X^{(i)}$ is special, without loss of generality, take the truncation function to be $h=x$. To this end,  we utilize \cite[Theorem A.1]{Aksamitetal2017} and the general Jacod decomposition in \cite[Theorem B.1]{Choulli&Yansori2022c},  and deduce the existence of a pair of processes, $(\beta,f)\in\mathcal{P}\times\widetilde{\mathcal{P}}$ such that  
\begin{equation*}
\begin{split}
&f>0 \quad P\otimes\mu^{(i)}-a.e.,\quad \beta^{tr}c^{(i)}\beta\cdot A+\sqrt{(f-1)^2\star\mu^{(i)}}\in\mathcal{A}^{+}_{loc}(P),\\&
\int|x(f(x)-1)| F^{(i)}(\D x)<+\infty\quad P\otimes\D A-a.e.,\\
&b^{(i)}+c^{(i)}\beta+\int x(f(x)-1) F^{(i)}(\D x)=0,\quad P\otimes\D A-a.e.
\end{split}
\end{equation*}
Combining Proposition \ref{Proposition:GRSJDPredictableCharacteristics}, Lemma \ref{Lemma: CompositionConditionalExpectation} and Corollary \ref{Corollary: TGJDPredictableCharacteristics}, it is clear that the right hand side of Condition (\ref{Condition: AbsoluteContinuous}) is exactly  $\widecheck{F}^{(i)}$ associated with the process $\widecheck{X}^{(i)}$.

So with $\beta$ unchanged, for $\widecheck{F}^{(i)}$, just define 
\[\widecheck{f}:=(f-1)\dfrac{\D F^{(i)}}{\D \widecheck{F}^{(i)}}+1.\]
Since the common-shock compensators are continuous,
$$\widecheck{\nu}^{(i)}(\{t\},\D x)=I_{\{\alpha_{t-}=i\}}\nu^{(i)}(\{t\},\D x).$$
Thus $\widecheck{f}=f$ at predictable jumps on $\{\alpha_-=i\}$, and the corresponding terms and conditions in the general Jacod decomposition are unchanged.
Again by Corollary \ref{Corollary: TGJDPredictableCharacteristics}, \cite[Theorem B.1]{Choulli&Yansori2022c} and the boundedness of $\dfrac{\D F^{(i)}}{\D \widecheck{F}^{(i)}}$, it is clear that 
$\mathcal{D}\left(\widecheck{X}^{(i)}\right)\neq\varnothing$.
This ends the proof.
\end{proof}
\begin{remark}
Notice that, when there are no common jumps between $X^{(i)}$ and $\gamma_i\cdot\widetilde{J}$, we always have
\begin{equation*}
\dfrac{\D F^{(i)}}{ M^{P}_{\mu^{(i)}}\left[I_{\{\gamma_i\Delta J=0\}}\big| \widetilde{\cal{P}}\right] \D F^{(i)}+\sum_{j}M^{P}_{\mu^{(ij)}}\left[I_{\{\gamma_i^{(j)}\Delta J_{j}\neq 0\}}\big| \widetilde{\cal{P}}\right] \D F^{(ij)}}=1.
\end{equation*}
\end{remark}

Now the probability measure is changed in an absolutely continuous way. 
\begin{theorem}
Suppose for each $i\in\{1,2,...,D\}$, $X^{(i)}$ is special. Then 
$$\mathcal{Z}_{loc}(\widehat{X},\mathbb{P}_1)\neq\varnothing\quad\Longleftrightarrow\quad\mathcal{Z}_{loc}\left(\widecheck{X}-\left(M^{P}_{\mu^{(J)}}\left[I_{\{\Delta \widebar{X}=0\}}\big| \widetilde{\cal{P}}\right]x\right)\star\nu^{(J)},\mathbb{P}_1\right)\neq\varnothing,$$
where \[\frac{\D \mathbb{P}_1}{\D P}:=\mathcal{E}_T\left(-I_{\mbox{\rm supp}\widecheck{F}^{(\perp)}}\star(\widehat{\mu}-\widehat{\nu})\right)\geq 0,\]
and \[\D F^{(J)}=\theta\D \widecheck{F}+\D \widecheck{F}^{(\perp)}\] is the Lebesgue decomposition of $F^{(J)}$.
\end{theorem}
\begin{proof}
	First, for $0\leq t\leq T$,
	\begin{equation*}
	\begin{split}
	0\leq&\mathcal{E}_t\left(-I_{\mbox{\rm supp}\widecheck{F}^{(\perp)}}\star(\widehat{\mu}-\widehat{\nu})\right)\leq\exp\left(\left(I_{\mbox{\rm supp}\widecheck{F}^{(\perp)}}\star\widehat{\nu}\right)_t\right)\\
	=&\exp\left(\int_0^t \int_{\mathbb{R}^d\backslash\{0\}}M^{P}_{\mu^{(J)}}\left[I_{\{\Delta \widebar{X}=0\}}\big| \widetilde{\cal{P}}\right](s,x)\widecheck{F}_s^{(\perp)}(\D x)\D A_s\right)\\
	\leq&\exp\left(\int_0^t \sum_i q_i(s)\D A_s\right)\leq\exp\left(T\max_i\sum_{j\neq i}q_{ij}\right).
	\end{split}
	\end{equation*}
	So $\mathcal{E}\left(-I_{\mbox{\rm supp}\widecheck{F}^{(\perp)}}\star(\widehat{\mu}-\widehat{\nu})\right)$ is a bounded local martingale, hence a uniformly integrable martingale on $[0,T]$.
	 
Moreover,
$$\left(|x|\theta M^{P}_{\mu^{(J)}}\left[I_{\{\Delta\widebar{X}=0\}}\big|\widetilde{\cal{P}}\right]\right)\star\widecheck{\nu}\leq |x|\star\nu^{(J)}\in\mathcal{A}_{loc}^+(P).$$
	According to Proposition \ref{Proposition:GRSJDPredictableCharacteristics}, \cite[Theorem III.3.24]{Jacod&Shiryaev2003}, the characteristics of $\widehat{X}$ under probability $\mathbb{P}_1$ are (take $h(x)=x$ by special semimartingale)
	\[\left(\widehat{b}^{\mathbb{P}_1},\widehat{c}^{\mathbb{P}_1},\widehat{F}^{\mathbb{P}_1}\right)=\left(\widehat{b}-(M^{P}_{\mu^{(J)}}\left[I_{\{\Delta \widebar{X}=0\}}\big| \widetilde{\cal{P}}\right]x).\widecheck{F}^{(\perp)},\widehat{c},(1+\theta M^{P}_{\mu^{(J)}}\left[I_{\{\Delta \widebar{X}=0\}}\big| \widetilde{\cal{P}}\right]).\widecheck{F}\right).\]
	Since $\widecheck{F}^{(\perp)}$ is singular with respect to $\widecheck{F}$, and
	\[\widehat{\mu}-\widehat{\nu}-(\widecheck{\mu}-\widecheck{\nu})\quad \mbox{is orthogonal to}\quad \widecheck{\mu}-\widecheck{\nu},\]  the characteristics of $\widecheck{X}-\left(M^{P}_{\mu^{(J)}}\left[I_{\{\Delta \widebar{X}=0\}}\big| \widetilde{\cal{P}}\right]x\right)\star\nu^{(J)}$ under $\mathbb{P}_1$ are unchanged. For more discussions here, please see \cite[Proof of Theorem III.3.24]{Jacod&Shiryaev2003}.
	
	Define \[Z:=\mathcal{E}\left(\left(\theta M^{P}_{\mu^{(J)}}\left[I_{\{\Delta \widebar{X}=0\}}\big| \widetilde{\cal{P}}\right]\right)\star(\widecheck{\mu}-\widecheck{\nu})\right)>0,\quad\frac{\D \mathbb{P}_2}{\D \mathbb{P}_1}:=Z_T.\]
	Since $\theta.\widecheck{F}\leq F^{(J)}$ and $\widecheck{\nu}$ is unchanged under $\mathbb{P}_1$,
	\begin{equation*}
	\begin{split}
	E_{\mathbb{P}_1}\left[\left(Z_{-}\theta M^{P}_{\mu^{(J)}}\left[I_{\{\Delta \widebar{X}=0\}}\big| \widetilde{\cal{P}}\right]\right)\star\widecheck{\mu}_T\right]
	&=E_{\mathbb{P}_1}\left[\left(Z_{-}\theta M^{P}_{\mu^{(J)}}\left[I_{\{\Delta \widebar{X}=0\}}\big| \widetilde{\cal{P}}\right]\right)\star\widecheck{\nu}_T\right]\\
	&\leq\max_i\sum_{j\neq i}q_{ij}\int_0^TE_{\mathbb{P}_1}[Z_{t-}]\D t\leq T\max_i\sum_{j\neq i}q_{ij}<\infty.
	\end{split}
	\end{equation*}
	Thus
	\[Z=1+\left(Z_{-}\theta M^{P}_{\mu^{(J)}}\left[I_{\{\Delta \widebar{X}=0\}}\big| \widetilde{\cal{P}}\right]\right)\star(\widecheck{\mu}-\widecheck{\nu})\]
	is a uniformly integrable $\mathbb{P}_1$-martingale on $[0,T]$.

	Again, the characteristics of $\widecheck{X}-\left(M^{P}_{\mu^{(J)}}\left[I_{\{\Delta \widebar{X}=0\}}\big| \widetilde{\cal{P}}\right]x\right)\star\nu^{(J)}$ under probability $\mathbb{P}_2$ are
	\[\left(\widehat{b}-(M^{P}_{\mu^{(J)}}\left[I_{\{\Delta \widebar{X}=0\}}\big| \widetilde{\cal{P}}\right]x).\widecheck{F}^{(\perp)},\widehat{c},(1+\theta M^{P}_{\mu^{(J)}}\left[I_{\{\Delta \widebar{X}=0\}}\big| \widetilde{\cal{P}}\right]).\widecheck{F}\right)=\left(\widehat{b}^{\mathbb{P}_1},\widehat{c}^{\mathbb{P}_1},\widehat{F}^{\mathbb{P}_1}\right).\]
	Hence
	\[\mathcal{Z}_{loc}(\widehat{X},\mathbb{P}_1)\neq\varnothing\quad\Longleftrightarrow\quad\mathcal{Z}_{loc}\left(\widecheck{X}-\left(M^{P}_{\mu^{(J)}}\left[I_{\{\Delta \widebar{X}=0\}}\big| \widetilde{\cal{P}}\right]x\right)\star\nu^{(J)},\mathbb{P}_2\right)\neq\varnothing\] by \cite[Theorem A.1]{Aksamitetal2017}.
	
	Then \cite[Proposition 2.5-(a)]{Aksamitetal2017} gives the result.
\end{proof}
\begin{corollary}
	Suppose for each $i\in\{1,2,...,D\}$, $X^{(i)}$ is special, and 
	\[\sum_{s\leq\cdot}I_{\{\Delta X^{(i)}(s)= 0\}}I_{\{\Delta\widehat{X}^{(i)}(s)\in\mbox{\rm supp}\widecheck{F}_s^{(i,\perp)}\}}=\sum_{s\leq\cdot}I_{\{\Delta X^{(i)}(s)= 0\}}I_{\{\gamma_i\Delta J\in\mbox{\rm supp}\widecheck{F}_s^{(i,\perp)}\}}\equiv 0 \ P-a.s..\] Then 
	$$\mathcal{Z}_{loc}(\widehat{X})\neq\varnothing\quad\Longleftrightarrow\quad\mathcal{Z}_{loc}\left(\widecheck{X}-\left(M^{P}_{\mu^{(J)}}\left[I_{\{\Delta \widebar{X}=0\}}\big| \widetilde{\cal{P}}\right]x\right)\star\nu^{(J)}\right)\neq\varnothing.$$
\end{corollary}
\begin{proof}
	For any $i\in\{1,2,...,D\}$, notice that $\{0\}\nsubseteq \mbox{\rm supp}\widecheck{F}^{(i,\perp)}$ by \cite[II.2.11]{Jacod&Shiryaev2003} and $J$ jumps finitely many times on any finite time horizon. Thus, the sum process in the condition is always well defined.
	
	Besides, the condition is equivalent to 
	\[E\left[\left(M^{P}_{\mu^{(Ji)}}\left[I_{\{\Delta X^{(i)}=0\}}\big| \widetilde{\cal{P}}\right]I_{\mbox{\rm supp}\widecheck{F}^{(i,\perp)}}\right)\star\nu_\infty^{(Ji)}\right]=E\left[\left(I_{\{\Delta X^{(i)}= 0\}}I_{\mbox{\rm supp}\widecheck{F}^{(i,\perp)}}\right)\star\mu_\infty^{(Ji)}\right]=0.\]
	The above expression also equals
	\begin{equation*}
	\begin{split}
	&E\left[\int_0^\infty \int_{\mbox{\rm supp}\widecheck{F}_t^{(i,\perp)}} M^{P}_{\mu^{(Ji)}}\left[I_{\{\Delta X^{(i)}=0\}}\big| \widetilde{\cal{P}}\right](t,x) F_t^{(Ji)}(\D x)\D A_t\right]\\
	=&E\left[\int_0^\infty \int_{\mbox{\rm supp}\widecheck{F}_t^{(i,\perp)}} M^{P}_{\mu^{(Ji)}}\left[I_{\{\Delta X^{(i)}=0\}}\big| \widetilde{\cal{P}}\right](t,x)\widecheck{F}_t^{(i,\perp)}(\D x)\D A_t\right]\\
	=&E\left[\int_0^\infty \int_{\mbox{\rm supp}\widecheck{F}_t^{(i,\perp)}} \widehat{F}^{(i)}_t(\D x)\D A_t\right]=E\left[I_{\mbox{\rm supp}\widecheck{F}_t^{(i,\perp)}}\star\widehat{\nu}^{(i)}_\infty\right]=E\left[I_{\mbox{\rm supp}\widecheck{F}_t^{(i,\perp)}}\star\widehat{\mu}^{(i)}_\infty\right]=0.
	\end{split}
	\end{equation*}
	This implies that \[I_{\mbox{\rm supp}\widecheck{F}_t^{(i,\perp)}}\star\widehat{\mu}^{(i)}_\infty\equiv0\ P-a.s.,\]
	and then $\mathbb{P}_1\sim P$. The corollary follows by  \cite[Proposition 2.5-(a)]{Aksamitetal2017}.
\end{proof}

Now, we elaborate our main result of this section. To this end, for any local martingale with finite variation $V$ and any bounded and optional process $\xi$, we put $\xi\odot{V}:=\sum \xi\Delta{V}-\left(\sum \xi\Delta{V}\right)^p$. For more details about this optional-integral, we refer the reader to \cite[Section IX.2]{HeWangYan1992} and \cite[Section VIII]{Dellacherie&Meyer1982}.
\begin{theorem}\label{Theorem: StableTSNSCondition}
The following assertions are equivalent.\\
{\rm (a)}$\mathcal{D}(\widehat{X})\neq\varnothing$.\\ %
{\rm (b)} There exists a constant $\theta\in[0,1]$, such that $\mathcal{D}(\widecheck{X}+\theta\underline{J})\neq\varnothing$.\\
{\rm (c)} For each $i\in\{1,2,...,D\}$, there exists a constant $\theta^{(i)}\in[0,1]$, such that $\mathcal{D}(\widecheck{X}^{(i)}+\theta^{(i)}\underline{J}^{(i)})\neq\varnothing$.\\
{\rm (d)} There exists an optional process $\theta$ with values in $(0,1)$, such that $\mathcal{Z}_{loc}(\widecheck{X}+\theta\odot\underline{J})\neq\varnothing$.\\
{\rm (e)} For each $i\in\{1,2,...,D\}$, there exists an optional process $\theta^{(i)}$ with values in $(0,1)$, such that $\mathcal{Z}_{loc}(\widecheck{X}^{(i)}+\theta^{(i)}\odot\underline{J}^{(i)})\neq\varnothing$.
\end{theorem}

\begin{proof}
Notice that the jump sums supported on the jumps of $J$ have continuous compensators, since $J$ has a continuous compensator. Hence $\widecheck{X}$ and $\underline{J}$ have no common jumps. 

 (a) implying (b) is trivial. Suppose (b) is satisfied. We use Proposition \ref{Proposition: Deflator} to write $Z=\mathcal{E}(M)\mathcal{E}(-V)\in \mathcal{D}(\widecheck{X}+\theta\underline{J})$ with $M$ and $V$ fulfilling the relevant conditions. In particular, $\Delta M>-1$ with its Jacod decomposition 
\[M=M_1+M_2,\quad\mbox{and}\quad [M_1,
\widecheck{X}]\equiv 0,\ [M_2,\underline{J}]\equiv 0,\ \{\Delta M_1\neq 0\}\subseteq\{\Delta \underline{J}\neq 0\},\ \Delta M_1>-1.\]

By Proposition \ref{Proposition: Deflator}, for all bounded and predictable
\[\varphi=(\varphi_1,\varphi_2,...,\varphi_d)^{tr}\quad\mbox{with}\quad \varphi^{tr}\Delta (\widecheck{X}+\theta\underline{J})>-1,\] there is a unique process $ A^{(\varphi,M)}\in\mathcal{A}_{loc}\cap\mathcal{P}$ such that 
\begin{equation*}
\varphi\cdot(\widecheck{X}+\theta\underline{J})+[\varphi\cdot(\widecheck{X}+\theta\underline{J}), M]=A^{(\varphi,M)}+\mbox{local martingale},\quad \mbox{and}\quad \frac{1}{1-\Delta V}\cdot V\succeq A^{(\varphi,M)}.
\end{equation*}
But
\begin{equation*}
\varphi\cdot(\widecheck{X}+\theta\underline{J})+[\varphi\cdot(\widecheck{X}+\theta\underline{J}),M]=\varphi\cdot\widecheck{X}+(\theta\varphi)\cdot\underline{J}+[\varphi\cdot\widecheck{X},M_2]+[(\theta\varphi)\cdot\underline{J}, M_1]
\end{equation*}
This implies that
\begin{equation*}
\varphi\cdot\widehat{X}+[\varphi\cdot\widehat{X},\theta M_1+M_2]=\varphi\cdot\widecheck{X}+\varphi\cdot\underline{J}+[\varphi\cdot\widecheck{X},M_2]+[\varphi\cdot\underline{J},\theta M_1]=A^{(\varphi,M)}+\mbox{local martingale}.
\end{equation*}
Therefore, $Z:=\mathcal{E}(\theta M_1+M_2)\mathcal{E}(-V)\in\mathcal{D}(\widehat{X})$ by noticing that \[\theta \Delta M_1>-1\quad\mbox{and}\quad \varphi^{tr}\Delta \widehat{X}>-1\Rightarrow\varphi^{tr}\Delta (\widecheck{X}+\theta\underline{J})>-1.\]

Combining Theorem \ref{Theorem:GRSJDNumerairePortfolio} and mimicking the proof above, it is easy to see (b) is equivalent to (c), (d) is equivalent to (e), and (d) implies (a).

The only remaining implication is (a) $\Rightarrow$ (d). The proof here is based on \cite[Proposition 3.1-(b)]{Choulli&Yansori2022b} and the fact that \[\mathcal{Z}_{loc}(X)\neq\varnothing\quad\Leftrightarrow\quad\mathcal{D}(X)\neq\varnothing\]
for any semimartingale $X$. To see this fact, we refer the readers to \cite[Theorem 2.8]{Choullietal2015} and \cite[Proposition 2.1]{Choulli&Yansori2022c}.

Suppose (a) is true. Then there exists a local martingale deflator, $Z=\mathcal{E}(M)\in\mathcal{Z}_{loc}(\widehat{X})$, with $\Delta M>-1$ and its Jacod decomposition
\[M=M_1+M_2,\quad\mbox{and}\quad [M_1,
\widecheck{X}]\equiv 0,\ [M_2,\underline{J}]\equiv 0,\ \{\Delta M_1\neq 0\}\subseteq\{\Delta \underline{J}\neq 0\},\ \Delta M_1>-1.\]
There exists a scalar predictable process $\varphi$ with $0<\varphi\leq 1$ such that
\begin{equation*}
\varphi\cdot\widehat{X}+[\varphi\cdot\widehat{X},M]=\varphi\cdot\widecheck{X}+\varphi\cdot\underline{J}+[\varphi\cdot\widecheck{X},M_2]+[\varphi\cdot\underline{J}, M_1]
\end{equation*}
is a $d$-dimensional local martingale.

Set $\theta:=(1+(\Delta M_1)^-)/2$. Then $\theta$ is optional and
$$\frac{1}{2}\leq\theta<1,\quad (\Delta M_1)^-<\theta,\quad \frac{\Delta M_1}{\theta}>-1.$$
Then by \cite[VIII.40-(c)(g)]{Dellacherie&Meyer1982},
\begin{equation*}
\varphi\cdot(\widecheck{X}+\theta\odot\underline{J})+[\varphi\cdot(\widecheck{X}+\theta\odot\underline{J}),\frac{1}{\theta}\odot M_1+M_2]=\varphi\cdot\widecheck{X}+(\theta\varphi)\odot\underline{J}+[\varphi\cdot\widecheck{X},M_2]+[\varphi\cdot\underline{J}, M_1]
\end{equation*}
is a $d$-dimensional local martingale. So $Z:=\mathcal{E}(\frac{1}{\theta}\odot M_1+M_2)\in\mathcal{Z}_{loc}(\widecheck{X}+\theta\odot\underline{J})$ by $\frac{1}{\theta}\Delta M_1>-1$ and the fact that $M_1$ jumps only at the totally inaccessible jump times of $J$. Hence (d) is satisfied.
\end{proof}
\begin{remark}
{\rm(a)} As a consequence of $\mathcal{D}(\widehat{X})\neq\varnothing$ (or $\mathcal{D}_{log}(\widehat{X})\neq\varnothing$), the infimum of possible $\theta$ in Theorem \ref{Theorem: StableTSNSCondition}-(b), $\theta'$, is a measure that reflects how the transition shock distorts the model, i.e. how far from $\widecheck{X}$ to $\widehat{X}$.
		\begin{center}
			\begin{tikzpicture}[>=Stealth, x=0.8\linewidth]
			\draw[->, thick] (0,0) -- (1.1,0);

			\draw[->, thick] (0,0) -- (1.1,0);

			\node[below] at (0,-0.15) {$\widecheck{X}$};
			\filldraw (0,0) circle (2pt);

			\node[below] at (1,-0.15) {$\widehat{X}$};
			\filldraw (1,0) circle (2pt);
			
			\node[below] at (0.75,-0.15) {$\widecheck{X}+\theta'\underline{J}$};
			\filldraw (0.75,0) circle (2pt);

			\draw[decorate, decoration={brace, amplitude=5pt, raise=2pt}, thick] 
			(0.77,0.5) -- (1,0.5) node[midway, above=5pt] {$\mathcal{D}(\widecheck{X}+\theta\underline{J})\neq \varnothing$};
			\draw[decorate, decoration={brace, amplitude=5pt, raise=2pt}, thick] 
			(0,0.5) -- (0.75,0.5) node[midway, above=5pt] {$\mathcal{D}(\widecheck{X}+\theta\underline{J})= \varnothing$};
			\end{tikzpicture}
		\end{center}
Nevertheless, in general we cannot hope for $\theta'\equiv 0$.\\
{\rm(b)} As a practical sufficient condition of $\mathcal{D}(\widehat{X})\neq\varnothing$, arbitrary $\theta$ in Theorem \ref{Theorem: StableTSNSCondition}-(b) is enough to pass from $\widecheck{X}$ to $\widehat{X}$. In particular, taking $\theta=0$ in Theorem \ref{Theorem: StableTSNSCondition}-(b) gives $\mathcal{D}(\widecheck{X})\neq\varnothing\Longrightarrow\mathcal{D}(\widehat{X})\neq\varnothing$.

\end{remark}
\section{The log-optimal portfolio }
This section addresses the log-optimal portfolio under regime switching in two aspects. On the one hand, for our general switching model, we fully characterize the log-optimal portfolio in various manners.  On the other hand, we quantify the impact of the stochastic switching on the portfolio by singling out all the factors intrinsic to the switching regime that really impact the optimal portfolio in a way or another. 

\subsection{Characterisation of the log-optimal portfolio}
Throughout the rest of the paper, for any market model represented by its return $X$ and admitting its log-optimal portfolio $\widetilde{\pi}$, we denote by $u_T(X)$ the optimal logarithmic utility from terminal wealth given by 
\begin{equation}\label{OptimalValue4Log}
u_T(X):=\sup_{\pi\in\Theta(X)} E\left[ \ln\left(\mathcal{E}_T({\pi}\cdot X)\right)\right]=E\left[ \ln\left(\mathcal{E}_T(\widetilde{\pi}\cdot X)\right)\right].
\end{equation}

We also define
$$\mathcal{D}_{log}(X):=\{Z\in\mathcal{D}(X):E\ln(Z_T)>-\infty\}.$$

Thanks to \cite[Theorem 2.1]{Choulli&Yansori2022c}, which we apply to our general switching model $\widehat{X}$ defined in (\ref{GRSJDDynamic}), we obtain the following results on the existence and characterization of the log-optimal portfolio for the model $\widehat{X}$ and establish the duality for it.
\begin{theorem}\label{Theorem:GRSJDLog-optimal}
	Suppose that $X^{(i)}$ is special for each $i\in\{1,2,...,D\}$. Then the following assertions hold.\\
{\rm(a)}  The model $\widehat{X}$ admits the log-optimal portfolio, denoted by $\widetilde{\pi}_\alpha$,  if and only if  for every $i\in\{1,2,...,D\}$, the model $I_{\{\alpha_{-}=i\}}\cdot(X^{(i)}+\gamma_i\cdot\widetilde{J})$ admits the log-optimal portfolio denoted by $\widetilde{\pi}_i$.\\
{\rm(b)}  If the log-optimal portfolio $\widetilde{\pi}_\alpha$ for $\widehat{X}$ exists, then  
\begin{equation}\label{U_T(X)}
		\begin{split}
		\widetilde{\pi}_\alpha=\sum_{i=1}^DI_{\{\alpha_{-}=i\}}\widetilde{\pi}_i,\quad\mbox{and}\quad u_T(\widehat{X})=E\left[\int_0^T \sum_{i=1}^D I_{\{\alpha(t-)=i\}}\widehat{\kappa}(t,i,\widetilde{\pi}_i(t))\D A_t\right].
		\end{split}
		\end{equation}
Here,  for any $i\in\{1,2,...,D\}$, any $t\in[0,T]$ and any $\pi\in{\cal{L}}(X^{(i)})$, $\widehat{\kappa}(t,i,\pi)$ is given by 
		\begin{equation}\label{TGJDRate} 
		\begin{split}
		\widehat{\kappa}(t,i,\pi):=&\pi^{tr}b^{(i)}_t-\frac{1}{2}\pi^{tr}c^{(i)}_t\pi+\int\pi^{tr}(x-h(x))F^{(i)}_t(\D x)\\
		+&\int \left(\ln(1+\pi^{tr}x)-\pi^{tr}x\right)M^{P}_{\mu^{(i)}}\left[I_{\{\gamma_i\Delta J=0\}}\big| \widetilde{\cal{P}}\right](t,x)F^{(i)}_t(\D x)\\
		+& \sum_{j=1}^Dq_{j}(t)\left(\ln (1+\pi^{tr}\gamma_i^{(j)}(t))-\pi^{tr}\gamma_i^{(j)}(t)\right) M^{P}_{\mu^{(Ji)}}\left[I_{\{\Delta X^{(i)}=0\}}\big| \widetilde{\cal{P}}\right](t,\gamma_i^{(j)}(t)) \\
		+&\sum_{j=1}^D \int M^{P}_{\mu^{(i)}}\left[I_{\{\gamma_i^{(j)}\Delta J_{j}\neq 0\}}\big| \widetilde{\cal{P}}\right](t,x)  	\left(\ln{(1+\pi^{tr}(x+\gamma_i^{(j)}(t)))}-\pi^{tr}(x+\gamma_i^{(j)}(t))\right)F_t^{(i)}(\D x).
		\end{split}
		\end{equation}

Furthermore, the following hold.
		\begin{equation}\label{GRSJDDuality}
		\begin{split}
		\frac{1}{\mathcal{E}(\widetilde{\pi}_\alpha\cdot \widehat{X})}&=\mathcal{E}(\widetilde{L})\mathcal{E}(-\widetilde{v}\cdot A)
		=\prod_{i=1}^D\mathcal{E}(\widetilde{L}^{(i)})\mathcal{E}(-\widetilde{v}^{(i)}\cdot A)\in\mathcal{D}_{log}(\widehat{X}),\\
		\mathcal{E}(\widetilde{L}^{(i)})\mathcal{E}(-\widetilde{v}^{(i)}\cdot A)&\in\mathcal{D}_{log}(I_{\{\alpha_{-}=i\}}\cdot(X^{(i)}+\gamma_i\cdot\widetilde{J})),\ \mbox{for any}\ i\in\{1,2,...,D\},
		\end{split}
		\end{equation}
where  $(\widetilde{L},\widetilde{v},\widetilde{L}^{(i)},\widetilde{v}^{(i)})$ are given by 
\begin{equation}\label{GRSJDLValpha}
		\begin{split}
		\widetilde{L}:=&-\widetilde{\pi}_\alpha\cdot (\widehat{X})^c-\frac{\widetilde{\Gamma}\widetilde{\pi}_\alpha^{tr}x}{1+\widetilde{\pi}_\alpha^{tr}x}\star(\widehat{\mu}-\widehat{\nu}),\quad\widetilde{v}:=\widetilde{\pi}_\alpha^{tr}(\widehat{b}-\widehat{c}\widetilde{\pi}_\alpha)+\int \left(\dfrac{\widetilde{\pi}_\alpha^{tr}x}{1+\widetilde{\pi}_\alpha^{tr}x}-\widetilde{\pi}_\alpha^{tr} h(x)\right)\widehat{F}(\D x),\\
		\widetilde{L}^{(i)}:=&-(I_{\{\alpha_{-}=i\}}\widetilde{\pi}_i)\cdot (\widehat{X}^{(i)})^c-\left(I_{\{\alpha_{-}=i\}}\frac{\widetilde{\Gamma}^{(i)}\widetilde{\pi}_i^{tr}x}{1+\widetilde{\pi}_i^{tr}x}\right)\star(\widehat{\mu}^{(i)}-\widehat{\nu}^{(i)}),\\
		\widetilde{v}^{(i)}:=&I_{\{\alpha_{-}=i\}}\left(\widetilde{\pi}_i^{tr}(\widehat{b}^{(i)}-\widehat{c}^{(i)}\widetilde{\pi}_i)+\int \left(\dfrac{\widetilde{\pi}_i^{tr}x}{1+\widetilde{\pi}_i^{tr}x}-\widetilde{\pi}_i^{tr} h(x)\right)\widehat{F}^{(i)}(\D x)\right),\\
		\widetilde{\Gamma}_t:=&\left(1+\int\left((1+\widetilde{\pi}_\alpha^{tr}(t)x)^{-1}-1\right)\widehat{\nu}(\{t\},\D x)\right)^{-1}=(1-\widetilde{v}(t)\Delta A_t)^{-1},\\
		\widetilde{\Gamma}^{(i)}_t:=&\left(1+\int\left((1+\widetilde{\pi}_i^{tr}(t)x)^{-1}-1\right)\widehat{\nu}^{(i)}(\{t\},\D x)\right)^{-1}=(1-\widetilde{v}^{(i)}(t)\Delta A_t)^{-1}.
		\end{split}
		\end{equation}
\end{theorem}

\begin{proof}
Put
$$\widetilde{V}:=\widetilde{v}\cdot A,\quad \widetilde{V}^{(i)}:=\widetilde{v}^{(i)}\cdot A,\quad i\in\{1,2,...,D\}.$$
	The duality (\ref{GRSJDDuality}) is a direct consequence of \citeauthor{Choulli&Yansori2022c}\cite[Equations (2.9), (2.11) and Theorem 2.1]{Choulli&Yansori2022c},  provided that
\begin{equation*}
\widetilde{L}=\sum_{i=1}^D \widetilde{L}^{(i)},\ \widetilde{v}=\sum_{i=1}^D \widetilde{v}^{(i)}\quad\text{and}\quad [\widetilde{L}^{(i)},\widetilde{L}^{(j)}]=[\widetilde{V}^{(i)},\widetilde{V}^{(j)}]=0,\  \forall i
\neq j.
\end{equation*}
	
According to \cite[Proposition 2.1 and Theorem 2.1]{Choulli&Yansori2022c}, the log-optimal portfolio, if it exists, coincides with the num\'eraire portfolio. Therefore, the last equality in (\ref{GRSJDDuality})  is due to Theorem \ref{Theorem:GRSJDNumerairePortfolio}.
	
	Then, we directly compute the expected utility from the optimal terminal wealth.
	
By \cite[proof of Theorem 2.1, Step 4]{Choulli&Yansori2022c},
\begin{equation*}
u_T(\widehat{X})=E\left[\int_0^T\left(\widetilde{\pi}_\alpha^{tr}\widehat{b}-\frac{1}{2}\widetilde{\pi}_\alpha^{tr}\widehat{c}\widetilde{\pi}_\alpha\right)(t)\D A_t+\left(\ln(1+\widetilde{\pi}_\alpha^{tr}x)-\widetilde{\pi}_\alpha^{tr}h(x)\right)\star\widehat{\nu}_T\right].
\end{equation*}
	Based on (\ref{GRSJDTriplet}), we can expand $\widehat{b}$ in terms of $\widebar{b}$ and $\widebar{F}$, use Lemma \ref{Lemma: GRSJDIntegalLevyMeasure} to compute the integral w.r.t. $\widehat{F}$ and split the expression out into each regime thanks to Lemma \ref{Lemma: SwitchingConditionalExpectation} and (\ref{Relationship}). This gives
	\begin{equation*}
	\begin{split}
	&u_T(\widehat{X})=	\sum_{i=1}^DE\left[\int_0^T I_{\{\alpha(t-)=i\}}\widehat{\kappa}(t,i,\widetilde{\pi}_i(t))\D A_t\right].
	\end{split}
	\end{equation*}
	
	In the similar way, it is obtained that
	\begin{equation*}
	u_T(I_{\{\alpha_{-}=i\}}\cdot(X^{(i)}+\gamma_i\cdot\widetilde{J}))=E\left[\int_0^T I_{\{\alpha(t-)=i\}}\widehat{\kappa}(t,i,\widetilde{\pi}_i(t))\D A_t\right].
	\end{equation*}
	So statement (a) follows immediately by \cite[Theorem 2.1]{Choulli&Yansori2022c} and Theorem \ref{Theorem:GRSJDNumerairePortfolio}. This proves the theorem.
\end{proof}
Below, we discuss special cases when the transition shock does not jump simultaneously as $X^{(i)}$.
\begin{corollary} The following assertions hold.\\
{\rm{(a)}} If $[\gamma_\alpha\cdot J,\widebar{X}]\equiv 0$ (i.e. no common jumps), then 
		\begin{equation*}
		M^{P}_{\mu^{(i)}}\left[I_{\{\gamma_i\Delta J=0\}}\big| \widetilde{\cal{P}}\right]\equiv 1,\quad M^{P}_{\mu^{(Ji)}}\left[I_{\{\Delta X^{(i)}=0\}}\big| \widetilde{\cal{P}}\right]\equiv 1,\quad M^{P}_{\mu^{(i)}}\left[I_{\{\gamma_i^{(j)}\Delta J_{j}\neq 0\}}\big| \widetilde{\cal{P}}\right]\equiv 0,
		\end{equation*} 
		and the operator $\widehat{\kappa}$ takes the following form
		\begin{equation}\label{TGJDRate1} 
		\begin{split}
	\widehat{\kappa}(t,i,\pi)=&\pi^{tr}b^{(i)}_t-\frac{1}{2}\pi^{tr}c^{(i)}_t\pi+\int \left(\ln(1+\pi^{tr}x)-\pi^{tr}h(x)\right)F^{(i)}_t(\D x)\\
	&+ \sum_{j=1}^Dq_{j}(t)\left(\ln (1+\pi^{tr}\gamma_i^{(j)}(t))-\pi^{tr}\gamma_i^{(j)}(t)\right).
		\end{split}
		\end{equation}
{\rm{(b)}} If $\gamma_\alpha\equiv0 $, or equivalently $ \widehat{X}\equiv \widebar{X}$, then
		\begin{equation*}
		M^{P}_{\mu^{(i)}}\left[I_{\{\gamma_i\Delta J=0\}}\big| \widetilde{\cal{P}}\right]\equiv 1,\quad M^{P}_{\mu^{(Ji)}}\left[I_{\{\Delta X^{(i)}=0\}}\big| \widetilde{\cal{P}}\right]\equiv 0,\quad M^{P}_{\mu^{(i)}}\left[I_{\{\gamma_i^{(j)}\Delta J_{j}\neq 0\}}\big| \widetilde{\cal{P}}\right]\equiv 0.
		\end{equation*} 
		and the operator $\widehat{\kappa}$ becomes
	\begin{equation}\label{TGJDRate2} 
	\begin{split}
	&\widehat{\kappa}(t,i,\pi)
	=\pi^{tr}b^{(i)}_t-\frac{1}{2}\pi^{tr}c^{(i)}_t\pi
	+\int \left(\ln(1+\pi^{tr}x)-\pi^{tr}h(x)\right)F^{(i)}_t(\D x).
	\end{split}
	\end{equation}	
\end{corollary}

For our comparison later, we also give the result about the log-optimal portfolio for the market $X^{(i)}$.
\begin{prop}\label{Proposition: GJDLog-optimal} Let $i\in\{1,...,D\}$. Then the log-optimal portfolio for $X^{(i)}$, denoted by $\widetilde{\pi}_i^*$, exists if and only if $\widetilde{\pi}_i^*$ is the num\'eraire portfolio for $X^{(i)}$ and $u_T(X^{(i)})<+\infty$.\\
 Furthermore, we have 
		\begin{equation} \label{GJDLogIntegrabilityCondition}
		\begin{split}
		u_T(X^{(i)})=&E\left[\int_0^T\left\{( b^{(i)}_t-\frac{1}{2}c^{(i)}_t\widetilde{\pi}_i^*(t))^{tr}\widetilde{\pi}_i^*(t)+\int \left(\ln(1+x^{tr}\widetilde{\pi}_i^*(t))-h(x)^{tr}\widetilde{\pi}_i^*(t)\right) F_t^{(i)}(\D x)\right\}dA_t\right],\\
		=:&E\left[\int_0^T\kappa(t,i,\widetilde{\pi}_i^*(t))\D A_t   \right] ,
			\end{split}
		\end{equation}
and 
		\begin{align}
		\frac{1}{\mathcal{E}(\widetilde{\pi}_i^*\cdot X^{(i)})}=\mathcal{E}(\widetilde{L}^{(i)})\mathcal{E}(-\widetilde{v}^{(i)}\cdot{A}) \in\mathcal{D}_{log}(X^{(i)}). \label{GJDDuality}
		\end{align}
		Here the pair $(\widetilde{L}^{(i)},\widetilde{v}^{(i)})$ is given by 
		\begin{equation}\label{GJDLVl}
		\begin{split}
		\widetilde{L}^{(i)}:=&-\widetilde{\pi}_i^*\cdot (X^{(i)})^c-\frac{\widetilde{\Gamma}^{(i)}\widetilde{\pi}_i^{*tr}x}{1+\widetilde{\pi}_i^{*tr}x}\star(\mu^{(i)}-\nu^{(i)}),\\
		\widetilde{v}^{(i)}:=&\widetilde{\pi}_i^{*tr}(b^{(i)}-c^{(i)}\widetilde{\pi}_i^*)+\int \left(\dfrac{\widetilde{\pi}_i^{*tr}x}{1+\widetilde{\pi}_i^{*tr}x}-\widetilde{\pi}_i^{*tr} h(x)\right)F^{(i)}(\D x),\\
		\widetilde{\Gamma}^{(i)}_t:=&\left(1+\int\left((1+\widetilde{\pi}_i^{*tr}(t)x)^{-1}-1\right)\nu^{(i)}(\{t\},\D x)\right)^{-1}=(1-\widetilde{v}^{(i)}(t)\Delta A_t)^{-1}.
		\end{split}
		\end{equation}
\end{prop}
\noindent The proof of the corollary is obvious from Theorem \ref{Theorem:GRSJDLog-optimal}, and it will be omitted herein.  In the specific market (\ref{RSJDdynamic}), with $A_t=t$, Theorem \ref{Theorem:GRSJDLog-optimal} and Proposition \ref{Proposition: GJDLog-optimal} take tractable forms which include three new features:\\
{\rm(a)} (\ref{TJDNumeraireEquation}) as an equality gives another expression of the optimal utility rate.\\
{\rm(b)} In the Markov-modulated market (\ref{RSJDdynamic}), the coefficients\\ $\rho(t,i),\sigma(t,i),\zeta(t,i),\gamma(t,i)$ are deterministic functions of $t$, and then so is the optimal investment proportion on the risky assets $\widetilde{\pi}_1^*(t), t\in [0,T]$. Hence the expectation in (\ref{GJDLogIntegrabilityCondition}) can be dropped.\\
{\rm(c)} $\kappa$ and $\widehat{\kappa}$ take their explicit forms based on the special setting in (\ref{RSJDdynamic}).\\
As a result, we obtain the following corollary.
\begin{corollary}\label{Corollary:RSJDLog-optimal}
	For the specific market model of (\ref{RSJDdynamic}), the rate $\widehat{\kappa}$ in Theorem \ref{Theorem:GRSJDLog-optimal} becomes
	\begin{equation*}
	\begin{split}
	\widehat{\kappa}(t,i,\pi)=& \pi^{tr}\left(\rho(t,i)-\zeta(t,i)\lambda(i)-\gamma(t,i) q-\frac{1}{2}\sigma(t,i)\sigma(t,i)^{tr}\pi\right) \\
	+ & \sum_k  \ln(1+\pi^{tr}\zeta^{(k)}(t,i))\widebar{\lambda}_k(t,i)
	+\sum_{j}  \ln(1+\pi^{tr}\gamma^{(j)}(t,i)) \widebar{q}_{j}(t) \\
	+&\sum_{k}\sum_{j}\ln(1+\pi^{tr}(\zeta^{(k)}(t,i)+\gamma^{(j)}(t,i)))\dfrac{\D \langle\widetilde{N}_{i k},\widetilde{J}_{j}\rangle(t)}{\D t},
	\end{split}
	\end{equation*}
	and it satisfies
	\begin{equation*}
	\widehat{\iota}(t,\alpha(t-),\widetilde{\pi}_\alpha(t))=\widehat{\kappa}(t,\alpha(t-),\widetilde{\pi}_\alpha(t)).
	\end{equation*}
	Here 
	\begin{equation*}
	\begin{split}
	\widehat{\iota}(t,i,\pi):=&\frac{1}{2}\pi^{tr}\sigma(t,i)\sigma(t,i)^{tr}\pi +\sum_k{f}\left(\pi^{tr}\zeta^{(k)}(t,i)\right)\widebar{\lambda}_k(t,i)+\sum_{j}f\left(\pi^{tr}\gamma^{(j)}(t,i) \right)\widebar{q}_{j}(t) \\
	&+\sum_{k}\sum_{j}f\left( \pi^{tr}(\zeta^{(k)}(t,i)+\gamma^{(j)}(t,i))\right)\dfrac{\D \langle\widetilde{N}_{i k},\widetilde{J}_{j}\rangle(t)}{\D t},\\
	\mbox{where $f$ is given by}&\quad f(y):=\ln(1+y)-{{y}\over{1+y}},\quad y>-1.
	\end{split}
	\end{equation*}
\end{corollary}
In the following we discuss more particular cases. 
\begin{corollary}
\samepage
	For the special model of (\ref{RSJDdynamic}), the following assertions hold.\\
{\rm(a)} If $[\gamma_\alpha\cdot J,\widebar{X}]\equiv0$, then 
	   the operators $\widehat{\kappa}$ and $\widehat{\iota}$  become
		\begin{equation*}
		\begin{split}
	\widehat{\kappa}(t,i,\pi)=& \pi^{tr}\left(\rho(t,i)-\zeta(t,i)\lambda(i)-\gamma(t,i) q-\frac{1}{2}\sigma(t,i)\sigma(t,i)^{tr}\pi\right) \\
	+ & \sum_k  \ln(1+\pi^{tr}\zeta^{(k)}(t,i))\lambda_k(i)
	+\sum_{j}  \ln(1+\pi^{tr}\gamma^{(j)}(t,i)) q_{j}(t) ,\\
	\widehat{\iota}(t,i,\pi)=&\frac{1}{2}\pi^{tr}\sigma(t,i)\sigma(t,i)^{tr}\pi
	+\sum_k\left(-\frac{\pi^{tr}\zeta^{(k)}(t,i)}{1+\pi^{tr}\zeta^{(k)}(t,i)}+\ln(1+\pi^{tr}\zeta^{(k)}(t,i)) \right) \lambda_k(i)\\
	+&\sum_{j}\left(  -\frac{\pi^{tr}\gamma^{(j)}(t,i)}{1+\pi^{tr}\gamma^{(j)}(t,i)}+\ln(1+\pi^{tr}\gamma^{(j)}(t,i)) \right)q_{j}(t).
		\end{split}
		\end{equation*}
{\rm(b)} If $\gamma_\alpha\equiv0 $, which means $ \widehat{X}\equiv \widebar{X}$, then operators $\widehat{\kappa}$ and $\widehat{\iota}$  become
		\begin{equation*}
	\begin{split}
	\widehat{\kappa}(t,i,\pi)=& \pi^{tr}\left(\rho(t,i)-\zeta(t,i)\lambda(i)-\gamma(t,i) q-\frac{1}{2}\sigma(t,i)\sigma(t,i)^{tr}\pi\right) \\
	&+ \sum_k  \ln(1+\pi^{tr}\zeta^{(k)}(t,i))\lambda_k(i).\\
	\widehat{\iota}(t,i,\pi)=&\frac{1}{2}\pi^{tr}\sigma(t,i)\sigma(t,i)^{tr}\pi
	+\sum_k\left(-\frac{\pi^{tr}\zeta^{(k)}(t,i)}{1+\pi^{tr}\zeta^{(k)}(t,i)}+\ln(1+\pi^{tr}\zeta^{(k)}(t,i)) \right) \lambda_k(i).
	\end{split}
	\end{equation*}

\end{corollary}
And for Proposition \ref{Proposition: GJDLog-optimal}, similarly we have the corollary below.
\begin{corollary}
	For the specific market model of (\ref{RSJDdynamic}), the rate $\kappa$ in Proposition \ref{Proposition: GJDLog-optimal} becomes
	\begin{align*}
	&\kappa(t,i,\pi)=\pi^{tr}\left( \rho(t,i)-\zeta(t,i)\lambda(i)-\frac{1}{2}\sigma(t,i)\sigma(t,i)^{tr}\pi\right)+\sum_k\ln(1+\pi^{tr}\zeta^{(k)}(t,i))\lambda_k(i),\\&  t\in [0,T], i\in\{1,2,...,D\},\pi\in\mathbb{R}^{d},
	\end{align*} 
	satisfying 
	\begin{equation*}
	\iota(t,i,\widetilde{\pi}_i^*(t))=\kappa(t,i,\widetilde{\pi}_i^*(t)),\quad\iota(t,\alpha(t-),\widetilde{\pi}_\alpha^*(t))=\kappa(t,\alpha(t-),\widetilde{\pi}_\alpha^*(t)),
	\end{equation*}
	where 
	\begin{equation*}
	\begin{split}
	\iota(t,i,\pi):=\frac{1}{2}\pi^{tr}\sigma(t,i)\sigma(t,i)^{tr}\pi
	+\sum_k\left(-\frac{\pi^{tr}\zeta^{(k)}(t,i)}{1+\pi^{tr}\zeta^{(k)}(t,i)}+\ln(1+\pi^{tr}\zeta^{(k)}(t,i)) \right) \lambda_k(i).
	\end{split}
	\end{equation*}
	In addition,
	\begin{equation*}
	u_T(X^{(i)})=\int_0^T \kappa(t,i,\widetilde{\pi}_i^*(t))\D t.
	\end{equation*}
\end{corollary}

\begin{remark}
	We emphasize some facts of the rates $\widehat{\kappa}$ and $\kappa$.\\
{\rm(a)} $\kappa$ or $\widehat{\kappa}$ is corresponding to the so-called growth rate in  \cite[Equation (3.4)]{Karatzas&Kardaras2007}.\\
{\rm(b)} Actually, 
		\begin{equation*}
		\widehat{\kappa}(t,i,\pi)=\pi^{tr}\left(\widehat{b}^{(i)}(t)-\frac{1}{2}\widehat{c}^{(i)}(t)\pi\right)+\int \left(\ln(1+\pi^{tr}x)-\pi^{tr}h(x)\right) \widehat{F}^{(i)}_t(\D x).
		\end{equation*}
		(recall $(\widehat{b}^{(i)},\widehat{c}^{(i)},\widehat{F}^{(i)})$ in Corollary \ref{Corollary: TGJDPredictableCharacteristics}.)\\
{\rm(c)} When $\gamma_i\equiv 0$, for $i\in\{1,2,...,D\}$, the following relations are valid.
		\begin{equation*}
		\begin{split}
		(\widehat{b}^{(i)},\widehat{c}^{(i)},\widehat{F}^{(i)})=&(b^{(i)},c^{(i)},F^{(i)}) \quad \text{on}\quad \{\alpha_{-}=i\},\\
		\kappa=\widehat{\kappa},\quad \widetilde{\pi}_\alpha=\widetilde{\pi}_\alpha^*,\quad
		&u_T(\widebar{X})=E\left[ \int_0^T\kappa(t,\alpha(t-),\widetilde{\pi}_\alpha^*(t))\D A_t  \right].
		\end{split}
		\end{equation*}

\end{remark}
\subsection{Quantification of the switching impact}
This subsection analyses the {\it natural} question of whether the random switching mechanism affects the optimal value of the logarithmic utility or not? and how this effect can be quantified? This task becomes feasible due to the results of the previous subsection, which determine --as explicitly as possible-- the optimal values of the logarithmic utilities for both the regime switching model $\widehat{X}$ and the non-switching model $X^{(1)}$.\\ 
Recall that, in this paper, we assume that  $\alpha(0)=1$ (i.e. the first regime is the state $1$). In other words, the switching model may randomly alternate away from this initial regime and switch among various states for several times during the timespan $[0,T]$, while the non-switching model remains in the initial state $\alpha(0)=1$ over the investment interval $[0,T]$. Thus, in this subsection, we study {\it the increment of the optimal expected utility value from terminal wealth} for the pair of models  $(\widehat{X},X^{(1)})$ given by  
\begin{equation}\label{increment4LOP}
\Delta_T(\widehat{X}):=u_T(\widehat{X})-u_T(X^{(1)}).\end{equation}
	
\begin{lemma}\label{Lemma: GRSJDTransitionShock}
	For $\pi\in\mathbb{R}^{d}$, the following equality holds whenever the expression makes sense, i.e., no ``$\infty-\infty$" case appears.
	\begin{equation}\label{operatorS}
	\begin{split}
	\mathcal{S}(\pi)&:=\kappa(\alpha_{-},\pi)-\widehat{\kappa}(\alpha_{-},\pi),\\
	&=\underbrace{\pi^{tr}\gamma_\alpha q -\sum_i\ln(1+\pi^{tr}\gamma_\alpha^{(i)})q_i}_{=:\mathcal{S}^{(1)}(\pi)=\text{direct shock}\geq 0}-\underbrace{ \sum_{i=1}^D\int M^{P}_{\widebar{\mu}}\left[I_{\{\gamma_\alpha^{(i)}\Delta J_i\neq 0\}}\big| \widetilde{\cal{P}}\right](x)\ln\dfrac{ 1+\pi^{tr}(x+\gamma_\alpha^{(i)})}{(1+\pi^{tr}x)(1+\pi^{tr}\gamma_\alpha^{(i)})} \widebar{F}(\D x)}_{\text{return-shock superposition}},\\
	&=:\mathcal{S}^{(1)}(\pi)-\mathcal{S}^{(2)}(\pi)
		\end{split}
	\end{equation}
	where
	\begin{equation*}
	\begin{split}
	\mathcal{S}^{(1)}(\pi):=&\underbrace{\pi^{tr}\gamma_\alpha q -\sum_i\ln(1+\pi^{tr}\gamma_\alpha^{(i)})q_i}_{\text{direct shock}\geq 0},\\
	\mathcal{S}^{(2)}(\pi):=&\underbrace{ \sum_i\int M^{P}_{\widebar{\mu}}\left[I_{\{\gamma_\alpha^{(i)}\Delta J_i\neq 0\}}\big| \widetilde{\cal{P}}\right](x)\ln\dfrac{ 1+\pi^{tr}(x+\gamma_\alpha^{(i)})}{(1+\pi^{tr}x)(1+\pi^{tr}\gamma_\alpha^{(i)})} \widebar{F}(\D x)}_{\text{return-shock superposition}}.
	\end{split}
	\end{equation*}
In particular, if for each $i\in\{1,2,...,D\}$, $X^{(i)}$ is special, then $\mathcal{S}(\widetilde{\pi}_\alpha)$ and $\mathcal{S}(\widetilde{\pi}_\alpha^*)$ are well-defined. Furthermore, $\mathcal{S}(\widetilde{\pi}_\alpha)$ takes values in$[-\infty,\infty)$ and $ \mathcal{S}(\widetilde{\pi}_\alpha^*)$ takes values in$(-\infty,\infty]$.
\end{lemma}
\noindent  The proof of this lemma is relegated to Appendix \ref{Appendix: GRSJDTransitionShock}. For the particular case defined in (\ref{RSJDdynamic}), the results of Lemma \ref{Lemma: GRSJDTransitionShock}  take a simplified form as follows. 
\begin{corollary} \label{Corollary: RSJDTransitionShock}
	In the specific market (\ref{RSJDdynamic}), with $A_t=t$, for $ t\in [0,T], \pi\in\mathbb{R}^{d}$,
	\begin{align*}
	&\mathcal{S}(\pi)
	= \underbrace{\pi^{tr}\gamma_\alpha q -\sum_{i=1}^D\ln(1+\pi^{tr}\gamma_\alpha^{(i)})q_i}_{\text{direct shock}\geq 0}-\underbrace{  \sum_k\sum_i\ln\frac{1+\pi^{tr}(\zeta_\alpha^{(k)}+\gamma_\alpha^{(i)})}{(1+\pi^{tr}\zeta_\alpha^{(k)})(1+\pi^{tr}\gamma_\alpha^{(i)})}  \frac{\D \langle\widetilde{N}_{\alpha k},\widetilde{J}_i\rangle}{\D t}}_{\text{return-shock superposition}}.\end{align*}
\end{corollary}
The proof of the corollary is immediate and will be omitted. Hereto, we discuss the results of Lemma \ref{Lemma: GRSJDTransitionShock} and Corollary \ref{Corollary: RSJDTransitionShock} for particular cases on the transition shock part. 
\begin{corollary} The following assertions hold.\\
{\rm{(a)}} If $[\gamma_\alpha\cdot J,\widebar{X}]\equiv0$ (i.e. no common jumps), then 
		the operator $\mathcal{S}$ becomes
		\begin{equation}\label{GRSJDTransitionShock1}
		\begin{split}
		\mathcal{S}(\pi)
		=&\underbrace{\pi^{tr}\gamma_\alpha q -\sum_{i=1}^Dq_i\ln(1+\pi^{tr}\gamma_\alpha^{(i)})}_{\text{direct shock}\geq 0}.
		\end{split}
		\end{equation}
{\rm{(b)}}  If $\gamma_\alpha\equiv0 $ (or equivalently $ \widehat{X}\equiv \widebar{X}$), then $\mathcal{S}\equiv 0$.
\end{corollary}
For the reader's convenience, we recall below the log-optimal portfolios and their associated optimal utility-rates for the three models  $\widehat{X}$, $\widebar{X}$ and $X^{(i)}$. Recall that we have $\widehat{X}=\widebar{X}+\gamma_\alpha\cdot\widetilde{J}$.
\begin{itemize}
	\item $\widetilde{\pi}_\alpha$:= num\'eraire (log-optimal) portfolio, and $\widehat{\kappa}(t,\alpha(t-),\widetilde{\pi}_\alpha(t)):=$ optimal utility-rate for $\widehat{X}$
	\item $\widetilde{\pi}_\alpha^*:=$ num\'eraire (log-optimal)  portfolio , and $\kappa(t,\alpha(t-),\widetilde{\pi}_\alpha^*(t)):=$ optimal utility-rate for $\widebar{X}$.
	\item $\widetilde{\pi}_i^*:=$ num\'eraire portfolio (log-optimal) for $X^{(i)}$, and $\kappa(t,i,\widetilde{\pi}_i^*(t)):=$ optimal utility-rate for $X^{(i)}$
\end{itemize}

\noindent Recall that in the proof of Proposition \ref{Proposition:GRSJDPredictableCharacteristics}, the jump measure $\widehat{\mu}$ is carefully decomposed into three parts with no common jumps among each other:
\begin{align*}
\widehat{\mu}(\D t, \D x)=&I_{\{\gamma_\alpha(t)\Delta J(t)= 0\}}\widebar{\mu}(\D t ,\D x)+I_{\{\Delta \widebar{X}(t)=0\}}\mu^{(J)}(\D t,\D x)+\sum_{i=1}^DI_{\{\gamma_\alpha^{(i)}(t)\Delta J_i(t)\neq 0\}}\widebar{\mu}^{(i)}(\D t ,\D x),\\
\widebar{\mu}^{(i)}([0,t]\times \mathscr{A}):=&\int_{[0,t]\times\mathbb{R}^d}I_\mathscr{A}(x+\gamma_\alpha^{(i)}(s))I_{\{x+\gamma_\alpha^{(i)}(s)\neq 0 \}}\widebar{\mu}(\D s,\D x),\ \forall \mathscr{A}\in \mathcal{B}(\mathbb{R}^d).
\end{align*}
Therefore, by combining (\ref{GRSJDLValpha}), Lemma \ref{Lemma: CompositionConditionalExpectation}, Proposition \ref{Proposition:GRSJDPredictableCharacteristics}, and Lemma \ref{Lemma: GRSJDIntegalLevyMeasure} the following lemma follows.
\begin{lemma} The pair $(\widetilde{L},\widetilde{V})$ defined in (\ref{GRSJDLValpha}) satisfies 
	\begin{equation*}
	\begin{split}
	\widetilde{L}=\widetilde{L}_1+\widetilde{L}_2\quad\mbox{with}\quad	[\widetilde{L}_1,\widetilde{L}_2]\equiv 0,\quad\mbox{and}\quad \widetilde{V}:=\widetilde{v}\cdot A=\widetilde{V}_1+\widetilde{V}_2.
	\end{split}
	\end{equation*}
Here the processes $\widetilde{L}_1$, $\widetilde{L}_2$, $\widetilde{V}_1$ and $\widetilde{V}_2$ are given by 
	\begin{equation*}
	\begin{split}
	\widetilde{L}_1:=&-\widetilde{\pi}_\alpha\cdot (\widehat{X})^c-\left(\frac{\widetilde{\Gamma}\widetilde{\pi}_\alpha^{tr}x}{1+\widetilde{\pi}_\alpha^{tr}x}\right)\star\left(I_{\{\gamma_\alpha\Delta J=0\}}.\widebar{\mu}-M^{P}_{\widebar{\mu}}\left[I_{\{\gamma_\alpha\Delta J=0\}}\big| \widetilde{\cal{P}}\right].\widebar{\nu}\right), \\
	\widetilde{L}_2:=&-\left(\frac{\widetilde{\pi}_\alpha^{tr}x}{1+\widetilde{\pi}_\alpha^{tr}x}\right)\star\left(I_{\{\Delta \widebar{X}=0\}}.\mu^{(J)}-M^{P}_{\mu^{(J)}}\left[I_{\{\Delta \widebar{X}=0\}}\big| \widetilde{\cal{P}}\right].\nu^{(J)}\right)\\&-\sum_i \left(\frac{\widetilde{\pi}_\alpha^{tr}x}{1+\widetilde{\pi}_\alpha^{tr}x}\right)\star\left(I_{\{\gamma_\alpha^{(i)}\Delta J_i\neq 0\}}.\widebar{\mu}^{(i)}-M^{P}_{\widebar{\mu}^{(i)}}\left[I_{\{\gamma_\alpha^{(i)}\Delta J_i\neq 0\}}\big| \widetilde{\cal{P}}\right].\widebar{\nu}^{(i)}\right), \\
	\widetilde{V}_1:=&\widetilde{v}_1\cdot A,\quad\widetilde{V}_2:=\widetilde{v}_2\cdot A,\\
	\widetilde{v}_1:=&\widetilde{\pi}_\alpha^{tr}\left(\widebar{b}-\widebar{c}\widetilde{\pi}_\alpha+\int(x-h(x))\widebar{F}(\D x)\right)+\int \left(\dfrac{\widetilde{\pi}_\alpha^{tr}x}{1+\widetilde{\pi}_\alpha^{tr}x}-\widetilde{\pi}_\alpha^{tr} x\right)M^{P}_{\widebar{\mu}}\left[I_{\{\gamma_\alpha\Delta J=0\}}\big| \widetilde{\cal{P}}\right](x)\widebar{F}(\D x),\\
	\widetilde{v}_2:=&\int \left(\dfrac{\widetilde{\pi}_\alpha^{tr}x}{1+\widetilde{\pi}_\alpha^{tr}x}-\widetilde{\pi}_\alpha^{tr} x\right)M^{P}_{\mu^{(J)}}\left[I_{\{\Delta \widebar{X}=0\}}\big| \widetilde{\cal{P}}\right](x)F^{(J)}(\D x)\\&+\sum_i\int\left(\dfrac{\widetilde{\pi}_\alpha^{tr}x}{1+\widetilde{\pi}_\alpha^{tr}x}-\widetilde{\pi}_\alpha^{tr} x\right) M^{P}_{\widebar{\mu}^{(i)}}\left[I_{\{\gamma_\alpha^{(i)}\Delta J_i\neq 0\}}\big| \widetilde{\cal{P}}\right](x)\widebar{F}^{(i)}(\D x).
	\end{split}\end{equation*} 
	Moreover $\widetilde{v}_2\leq 0$.
\end{lemma}
With regard to the terms defined above, we just cut down the whole terms  $\widetilde{L}$ and $\widetilde{V}$ in (\ref{GRSJDLValpha}) into two parts:
\begin{itemize}
	\item $\widetilde{L}_1$ and  $\widetilde{V}_1$: representing the part of $\widebar{X}$ without joint jumps with $\gamma_\alpha\cdot\widetilde{J}$.
	\item  $\widetilde{L}_2$ and  $\widetilde{V}_2$: matching the part from $\gamma_\alpha\cdot\widetilde{J}$.
\end{itemize}
Let   $\widetilde{\mathbb{P}}$ be the probability measure on $(\Omega,\mathcal{F})$ given by 
\begin{equation}\label{Ptilde}
\frac{\D \widetilde{\mathbb{P}}}{\D P}:=\dfrac{\mathcal{E}_T(\widetilde{L}_2)}{E[\mathcal{E}_T(\widetilde{L}_2)]},\end{equation}
and consider its relative-entropy 
\begin{equation*}
\mathcal{H}(P|\widetilde{\mathbb{P}}):=E_{\widetilde{\mathbb{P}}}\left[\frac{\D P}{\D \widetilde{\mathbb{P}}}\ln\frac{\D P}{\D \widetilde{\mathbb{P}}}\right]=E[\ln\frac{\D P}{\D \widetilde{\mathbb{P}}}]=E[-\ln\frac{\D \widetilde{\mathbb{P}}}{\D P}]=E[-\ln\mathcal{E}_T(\widetilde{L}_2)]+\ln E[\mathcal{E}_T(\widetilde{L}_2)]\geq 0\\
\end{equation*}

{

\begin{theorem}\label{Theorem: GRSJDSwitchingEffect} Suppose that the conditions in Lemma \ref{Lemma: GRSJDSwitchingEffect2} are satisfied. Then the following hold.\\
{\rm (a)} The quantity $\Delta_T(\widehat{X})$ can be decomposed as follows.
\begin{equation}\label{GRSJDSwitchingEffect1}\begin{split}
\Delta_T(\widehat{X})=&\underbrace{E\left[\int_0^T \kappa(t,\alpha(t-),\widetilde{\pi}_\alpha^*(t))-\kappa(t,1,\widetilde{\pi}_1^*(t))\D A_t\right]}_{\text{switching effect}}+\underbrace{\mathcal{H}(P|\widetilde{\mathbb{P}})-\ln E[\mathcal{E}_T(\widetilde{L}_2)]}_{
			\text{information premium}\geq 0}\\
+&\underbrace{E\left[\int_0^T \widehat{\mathcal{P}}^{(N)}(t)\D{A}_t\right]}_{\text{num\'eraire change premium}\geq 0}- \underbrace{E\left[\int_0^T\left(\mathcal{S}_t(\widetilde{\pi}_\alpha^*)-\mathcal{S}_t(\widetilde{\pi}_\alpha)-\widetilde{v}_2(t) \right)\D{A}_t\right]}_{\text{Transition-Shock-Risk}\geq 0}+\underbrace{	E\left[\int_0^T\widehat{\mathcal{P}}^{(J)}(t)\D{A}_t \right]}_{\text{Common-Jumps-Premium}\geq 0}.
\end{split}
\end{equation}
Here the quantities $\widehat{\mathcal{P}}^{(N)}$ and $\widehat{\mathcal{P}}^{(J)}$ are given by
		\begin{equation*}
		\begin{split}
	\widehat{\mathcal{P}}^{(N)}:=&(\widetilde{\pi}_\alpha-\widetilde{\pi}_\alpha^*) ^{tr}\widehat{b}-\frac{1}{2}\widetilde{\pi}_\alpha^{tr}\widehat{c}\widetilde{\pi}_\alpha + \frac{1}{2}\widetilde{\pi}_\alpha^{*tr}\widehat{c}\widetilde{\pi}_\alpha^*+
	\int \left(\ln \dfrac{1+\widetilde{\pi}_\alpha^{tr}x}{1+\widetilde{\pi}_\alpha^{*tr}x}-(\widetilde{\pi}_\alpha-\widetilde{\pi}_\alpha^*)^{tr}h(x)\right)\widehat{F}(\D x) ,\\
	\widehat{\mathcal{P}}^{(J)}:=&\int  M^{P}_{\widebar{\mu}}\left[I_{\{\gamma_\alpha\Delta J\neq 0\}}\big| \widetilde{\cal{P}}\right](x)\left(\widetilde{\pi}_\alpha^{tr}x-\ln(1+\widetilde{\pi}_\alpha^{tr}x) \right)\widebar{F}(\D x).
     	\end{split}
		\end{equation*}
{\rm (b)} The quantity $\Delta_T(\widehat{X})$ can be decomposed as follows.
		\begin{equation}\label{GRSJDSwitchingEffect2}
		\begin{split}
		\Delta_T(\widehat{X})=&\underbrace{E\left[\int_0^T \kappa(t,\alpha(t-),\widetilde{\pi}_\alpha^*(t))-\kappa(t,1,\widetilde{\pi}_1^*(t))\D A_t\right]}_{\text{switching effect}}\\
	&+\underbrace{E\left[ \int_0^T\widehat{\mathcal{P}}^{(N)}(t) \D A_t\right]}_{\text{num\'eraire change premium}}-\underbrace{E\left[ \int_0^T\mathcal{S}^{(1)}(t,\widetilde{\pi}_\alpha^*(t)) \D A_t\right]}_{\text{direct shock}}+\underbrace{E\left[ \int_0^T\mathcal{S}^{(2)}(t,\widetilde{\pi}_\alpha^*(t)) \D A_t\right]}_{\text{return-shock superposition}}.
		\end{split}
		\end{equation}
\end{theorem}   
\noindent The proof of theorem is relegated to Appendix \ref{Appendix: GRSJDSwitchingEffect}.
\begin{remark} {\rm (a)} By combining (\ref{PremiumShockIdentity}) with the fact $\mathcal{H}(P|\widetilde{\mathbb{P}})-\ln E[\mathcal{E}_T(\widetilde{L}_2)]\leq E\left[-\widetilde{V}_{2,T}\right]$, we deduce that 
$\widehat{\mathcal{P}}^{(N)}\leq \mathcal{S}(\widetilde{\pi}_\alpha^*)-\mathcal{S}(\widetilde{\pi}_\alpha).$
This implies that 
\begin{equation*}
\mbox{``information premium"}+\mbox{``num\'eraire change premium"}\ \leq\ \mbox{``total risk from transition shock"}.\end{equation*}
{\rm (b)} It is also easy to check that 
 \begin{equation*}	\begin{split}
\mbox{``Information-Premium"}\ +\mbox{``Common-Jumps-Premium"}=&\mbox{``total risk from transition shock"}-\mbox{``direct shock"}\\
&+\mbox{``return-shock superposition"} \geq 0.
\end{split}
	\end{equation*}
\end{remark}

\begin{corollary}
	Let $\widetilde{\pi}_\alpha$ be a num\'eraire portfolio for $\widehat{X}$. Suppose for each $i\in\{1,2,...,D\}$, $X^{(i)}$ admits the log-optimal portfolio. Then so does $\widehat{X}$ if 
	\begin{equation*}
	E\left[\int_0^T \widehat{\mathcal{P}}_t^{(J)} \D A_t\right]<\infty.
	\end{equation*}
In this case, the condition is equivalent to
	\[E\left[H^{(0)}_T(\mathfrak{L})\right]=\mathcal{H}(P|\mathbb{P}^{(J)})-\ln E[\mathcal{E}_T(\mathfrak{L})]=E\left[\int_0^T \widehat{\mathcal{P}}_t^{(J)} \D A_t\right]<\infty,\]
	where
	\[\mathfrak{L}:=\left(\widetilde{\pi}_\alpha^{tr}x\right)\star\left(I_{\{\gamma_\alpha\Delta J\neq 0\}}.\widebar{\mu}-M^{P}_{\widebar{\mu}}\left[I_{\{\gamma_\alpha\Delta J\neq 0\}}\big| \widetilde{\cal{P}}\right].\widebar{\nu}\right)\] is a uniformly integrable martingale, and 
	$\frac{\D \mathbb{P}^{(J)}}{\D P}:=\dfrac{\mathcal{E}_T(\mathfrak{L})}{E[\mathcal{E}_T(\mathfrak{L})]}$.
\end{corollary}
\begin{proof}
By $\ln(1+y)-y\leq0$ for $y>-1$, (\ref{TGJDRate}) and the num\'eraire property,
\[0\leq\widehat{\kappa}(t,\alpha(t-),\widetilde{\pi}_\alpha(t))\leq\kappa(t,\alpha(t-),\widetilde{\pi}_\alpha(t))+\widehat{\mathcal{P}}^{(J)}_t\leq\kappa(t,\alpha(t-),\widetilde{\pi}_\alpha^*(t))+\widehat{\mathcal{P}}^{(J)}_t.\]
Hence
\[E\left[\int_0^T\widehat{\kappa}(t,\alpha(t-),\widetilde{\pi}_\alpha(t))\D A_t\right]\leq\sum_i u_T(X^{(i)})+E\left[\int_0^T\widehat{\mathcal{P}}^{(J)}_t\D A_t\right]<\infty,\]
and \cite[Theorem 2.1-(e)]{Choulli&Yansori2022c} gives the conclusion.

The latter assertion is from \cite[Lemma B.1 and Proposition B.2]{Choulli&Yansori2022a}, and 
\[H^{(0)}(\mathfrak{L})=\left(\widetilde{\pi}_\alpha^{tr}x-\ln(1+\widetilde{\pi}_\alpha^{tr}x) \right)\star\left(I_{\{\gamma_\alpha\Delta J\neq 0\}}.\widebar{\mu}\right).\]
\end{proof}
 
\begin{remark}
{\rm(a)} When the transition shock is excluded, i.e. $\gamma_\alpha\equiv0$, the operator $\mathcal{S}$ will be $0$.\\
{\rm(b)} We see that the joint jumps between the transition shock and inherent jumps of the asset returns appear here and there, triggering costs or premiums in the optimal utility. This delicately displays how the shocks from regime switching interact with the asset return. But this phenomenon was ignored by previous literatures (for instance, under the continuous asset price cases, it is not easy to reflect the interplay between the regime switching and the asset return in a technically appropriate model setting).\\
{\rm(c)} Our model in (\ref{GRSJDDynamic}) suitably helps us to study this phenomenon and quantify its influence in a rather clear and reasonable way. In an economic sense, the change of macroeconomic or market conditions may happen at the same time as the abrupt events impacting the individual assets, and this possibility makes things more complicated.

\end{remark}
\begin{corollary}
{\rm(a)} If there are no common jumps between $\gamma_\alpha\cdot J$ and $\widebar{X}$, then (\ref{GRSJDSwitchingEffect1}) becomes
		\begin{equation*}
	\begin{split}
	\Delta_T(\widehat{X})
	=&\underbrace{E\left[\int_0^T \kappa(t,\alpha(t-),\widetilde{\pi}_\alpha^*(t))-\kappa(t,1,\widetilde{\pi}_1^*(t))\D A_t\right]}_{\text{switching effect}}+\underbrace{\mathcal{H}(P|\widetilde{\mathbb{P}})-\ln E[\mathcal{E}_T(\widetilde{L}_2)]}_{
		\text{information premium}}\\
	+&\underbrace{E\left[\int_0^T \widehat{\mathcal{P}}^{(N)}(t)\D A_t\right]}_{\text{num\'eraire change premium}}-\underbrace{E\left[\int_0^T\left(\mathcal{S}_t(\widetilde{\pi}_\alpha^*)-\mathcal{S}_t(\widetilde{\pi}_\alpha)-\widetilde{v}_2(t)\right)\D A_t\right]}_{\text{total risk from transition shock}}.
	\end{split}
	\end{equation*}
	And (\ref{GRSJDSwitchingEffect2}) becomes
	\begin{equation*}
\begin{split}
\Delta_T(\widehat{X})=&\underbrace{E\left[  \int_0^T\kappa(t,\alpha(t-),\widetilde{\pi}_\alpha^*(t))-\kappa(t,1,\widetilde{\pi}_1^*(t)) \D A_t\right] }_{\text{switching effect}}\\
&+\underbrace{E\left[\int_0^T \widehat{\mathcal{P}}^{(N)}(t)\D A_t\right]}_{\text{num\'eraire change premium}}-\underbrace{E\left[\int_0^T\mathcal{S}^{(1)}(t,\widetilde{\pi}_\alpha^*(t))\D A_t\right]}_{\text{direct shock}}.
	\end{split}
		\end{equation*}
{\rm(b)} If $\gamma_\alpha\equiv0 $, which means $ \widehat{X}\equiv \widebar{X}$,  then 
		\begin{equation*}
		\begin{split}
		&\Delta_T(\widehat{X})
		=\underbrace{E\left[\int_0^T \kappa(t,\alpha(t-),\widetilde{\pi}_\alpha^*(t))-\kappa(t,1,\widetilde{\pi}_1^*(t))\D A_t\right]}_{\text{switching effect}}.\\
		\end{split}
		\end{equation*}

\end{corollary}
}

\subsection{The level and Parrondo effects}
Throughout this subsection, we assume that 
\begin{equation}\label{Assumptions_A}
A\ \mbox{is deterministic},\quad 0<A_T<\infty,\quad \mbox{and}\quad  \sum_{i=1}^DE\left[\int_0^T|\kappa(t,i,\widetilde{\pi}_i^*(t))|\D A_t\right]<\infty.
\end{equation}

The switching effect represents the impact caused by the switching coefficients. This effect can be studied more. Compared with the non-switching market, the regime switching market adds in more possible regimes (states). These regimes are characterised by their coefficient groups. If these states naturally have higher or lower optimal utility rates than the single constant non-switching state $\alpha(0)$, then the optimal utilities are supposed to be different. This kind of difference is apart from the impacts caused by the randomly switching behaviour. This kind of impact due to unbalanced levels of the optimal utility rates is called a ``level effect". It should not be attributed to the pure randomly switching mechanism. \footnote{While as we shall see, its quantity is also related to the switching mechanism.} Therefore, we hope to set each state to be ``equivalent" in the sense of the optimal utility rate, which means that
\begin{equation}
E\left[\int_0^T \kappa(t,1,\widetilde{\pi}_1^*(t)) \D A_t\right]=E\left[\int_0^T \kappa(t,2,\widetilde{\pi}_2^*(t)) \D A_t\right]=...=E\left[\int_0^T \kappa(t,D,\widetilde{\pi}_D^*(t)) \D A_t\right], \label{OptimalUtilityEquivalenceCondition1}
\end{equation}\\

The condition (\ref{OptimalUtilityEquivalenceCondition1}) represents that, if without switching, then each state itself will generate the same optimal logarithmic utility during time $[0,T]$ due to (\ref{GJDLogIntegrabilityCondition}), i.e., 
\[ u_T(X^{(1)})=u_T(X^{(2)})=...=u_T(X^{(D)}).\]

If (\ref{OptimalUtilityEquivalenceCondition1}) is not satisfied in the market, one can always define (recall that $\alpha(0)=1$)

\begin{align*}
&\widebar{\kappa}(t,i,\widetilde{\pi}_i^*(t)):=\kappa(t,i,\widetilde{\pi}_i^*(t))-\frac{1}{A_T}\left(u_T(X^{(i)})-u_T(X^{(1)})\right),\ i\in \{1,2,...,D\},t\in[0,T],
\end{align*}

then (\ref{OptimalUtilityEquivalenceCondition1}) is satisfied by $\widebar{\kappa}(t,i,\widetilde{\pi}_i^*(t))$ as

\begin{equation}
E\left[\int_0^T \widebar{\kappa}(t,1,\widetilde{\pi}_1^*(t)) \D A_t\right]=E\left[\int_0^T \widebar{\kappa}(t,2,\widetilde{\pi}_2^*(t)) \D A_t\right]=...=E\left[\int_0^T \widebar{\kappa}(t,D,\widetilde{\pi}_D^*(t)) \D A_t\right]. 
\label{OptimalUtilityEquivalenceCondition2}
\end{equation}

By Fubini, the above equation is equivalent to

\begin{equation}
\int_0^T E\left[\widebar{\kappa}(t,1,\widetilde{\pi}_1^*(t))\right] \D A_t=\int_0^T E\left[\widebar{\kappa}(t,2,\widetilde{\pi}_2^*(t)) \right]\D A_t=...=\int_0^T E\left[\widebar{\kappa}(t,D,\widetilde{\pi}_D^*(t))\right] \D A_t. 
\label{OptimalUtilityEquivalenceCondition3}
\end{equation}

Note that the transform from $\kappa(t,i,\widetilde{\pi}_i^*(t))$ to $\widebar{\kappa}(t,i,\widetilde{\pi}_i^*(t))$ is just a vertical translation of $\kappa(t,i,\widetilde{\pi}_i^*)$, without hurting the shape of $\kappa(t,i,\widetilde{\pi}_i^*(t))$. This property is critical as we shall see later.

Then the switching optimal utility rate \[\kappa(t,\alpha(t-),\widetilde{\pi}_\alpha^*(t))=\sum_i I_{\{\alpha(t-)=i\}}\widebar{\kappa}(t,i,\widetilde{\pi}_i^*(t))+\sum_i I_{\{\alpha(t-)=i\}}\frac{1}{A_T}\left(u_T(X^{(i)})-u_T(X^{(1)})\right).\] can be this way decomposed into two components as above.

These two components are\\
{\rm(a)} One rate switches among each rate $\widebar{\kappa}(t,i,\widetilde{\pi}_i^*(t)),i\in\{1,2,...,D\}$ equivalent to state 1 in the sense of optimal logarithmic utility. These rates  $\widebar{\kappa}(t,i,\widetilde{\pi}_i^*(t))$ have the same shape as $\kappa(t,i,\widetilde{\pi}_i^*(t))$.\\
{\rm(b)} The other integrand randomly switches between a group of constant levels: \[\frac{1}{A_T}\left(u_T(X^{(i)})-u_T(X^{(1)})\right),i\in\{1,2,...,D\}.\]

Now that we have excluded the transition shock on asset returns and the level effect. Then any remaining impact of the optimal logarithmic utility has to be interpreted as purely caused by the randomly switching behaviour itself. We call this effect ``pure switching effect".
So finally we define the pure switching effect as
\begin{align*}
E\left[ \int_0^T \widebar{\kappa}(t,\alpha(t-),\widetilde{\pi}_\alpha^*(t))-\kappa(t,1,\widetilde{\pi}_1^*(t))\D A_t  \right]= \int_0^T E\left[\widebar{\kappa}(t,\alpha(t-),\widetilde{\pi}_\alpha^*(t))-\kappa(t,1,\widetilde{\pi}_1^*(t))\right]\D A_t  .
\end{align*}
The definition above makes sure that the pure switching effect is a ``fair" comparison between the markets with and without randomly switching. 

We would also call this effect ``Parrondo effect" as a similar phenomenon was discovered by \cite{Harmer&Abbott1999a,Harmer&Abbott1999b}. Relevant researches about Parrondo effect can be seen in \cite{Harmer&Abbott2002} and the references therein. But here we first mathematically define Parrondo effect in the continuous-time market.

Conditioning on the regime $\alpha_{-}$ (see \cite[Problem 1.5]{HeWangYan1992}), we obtain
\begin{equation*}
\begin{split}
&E\left[\widebar{\kappa}(t,\alpha(t-),\widetilde{\pi}_\alpha^*(t))-\kappa(t,1,\widetilde{\pi}_1^*(t))\right]\\
=&\sum_iP(\alpha(t-)=i)E\left[\widebar{\kappa}(t,i,\widetilde{\pi}_i^*(t))-\widebar{\kappa}(t,1,\widetilde{\pi}_1^*(t))\right]\\
+&\sum_iP(\alpha(t-)=i)\left(E\left[\widebar{\kappa}(t,i,\widetilde{\pi}_i^*(t))-\widebar{\kappa}(t,1,\widetilde{\pi}_1^*(t))|\alpha(t-)=i\right]-E\left[\widebar{\kappa}(t,i,\widetilde{\pi}_i^*(t))-\widebar{\kappa}(t,1,\widetilde{\pi}_1^*(t))\right]\right)\\
=&\sum_i e_1^{tr} \exp{(Qt)}e_iE\left[\widebar{\kappa}(t,i,\widetilde{\pi}_i^*(t))-\widebar{\kappa}(t,1,\widetilde{\pi}_1^*(t))\right]\\
+&\sum_i e_1^{tr}\exp{(Qt)}e_i\left(E\left[\widebar{\kappa}(t,i,\widetilde{\pi}_i^*(t))-\widebar{\kappa}(t,1,\widetilde{\pi}_1^*(t))|\alpha(t-)=i\right]-E\left[\widebar{\kappa}(t,i,\widetilde{\pi}_i^*(t))-\widebar{\kappa}(t,1,\widetilde{\pi}_1^*(t))\right]\right).
\end{split}
\end{equation*}

Therefore, we can further decompose the switching effect and obtain the next result.
\begin{corollary} Suppose that (\ref{Assumptions_A}) holds. Then we have 
	\begin{align*}
	&\Delta_T(\widehat{X})\\
	=& \underbrace{E\left[ \int_0^T \widehat{\mathcal{P}}^{(N)}(t) \D A_t\right] }_{\text{num\'eraire change premium for transition shock}}-\underbrace{E\left[ \int_0^T\mathcal{S}_t(\widetilde{\pi}_\alpha^*) \D A_t\right] }_{\text{transition shock}} \notag \\
	+ &\underbrace{\sum_i\frac{1}{A_T}\left(u_T(X^{(i)})-u_T(X^{(1)})\right) \int_0^Te_1^{tr} \exp{(Qt)}e_i\D A_t }_{\text{level effect}}\\
	+&\underbrace{\sum_i \int_0^Te_1^{tr} \exp{(Qt)}e_iE\left[\widebar{\kappa}(t,i,\widetilde{\pi}_i^*(t))-\widebar{\kappa}(t,1,\widetilde{\pi}_1^*(t))\right]\D A_t}_{\text{mixing effect}}\\
	+&\underbrace{\sum_i \int_0^Te_1^{tr}\exp{(Qt)}e_i\left(E\left[\widebar{\kappa}(t,i,\widetilde{\pi}_i^*(t))-\widebar{\kappa}(t,1,\widetilde{\pi}_1^*(t))|\alpha(t-)=i\right]-E\left[\widebar{\kappa}(t,i,\widetilde{\pi}_i^*(t))-\widebar{\kappa}(t,1,\widetilde{\pi}_1^*(t))\right]\right)\D A_t}_{\text{switching-market dependence}}.
	\end{align*}
\end{corollary}
\begin{remark}
{\rm(a)} In the market with PII $X^{(i)}$, the characteristics are deterministic and so are the optimal utility rates. Then the switching-market correlation is $0$. And the expectations can be dropped. The switching-market correlation disappears if for each $i\in\{1,2,...,D\},(b^{(i)},c^{(i)},F^{(i)})$ is independent of $\alpha$.\\
{\rm(b)} Intuitively, from Equation (\ref{OptimalUtilityEquivalenceCondition3}),
		the mixing effect is to let the nature stochastically ``choose" the time interval for each optimal utility rate $E\left[\kappa(t,i,\widetilde{\pi_i}^*(t))\right],i\in \{1,2,...,D\}$ to integrate over $[0,T].$  This random selection of time intervals, $\{t:t\in [0,T],\alpha(t-)=i\}$, ``distorts" the distribution of $\kappa(t,i,\widetilde{\pi}_i^*(t)),i\in \{1,2,...,D\}$ over time t, and then affects the total integral. Thus, the shape of the rate $\kappa(t,i,\widetilde{\pi}_i^*(t))$ over time $t$ is supposed to play a critical role in how the randomly switching behaviour affects the optimal utility. 

\end{remark}
Consider the case when the market is time-homogeneous with respect to $A$, which means that the coefficients only depend on the state variable $\alpha$. Then $\widetilde{\pi}_\alpha^*(t)$, determined by these coefficients, is also time-homogeneous. And the rate $\kappa(t,\alpha(t-),\widetilde{\pi}_\alpha^*(t))$ only randomly alternates among several constant deterministic functions $\kappa(i,\widetilde{\pi}_i^*)$ for $i\in\{1,2,...,D\}$. 

Consequently, the quantity of the integral of this rate, $\int_0^T\kappa(t,\alpha(t-),\widetilde{\pi}_\alpha^*(t))\D A_t$, only relies on the total $A$-time spent by the state $\alpha$ staying on each state $i\in\{1,2,...,D\}$ in $[0,T]$. For the extreme case when $\kappa(i,\widetilde{\pi}_i^*)$ are the same for each $i\in\{1,2,...,D\}$, the randomly switching does not have any effect since the integral of $\kappa(t,\alpha(t-),\widetilde{\pi}_\alpha^*(t))$ over $[0,T]$ is always the area of the rectangle with width $A_T$ and height $\kappa(1,\widetilde{\pi}_1^*)$.

On the other hand, if the coefficients above are time-nonhomogeneous, then for each $i\in\{1,2,...,D\}$, the integrand rate $E\left[\kappa(t,i,\widetilde{\pi}_i^*(t))\right]$ takes different values on different time points $t\in[0,T]$. In this case, not only how long in total but also when the state $\alpha$ stays on each regime $i\in\{1,2,...,D\}$ during $[0,T]$ matter. And in this case the distribution of $E\left[\kappa(t,\alpha(t-),\widetilde{\pi}_\alpha^*(t))\right]$ can be quite sophisticated. Fortunately, its expectation is analytically tractable (at least partially) in our market setting.
\begin{remark}\label{Remark:MMPIIJDOrthogonalDecomposition}
In this remark, the function spaces use $\D A$ on $[0,T]$, counting measure on $\{1,2,...,D\}$, and $P$ on $\Omega$ when present.
	There is another viewpoint of the Parrondo effect. Consider the vector space
\[
	L^1(\Omega\times[0,T]\times\{1,2,...,D\}):=\{\vartheta \text{\ is measurable},\quad\sum_iE\left[\int_0^T|\vartheta(t,i)|\D A_t\right]<\infty\},
\]
with its dual space
\[
L^\infty(\Omega\times[0,T]\times\{1,2,...,D\}):=\{\vartheta \text{\ is measurable and esssup } |\vartheta|<\infty \}.
\]
For 
\[(\vartheta_1,\vartheta_2)\in L^1(\Omega\times[0,T]\times\{1,2,...,D\})\times L^\infty(\Omega\times[0,T]\times\{1,2,...,D\}),\]
Define the linear functional in the sense of Riesz representation
 \[<\vartheta_1,\vartheta_2>:=\sum_iE\left[\int_0^T\vartheta_1(t,i)\vartheta_2(t,i)\D A_t\right].\]
	$L^1(\Omega\times[0,T]\times\{1,2,...,D\})$ has a subspace \[\mathfrak{0}:=\{\vartheta\in L^1(\Omega\times[0,T]\times\{1,2,...,D\}), E\left[\int_0^T \vartheta(t,i) \D A_t\right]=0,\ \forall i\in\{1,2,...,D\}\}.\]
	and its orthogonal complement \[\mathfrak{0}^{\perp}=\{\vartheta\in L^\infty(\Omega\times[0,T]\times\{1,2,...,D\}),\vartheta(\omega,t,i)\equiv\vartheta(i),(\omega,t,i)\in\Omega\times[0,T]\times \{1,2,...,D\}\}.\] 
	
	Suppose that there is a $\vartheta\in L^1(\Omega\times[0,T]\times\{1,2,...,D\})$, with its unique orthogonal decomposition \[\vartheta=\vartheta^{\mathfrak{0}}+\vartheta^{\mathfrak{0}^{\perp}}\quad \mbox{satisfying 
 }\quad (\vartheta^{\mathfrak{0}},\vartheta^{\mathfrak{0}^{\perp}})\in \mathfrak{0}\times\mathfrak{0}^{\perp}.\]
	Then it is apparent that 
	\[\vartheta^{\mathfrak{0}}=\vartheta-\frac{1}{A_T}E\left[\int_0^T \vartheta(u,i)\D A_u\right]\quad\text{and}\quad\vartheta^{\mathfrak{0}^{\perp}}=\frac{1}{A_T}E\left[\int_0^T \vartheta(u,i)\D A_u\right].\]
	Therefore, $\vartheta^{\mathfrak{0}}$ carries all of the information about the ``shape" of $\vartheta$ since it has the same shape as  $\vartheta$ (vertical translation). But $\vartheta^{\mathfrak{0}}$ ``loses" the information about the quantity of expected integral of $\vartheta$, since it always has $0$ expected integral.
	
	On the other hand,  $\vartheta^{\mathfrak{0}^{\perp}}$ carries all of the information about the quantity of expected integral of $\vartheta$ since it has the same expected integral as $\vartheta$. But $\vartheta^{\mathfrak{0}^{\perp}}$ ``loses" the information about the ``shape" of $\vartheta$, since it is just flat over $t$.
	
	Then for any \[\vartheta(t,i)= \kappa(t,i,\widetilde{\pi}_i^*(t))-\kappa(t,1,\widetilde{\pi}_1^*(t))\in L^2(\Omega\times[0,T]\times\{1,2,...,D\}),\]
	it admits the orthogonal decomposition \[\widebar{\kappa}(t,i,\widetilde{\pi}_i^*(t))-\kappa(t,1,\widetilde{\pi}_1^*(t))+\frac{1}{A_T}E\left[\int_0^T [\kappa(u,i,\widetilde{\pi}_i^*(u))-\kappa(u,1,\widetilde{\pi}_1^*(u))] \D A_u\right],\]
	where
	\[\widebar{\kappa}(t,i,\widetilde{\pi}_i^*(t))-\kappa(t,1,\widetilde{\pi}_1^*(t))\in \mathfrak{0},\]
	and 
	\[\frac{1}{A_T}E\left[\int_0^T [\kappa(u,i,\widetilde{\pi}_i^*(u))-\kappa(u,1,\widetilde{\pi}_1^*(u))] \D A_u\right]\in \mathfrak{0}^{\perp}.\]  
	
	$\mathfrak{0}^{\perp}$ is seen as a ``state level" space since it keeps the same level along the time horizon, and different levels across different states. Its orthogonal complement is naturally named ``pure switching". This is described as that it keeps its shape over the time axis, and generates the same expected integral over time horizon at each state when there is no switching.

In the analogous geometrical viewpoint, we can work on $L^1([0,T]\times\{1,2,...,D\})$ and its dual space  $L^\infty([0,T]\times\{1,2,...,D\})$. Take \[\vartheta_2(t,i):= e_1^{tr} \exp{(Qt)}e_i\in L^\infty([0,T]\times\{1,2,...,D\}).\]
For each of the following choices, assume that $\vartheta_1\in L^1([0,T]\times\{1,2,...,D\})$.

The mixing effect and switching-market correlation are represented as $<\vartheta_1,\vartheta_2>$
for \[\quad\vartheta_1(t,i):=E\left[\widebar{\kappa}(t,i,\widetilde{\pi}_i^*(t))-\widebar{\kappa}(t,1,\widetilde{\pi}_1^*(t))\right],\quad \vartheta_2(t,i):= e_1^{tr} \exp{(Qt)}e_i,\]
or 
\[\vartheta_1(t,i):=E\left[\widebar{\kappa}(t,i,\widetilde{\pi}_i^*(t))-\widebar{\kappa}(t,1,\widetilde{\pi}_1^*(t))|\alpha(t-)=i\right]-E\left[\widebar{\kappa}(t,i,\widetilde{\pi}_i^*(t))-\widebar{\kappa}(t,1,\widetilde{\pi}_1^*(t))\right].\]
If $\vartheta_1\in L^2([0,T]\times\{1,2,...,D\})$ and $\vartheta_1\neq0$, the signs of these terms are determined by the angle between $\vartheta_1$ and $\vartheta_2$ in $L^2([0,T]\times\{1,2,...,D\})$. If $\vartheta_1=0$, then $<\vartheta_1,\vartheta_2>=0$.
\end{remark}

\appendix
\section{Some useful results}
Here we summarise the main results obtained in \cite{Choulli&Yansori2022b,Choulli&Yansori2022c}, where the authors characterise the num\'eraire portfolio and deflators in any general semimartingale market.

\begin{prop} (\cite[Proposition 3.1 (a)]{Choulli&Yansori2022b})\label{Proposition: Deflator}
	In the market represented by a $d$-dimensional return semimartingale $X$ with its predictable characteristics $(b^X,c^X,F^X)$, we have 
	
	A process $Z\in\mathcal{D}(X)$ if and only if there exist $M\in\mathcal{M}_{loc}$ and $ V\in \mathcal{A}^+_{loc}\cap\mathcal{P}$, such that $Z=\mathcal{E}(M)\mathcal{E}(-V),M(0)=V(0)=0$, $\Delta M>-1,\Delta V<1$, and for any bounded and predictable
	\begin{equation*}
	\varphi=(\varphi_1,\varphi_2,...,\varphi_d)^{tr}\quad\mbox{with} \quad\varphi^{tr}\Delta X>-1,
	\end{equation*}
	$\varphi\cdot X+[\varphi\cdot X,M]$ is special, i.e., there exists a  process $ A^{(\varphi,M)}\in\mathcal{A}_{loc}\cap\mathcal{P}$, such that
	\begin{equation*}
	\varphi\cdot X+[\varphi\cdot X,M]=A^{(\varphi,M)}+\mbox{local martingale},
	\end{equation*}
	satisfying 
	\begin{equation*}
	\frac{1}{1-\Delta V}\cdot V\succeq A^{(\varphi,M)}.
	\end{equation*}
\end{prop}

\begin{prop}\label{Proposition: Numeraire} 
	In the market represented by a $d$-dimensional return semimartingale $X$ with its predictable characteristics $(b^X,c^X,F^X)$, we have\\
{\rm(a)} (\cite[Proposition 2.1 and Equation (2.5)]{Choulli&Yansori2022c}) The num\'eraire portfolio for the market $X$ exists if and only if there is a portfolio $\widetilde{\pi}\in L(X)\cap\mathcal{L}(X)$ and it satisfies 
		\begin{equation*}
		(\pi-\widetilde{\pi})^{tr}(b^X-c^X\widetilde{\pi})+\int \left(\dfrac{(\pi-\widetilde{\pi})^{tr}x}{1+\widetilde{\pi}^{tr}x}-(\pi-\widetilde{\pi})^{tr} h(x)\right)F^X(\D x)\leq 0,\  \forall \pi\in \mathcal{L}(X).
		\end{equation*}
		Then $\widetilde{\pi}$ is exactly the unique num\'eraire portfolio for the market $X$.\\
{\rm(b)} (\cite[Proposition 2.1]{Choulli&Yansori2022c}) The  num\'eraire portfolio for the market $X$ exists if and only if $\mathcal{D}(X)\neq\varnothing$.

\end{prop}

\section{Proof of Proposition \ref{Proposition:GRSJDPredictableCharacteristics} and some related results}\label{Appendix: PredictableCharacteristics}
The feature of common jumps brings hurdles in computing $\widehat{F}$, the conditional distribution of the jump size for $\widehat{X}$. From \cite[Proposition 7.14]{Bjork2021}, the jump measure of the sum of several jump measures is the sum of each jump measure, if they have no jumps in common. So we need to carefully split the common jump part out, and then try to derive an expression of $\widehat{F}$.

Following the proof in  \cite[II.2.32]{Jacod&Shiryaev2003}, we have
\begin{proof}[Proof of Proposition \ref{Proposition:GRSJDPredictableCharacteristics}]
	Recall that the predictable characteristics of $\widebar{X} $ are\\ $(\widebar{b},\widebar{c},\widebar{F},\widebar{A})=(\sum_iI_{\{\alpha_{-}=i\}}b^{(i)},\sum_iI_{\{\alpha_{-}=i\}}c^{(i)},\sum_iI_{\{\alpha_{-}=i\}}F^{(i)},A)$. Then clearly $\widehat{c}=\widebar{c}=\sum_iI_{\{\alpha_{-}=i\}}c^{(i)}$.
	
	For the jump measure, we split $\widehat{X}$ into three parts with no mutually common jumps,
	\begin{align*}
	\widehat{X}=&\widebar{X}	-\sum_{s\leq\cdot}\Delta \widebar{X}(s)I_{\{\gamma_\alpha(s)\Delta J(s)\neq 0\}}+I_{\{\Delta \widebar{X}=0\}}\gamma_\alpha\cdot \widetilde{J}+\sum_{s\leq\cdot}\Delta \widebar{X}(s)I_{\{\gamma_\alpha(s)\Delta J(s)\neq 0\}}+I_{\{\Delta \widebar{X}\neq 0\}}\gamma_\alpha\cdot \widetilde{J}.
	\end{align*}
	Accordingly, the possible jumps of $\widehat{X}$ can be separated into these three forms, and the jump measures can be directly added together by \cite[Proposition 7.14]{Bjork2021}, i.e., 
	\begin{equation*}
	\begin{split}
	\widehat{\mu}(\D t ,\D x)&:=\sum_s I_{\{\Delta \widehat{X}(s)\neq0\}}\delta_{(s,\Delta \widehat{X}(s))}(\D t ,\D x)\\
	&=\sum_s\{I_{\{\gamma_\alpha(s)\Delta J(s)= 0\}}I_{\{\Delta \widebar{X}(s)\neq 0\}}\delta_{(s,\Delta \widebar{X}(s))}(\D t ,\D x)+I_{\{\Delta \widebar{X}(s)=0\&\ \gamma_\alpha(s)\Delta J(s)\neq 0\}}\delta_{(s,\gamma_\alpha(s)\Delta J(s))}(\D t ,\D x)\\
	&\hskip 1cm +I_{\{\Delta \widebar{X}(s)+\gamma_\alpha(s)\Delta J(s)\neq 0\}}I_{\{\gamma_\alpha(s)\Delta J(s)\neq 0\}}I_{\{\Delta \widebar{X}(s)\neq 0\}}\delta_{(s,\Delta \widebar{X}(s)+\gamma_\alpha(s)\Delta J(s))}(\D t ,\D x)\}.
	\end{split}
	\end{equation*}
	We zip the first two terms above using $\widebar{\mu}$ and $\mu^{(J)}$ to write
	\begin{equation*}
	\begin{split}
	\widehat{\mu}(\D t ,\D x)=&I_{\{\gamma_\alpha(t)\Delta J(t)=0\}}\widebar{\mu}(\D t ,\D x)+I_{\{\Delta \widebar{X}(t)=0\}}\mu^{(J)}(\D t,\D x)\\
	&+\sum_s I_{\{\Delta \widebar{X}(s)+\gamma_\alpha(s)\Delta J(s)\neq 0\}}I_{\{\gamma_\alpha(s)\Delta J(s)\neq 0\}}I_{\{\Delta \widebar{X}(s)\neq 0\}}\delta_{(s,\Delta \widebar{X}(s)+\gamma_\alpha(s)\Delta J(s))}(\D t ,\D x).
	\end{split}
	\end{equation*}
	The last common jump term above can be separated further using the fact that $J=(J_1,J_2,...,J_D)^{tr}$ has no common jumps among its components.  And
	\begin{equation*}
	\gamma_\alpha\Delta J\equiv \gamma_\alpha^{(i)}\quad \text{on}\quad \{\gamma_\alpha^{(i)}\Delta J_i\neq 0\}.
	\end{equation*}
	Then
	\begin{equation*}
	\begin{split}
	\widehat{\mu}(\D t ,\D x)=&I_{\{\gamma_\alpha(t)\Delta J(t)= 0\}}\widebar{\mu}(\D t ,\D x)+I_{\{\Delta \widebar{X}(t)=0\}}\mu^{(J)}(\D t,\D x)\\
	&+\sum_s \sum_iI_{\{\Delta \widebar{X}(s)+\gamma_\alpha^{(i)}(s)\neq 0\}}I_{\{\gamma_\alpha^{(i)}(s)\Delta J_i(s)\neq 0\}}I_{\{\Delta \widebar{X}(s)\neq 0\}}\delta_{(s,\Delta \widebar{X}(s)+\gamma_\alpha^{(i)}(s))}(\D t ,\D x).
	\end{split}
	\end{equation*}
	Notice that
	\begin{align*}
	\widebar{\nu}:=&(\widebar{\mu})^{p}=\widebar{F}\otimes \widebar{A}=(\sum_i I_{\{\alpha_{-}=i\}}F^{(i)})\otimes A,\\
	\nu^{(J)}(\D t, \D x):=&(\mu^{(J)})^{p}(\D t, \D x)=\sum_iq_i(t)\delta_{\gamma_\alpha^{(i)}(t)}(\D x)\D A_t.
	\end{align*}
	Let the integrable functional $H\in\mathcal{P}\otimes\mathcal{B}(\mathbb{R}^d)$, and the composition functional $H^i(\omega,t,x):=H(\omega,t,x+\gamma_\alpha^{(i)}(\omega,t))I_{\{x+\gamma_\alpha^{(i)}(\omega,t) \neq 0\}}\in\mathcal{P}\otimes\mathcal{B}(\mathbb{R}^d)$.  We have
	\begin{equation*}
	\begin{split}
	H\star\widehat{\mu}=&\left( I_{\{\gamma_\alpha\Delta J= 0\}}H\right)\star \widebar{\mu}+\left(I_{\{\Delta \widebar{X}=0\}}H\right)\star\mu^{(J)}+\sum_i\left(I_{\{\gamma_\alpha^{(i)}\Delta J_i\neq 0\}}H^i\right)\star\widebar{\mu}.
	\end{split}
	\end{equation*}
	Using  \cite[Theorem 11.9]{HeWangYan1992},
	\begin{equation*}
	\begin{split}
	(H\star\widehat{\mu})^{p}=&\left( M^{P}_{\widebar{\mu}}\left[I_{\{\gamma_\alpha\Delta J= 0\}}\big| \widetilde{\cal{P}}\right]H\right)\star \widebar{\nu}+ \left(M^{P}_{\mu^{(J)}}\left[I_{\{\Delta \widebar{X}=0\}}\big| \widetilde{\cal{P}}\right]H\right)\star\nu^{(J)}+\sum_i\left(M^{P}_{\widebar{\mu}}\left[I_{\{\gamma_\alpha^{(i)}\Delta J_i\neq 0\}}\big| \widetilde{\cal{P}}\right]H^i\right)\star\widebar{\nu}\\
	=&\left( M^{P}_{\widebar{\mu}}\left[I_{\{\gamma_\alpha\Delta J=0\}}\big| \widetilde{\cal{P}}\right]H\right)\star \widebar{\nu}+ \left(M^{P}_{\mu^{(J)}}\left[I_{\{\Delta \widebar{X}=0\}}\big| \widetilde{\cal{P}}\right]H\right)\star\nu^{(J)}\\&+\sum_i\int_{[0,\cdot]\times\mathbb{R}^d}M^{P}_{\widebar{\mu}}\left[I_{\{\gamma_\alpha^{(i)}\Delta J_i\neq 0\}}\big| \widetilde{\cal{P}}\right](t,x-\gamma_\alpha^{(i)}(t))H(t,x)\widebar{\nu}^{(i)}(\D t,\D x),
	\end{split}
	\end{equation*}
	where
	\begin{equation*}
	\widebar{\nu}^{(i)}([0,t]\times \mathscr{A}):=\int_{[0,t]\times\mathbb{R}^d}I_\mathscr{A}(x+\gamma_\alpha^{(i)}(s))I_{\{x+\gamma_\alpha^{(i)}(s)\neq 0 \}}\widebar{\nu}(\D s,\D x),\ \forall \mathscr{A}\in \mathcal{B}(\mathbb{R}^d).
	\end{equation*}
	The second equality above is the general transformation rule in computing integral.
	And next
	\begin{align*}
	\widehat{F}_t(\D x)=&  M^{P}_{\widebar{\mu}}\left[I_{\{\gamma_\alpha\Delta J=0\}}\big| \widetilde{\cal{P}}\right](t,x)\widebar{F}_t(\D x)+ M^{P}_{\mu^{(J)}}\left[I_{\{\Delta \widebar{X}=0\}}\big| \widetilde{\cal{P}}\right](t,x)\sum_iq_i(t)\delta_{\gamma_\alpha^{(i)}(t)}(\D x)\notag\\
	&+\sum_i M^{P}_{\widebar{\mu}}\left[I_{\{\gamma_\alpha^{(i)}\Delta J_i\neq 0\}}\big| \widetilde{\cal{P}}\right](t,x-\gamma_\alpha^{(i)}(t))\widebar{F}^{(i)}_t(\D x)\\
	=&  M^{P}_{\widebar{\mu}}\left[I_{\{\gamma_\alpha\Delta J=0\}}\big| \widetilde{\cal{P}}\right](t,x)\sum_i I_{\{\alpha(t-)=i\}}F^{(i)}_t(\D x)+ M^{P}_{\mu^{(J)}}\left[I_{\{\Delta \widebar{X}=0\}}\big| \widetilde{\cal{P}}\right](t,x)\sum_iq_i(t)\delta_{\gamma_\alpha^{(i)}(t)}(\D x)\notag \\ 
	&+\sum_i M^{P}_{\widebar{\mu}}\left[I_{\{\gamma_\alpha^{(i)}\Delta J_i\neq 0\}}\big| \widetilde{\cal{P}}\right](t,x-\gamma_\alpha^{(i)}(t))\widebar{F}^{(i)}_t(\D x).
	\end{align*}
	Moreover, when $X^{(i)}$ is special for each $i\in\{1,2,...,D\}$, so is $\widebar{X}$ by applying \cite[I.3.5 and I.4.34]{Jacod&Shiryaev2003} to the canonical decompositions of $X^{(i)}$. And since $\gamma_\alpha\cdot \widetilde{J}$ is a local martingale, $\widehat{X}$ is also special. Then from \cite[II.2.29]{Jacod&Shiryaev2003}, $|x-h|\star \widebar{\mu}, |x-h|\star \widehat{\mu}\in{\cal A}_{loc}^+$.
	
	So comparing the unique canonical representations of $\widehat{X}$ and $\widebar{X}$,
	\begin{align*}
	\widehat{X}=&(\widehat{X})^c+\widehat{B}+h\star(\widehat{\mu}-\widehat{\nu})+(x-h)\star \widehat{\mu}\\
	=&(\widebar{X})^c+\widebar{B}+h\star(\widebar{\mu}-\widebar{\nu})+(x-h)\star \widebar{\mu}+x\star(\mu^{(J)}-\nu^{(J)}),\\
	\widehat{B}=&\widebar{B}+\left((x-h)\star \widebar{\mu}-(x-h)\star \widehat{\mu}\right)^{p}=\widebar{B}+(x-h)\star \widebar{\nu}-(x-h)\star \widehat{\nu},\\
	\widehat{b}=&\sum_i I_{\{\alpha_{-}=i\}}\left(b^{(i)}+\int (x-h(x))F^{(i)}(\D x)\right) -\int (x-h(x))\widehat{F}(\D x).
	\end{align*}
\end{proof}
The next lemma is useful throughout this paper and it is a direct consequence of Proposition \ref{Proposition:GRSJDPredictableCharacteristics}. 
\begin{lemma}\label{Lemma: GRSJDIntegalLevyMeasure}
	Suppose that $X^{(i)}$ is special for each $i\in\{1,2,...,D\}$. Then for any nonnegative or integrable $W\in\widetilde{\cal{O}}$ with $W(\omega,t,0)=0,$ we have 
	\begin{align*}
	\int W(t,x)\widehat{F}_t(\D x)=&\sum_i\int \left(M^{P}_{\widebar{\mu}}\left[I_{\{\gamma_\alpha\Delta J=0\}}\big| \widetilde{\cal{P}}\right](t,x)I_{\{\alpha(t-)=i\}}W(t,x)\right)F^{(i)}_t(\D x)\\&+ \sum_iq_i(t)M^{P}_{\mu^{(J)}}\left[I_{\{\Delta \widebar{X}=0\}}\big| \widetilde{\cal{P}}\right](t,\gamma_\alpha^{(i)}(t))W(t,\gamma_\alpha^{(i)}(t))\\ 
	&+\sum_i\sum_{j}\int\left( M^{P}_{\widebar{\mu}}\left[I_{\{\gamma_\alpha^{(i)}\Delta J_i\neq 0\}}\big| \widetilde{\cal{P}}\right](t,x) W(t,x+\gamma_{j}^{(i)}(t))I_{\{\alpha(t-)=j\}}\right)F^{(j)}_t(\D x).
	\end{align*}
\end{lemma}
According to the proof of Proposition \ref{Proposition:GRSJDPredictableCharacteristics}, the jump measure $\widehat{\mu}$ is carefully decomposed into three parts with no common jumps among each other:
\begin{align*}
\widehat{\mu}(\D t, \D x)=&I_{\{\gamma_\alpha(t)\Delta J(t)= 0\}}\widebar{\mu}(\D t ,\D x)+I_{\{\Delta \widebar{X}(t)=0\}}\mu^{(J)}(\D t,\D x)+\sum_iI_{\{\gamma_\alpha^{(i)}(t)\Delta J_i(t)\neq 0\}}\widebar{\mu}^{(i)}(\D t ,\D x),\\
\widebar{\mu}^{(i)}([0,t]\times \mathscr{A}):=&\int_{[0,t]\times\mathbb{R}^d}I_\mathscr{A}(x+\gamma_\alpha^{(i)}(s))I_{\{x+\gamma_\alpha^{(i)}(s)\neq 0 \}}\widebar{\mu}(\D s,\D x),\ \forall \mathscr{A}\in \mathcal{B}(\mathbb{R}^d).
\end{align*}

Inspecting the proof of Proposition \ref{Proposition:GRSJDPredictableCharacteristics}, actually we proved
\begin{lemma} \label{Lemma: CompositionConditionalExpectation}
	$ M^{P}_{\widebar{\mu}}\left[I_{\{\gamma_\alpha^{(i)}\Delta J_i\neq 0\}}\big| \widetilde{\cal{P}}\right](t,x-\gamma_\alpha^{(i)}(t))=M^{P}_{\widebar{\mu}^{(i)}}\left[I_{\{\gamma_\alpha^{(i)}\Delta J_i\neq 0\}}\big| \widetilde{\cal{P}}\right](t,x)$ $P\otimes\widebar{\nu}^{(i)}-a.e.$.
\end{lemma} 

Based on Proposition \ref{Proposition:GRSJDPredictableCharacteristics} and Proposition \ref{Corollary:RSJDPredictableCharacteristics}, it is found that the common jumps between $\widebar{X}$ and
$\gamma_\alpha\cdot\widetilde{J}$ lead to the special structure of the jump measure of $\widehat{X}$. We point out here that this is indeed reflecting the dependence between $\alpha$ and $X^{(i)}$. Please see the next proposition.

\begin{prop}\label{Propostion: IDPSwitchingNoCommonJump}
	If $\alpha$ is independent of $\{X^{(i)},i\in\{1,2,...,D\}\}$, then $J_{j}$ has no common jumps with $X^{(i)}$ for any $i,j\in\{1,2,...,D\}$.
\end{prop}
\begin{proof}
	
	$\alpha$ being independent of $X^{(i)}$ implies that $J$ is independent of $X^{(i)}$.
	Then the fact that $J_{j}$ has no common jumps with $X^{(i)}$ is just an immediate consequence of \citeauthor{Bremaud2020}\cite[Theorem 1.2.4]{Bremaud2020}. 
	
	For details of proof, we define an intensity measure on ${\cal{B}}([0,\infty))$: $M_{J_i}(B):=E\left[\int_{B\times\mathbb{R}}\mu^{J_i}(\D t,\D x)\right],\ \forall B \in {\cal{B}}([0,\infty))$, where $\mu^{J_i}$ is the jump measure associated with the process $J_i$,
	\begin{align*}
	\mu^{J_i}(\D t,\D x):=\sum_s I_{\{\Delta J_i(s) \neq 0\}}\delta_{(s,\Delta J_i(s))}(\D t,\D x)=\delta_1(\D x)\D J_i(t),
	\end{align*}
	with the compensator measure $\nu^{J_i}(\D t,\D x):=\delta_1(\D x)q_i\D A_t$.
	
	Since $\Omega\times B \times \mathbb{R}\in {\cal P}\otimes{\cal B(\mathbb{R})}$,
	\begin{align*}
	M_{J_i}(B)=E\left[\int_{B\times\mathbb{R}}\nu^{J_i}(\D t,\D x)\right]=E\left[\int_Bq_i\D A_t\right]=\int_BE\left[q_{\alpha(t-)i}I_{\{\alpha(t-)\neq i\}}\right]\D t.
	\end{align*}
	Then it is clear that $M_{J_i}$ is diffusive. 
	Also define 
	\begin{align*}
	M_{X^{(i)}}(B):=E\left[\int_{B\times\mathbb{R}^d}\mu^{(i)}(\D t,\D x)\right]=E\left[\int_{B\times\mathbb{R}^d}\nu^{(i)}(\D t,\D x)\right],\ \forall B \in {\cal{B}}([0,\infty)).
	\end{align*}
	By \cite[Corollary 1.2.3]{Bremaud2020}(Campbell's formula),
	\begin{align*}
	&E\left[\int_{[0,\infty)\times\mathbb{R}^d}\left(I_{\{t\}\times\mathbb{R}}\star\mu^{J_i}\right)\mu^{(i)}(\D t, \D x)\right]=E\left[\int_{[0,\infty)\times\mathbb{R}^d} M_{J_i}(\{t\})\nu^{(i)}(\D t, \D x)\right]= \int_{[0,\infty)}M_{J_i}(\{t\})M_{X^{(i)}}(\D t)=0,
	\end{align*}
	which implies that $\int_{[0,\infty)\times\mathbb{R}^d}\left(I_{\{t\}\times\mathbb{R}}\star\mu^{J_i}\right)\mu^{(i)}(\D t, \D x)=0,P-a.s.$, i.e., $J_i$ and $X^{(i)}$ have no common time points of jumps. Similarly, one can prove that $J$ has no common jumps with any $X^{(j)}$, and this completes the proof. 
\end{proof}

\begin{remark}
	We list the meaning of different jump measures introduced until now or to be introduced later for the reader's quick reference.
	\begin{itemize}
		\item $(\mu^{(i)},\nu^{(i)},F^{(i)})$: jump measure associated with $X^{(i)}$.
		\item  $(\widebar{\mu},\widebar{\nu},\widebar{F})$: jump measure associated with $\widebar{X}$.
		\item  $(\widehat{\mu},\widehat{\nu},\widehat{F})$: jump measure associated with $\widehat{X}$.
		\item  $(\mu^{(J)},\nu^{(J)},F^{(J)})$: jump measure associated with $\gamma_\alpha\cdot \widetilde{J}$.
		\item  $(\widebar{\mu}^{(i)},\widebar{\nu}^{(i)},\widebar{F}^{(i)})$: $(\widebar{\mu},\widebar{\nu},\widebar{F})$ with jump size shifted by $\gamma_\alpha^{(i)}$.
		\item  $(\mu^{(Ji)},\nu^{(Ji)},F^{(Ji)})$ (defined in Corollary \ref{Corollary: TGJDPredictableCharacteristics}): jump measure associated with $\gamma_i\cdot \widetilde{J}$.
		\item  $(\mu^{J_i},\nu^{J_i},F^{J_i})$ (defined in Proposition \ref{Propostion: IDPSwitchingNoCommonJump}): jump measure associated with $J_i$.
		\item  $(\underline{\mu}^{(Ji)},\underline{\nu}^{(Ji)},\underline{F}^{(Ji)})$ (will be defined in Lemma \ref{Lemma: ConjugateConditionalExpectation}): jump measure associated with $\gamma_\alpha^{(i)}\cdot \widetilde{J}_i$.
	\end{itemize}
\end{remark}
\section{Portfolio admissibility for $\widehat{X}$ versus that for $X^{(i)}$}\label{Appendix: PortfolioAdmissibility}
Let $Y$ be a $d$-dimensional semimartingale with jump compensator $\nu^Y(\D t,\D x)=F^Y_t(\D x)\D A^Y_t$. Following \cite[Equation (3.1)]{Choulli&Yansori2022a}, the admissible portfolio set can be written as
\begin{equation*}
\mathcal{L}(Y)=\{\pi:\pi\ \mbox{is $\mathbb F$-predictable},\quad 1+x^{tr}\pi>0,\quad P\otimes\D A^Y\otimes F^Y\mbox{-a.e.}\}.
\end{equation*}
The corresponding predictable set-valued process is given by
\begin{equation}\label{NaturalConstraint}
\mathcal{L}_{(\omega,t)}(Y):=\{\pi\in\mathbb{R}^d:\pi^{tr}x>-1,\quad F^Y_{(\omega,t)}\mbox{-a.e.}\}.
\end{equation}
For the market (\ref{RSJDdynamic}), $\mathcal{L}_{(\omega,t)}(\widehat{X})$ is given by
\begin{align*}
&\left(\bigcap_{k:\widebar{\lambda}_k(t,\alpha(t-))>0}\{\pi\in\mathbb{R}^d|\pi^{tr}\zeta_\alpha^{(k)}(\omega,t)>-1\}\right)\bigcap\left(\bigcap_{i:\widebar{q}_i(t)>0}\{\pi\in\mathbb{R}^d|\pi^{tr}\gamma_\alpha^{(i)}(\omega,t)>-1\}\right)\\&\bigcap\left(\bigcap_{k,i:\frac{\D\langle\widetilde{N}_{\alpha k},\widetilde{J}_i\rangle(t)}{\D t}>0}\{\pi\in\mathbb{R}^d|\pi^{tr}(\zeta_\alpha^{(k)}(\omega,t)+\gamma_\alpha^{(i)}(\omega,t))>-1\}\right).
\end{align*}

The natural constraint set (\ref{NaturalConstraint}) in the market (\ref{RSJDdynamic}) as finite intersections of open half-spaces on $\mathbb{R}^d$ is open, and contains the origin of $\mathbb{R}^d$. 

From \cite[Example 2.7]{Karatzas&Kardaras2007} or \cite[Equation (5.3)]{Choulli&Yansori2022a}, a predictable portfolio $\pi$ belongs to $\mathcal{L}(Y)$ if and only if for $P\otimes\D A^Y$-a.e. $(\omega,t)$, $\pi(\omega,t)\in\mathcal{L}_{(\omega,t)}(Y)$.

\subsection{Admissibility in the general switching market}
The following two lemmas handle the admissibility of portfolios in the switching market $\widehat{X}$ in (\ref{GRSJDDynamic}).

\begin{lemma}\label{Lemma: SwitchingNaturalConstraint}
	Recall the definition of the natural constraint as a set-valued process in (\ref{NaturalConstraint}), we have
	\begin{equation*}
	\mathcal{L}_{(\omega,t)}(\widebar{X})=\mathcal{L}_{(\omega,t)}(X^{(i)})
	\end{equation*}on the set \ $ \{(\omega,t)\in\Omega\times [0,\infty)|\alpha(\omega,t-)=i\}$.
\end{lemma}
\begin{proof}
	\begin{align*}
	\mathcal{L}_{(\omega,t)}(\widebar{X})=&\{\pi\in\mathbb{R}^d| \pi^{tr}x>-1,\widebar{F}_{(\omega,t)}-\text{a.e.}\}
	\end{align*}
	And notice that when $(\omega,t)\in\{(\omega,t)|\alpha(\omega,t-)=i\}, \widebar{F}_{(\omega,t)}= F_{(\omega,t)}^{(i)}$.
\end{proof}
\begin{lemma}\label{Lemma: GRSJDIntegrability}
	In the market ($\ref{GRSJDDynamic}$), we have the following properties.\\
{\rm(a)} $L(\widehat{X})=\bigcap_i L(I_{\{\alpha_{-}=i\}}\cdot(X^{(i)}+\gamma_i\cdot \widetilde{J}))$.\\
{\rm(b)} Suppose $\pi_\alpha=\sum_i I_{\{\alpha_{-}=i\}}\pi_i$. Then $\pi_\alpha\in L(\widehat{X})$ iff $\pi_i\in  L(I_{\{\alpha_{-}=i\}}\cdot(X^{(i)}+\gamma_i\cdot \widetilde{J}))$ for $i\in\{1,2,...,D\}$.\\
{\rm(c)} On the set  $ \{(\omega,t)\in\Omega\times [0,\infty)|\alpha(\omega,t-)=i\}$,
		\[\mathcal{L}_{(\omega,t)}(\widehat{X})=\mathcal{L}_{(\omega,t)}(I_{\{\alpha_{-}=i\}}\cdot(X^{(i)}+\gamma_i\cdot \widetilde{J}))=\mathcal{L}_{(\omega,t)}(X^{(i)}+\gamma_i\cdot \widetilde{J}).\]
\end{lemma}
\begin{proof}
	Statement (c) is from similar argument in Lemma \ref{Lemma: SwitchingNaturalConstraint}. Statements (a) and (b) are obvious by using \citeauthor{Karatzas&Kardaras2007}\cite[Theorem 6.2]{Karatzas&Kardaras2007} and noticing that 
	\begin{align*}
	(\widehat{b},\widehat{c},\widehat{F}(\D x))=\sum_i I_{\{\alpha_{-}=i\}}(\widehat{b}^{(i)},\widehat{c}^{(i)},\widehat{F}^{(i)}(\D x)),
	\end{align*}
	where $(\widehat{b}^{(i)},\widehat{c}^{(i)},\widehat{F}^{(i)})$ is the predictable characteristics of $I_{\{\alpha_{-}=i\}}\cdot (X^{(i)}+\gamma_i\cdot\widetilde{J})$ as in Corollary \ref{Corollary: TGJDPredictableCharacteristics}.
\end{proof}

\section{Formal decomposition for switching impact}
This section has three subsections. The first subsection discusses some conjugate properties due to the superposition of the jumps between $\gamma_{\alpha}\cdot{J}$ and $\widebar{X}$. The second subsection gives the proof of  Lemma \ref{Lemma: GRSJDTransitionShock}. 
\subsection{Conjugate properties between $M^{P}_{\widebar{\mu}}\left[I_{\{\gamma_\alpha\Delta J\neq 0\}}\big| \widetilde{\cal{P}}\right]$  and $M^{P}_{\mu^{(J)}}\left[I_{\{\Delta \widebar{X}\neq 0\}}\big| \widetilde{\cal{P}}\right]$}

Some conjugate properties between two jump measures are given as below.
\begin{lemma}\label{Lemma: ConjugateConditionalExpectation}
	Suppose that $X^{(i)}$ is special for each $i\in\{1,2,...,D\}$. Consider arbitrary functionals $U\in\widetilde{\cal{O}}$ and $H\in\widetilde{\cal{P}}$ with $ H(\omega,t,0)=0$, and both  $U$ and $H$ are either nonnegative or integrable. Define the auxiliary jump measure for each $i\in\{1,2,...,D\}$,
	\[\underline{\mu}^{(Ji)}(\D t,\D x):=\sum_s I_{\{\gamma_\alpha^{(i)}(s)\Delta J_i(s)\neq0\}} \delta_{( s, \gamma_\alpha^{(i)}(s)\Delta J_i(s))}(\D t,\D x).\]
	Then the following equalities hold.\\
	\begin{equation}\label{equality1}
	\begin{split}
	\int M^{P}_{\widebar{\mu}}\left[I_{\{\gamma_\alpha\Delta J\neq 0\}}\big| \widetilde{\cal{P}}\right](x)\widebar{F}(\D x)&=\sum_iq_i M^{P}_{\mu^{(J)}}\left[I_{\{\Delta \widebar{X}\neq 0\}}\big| \widetilde{\cal{P}}\right](\gamma_\alpha^{(i)})I_{\{\gamma_\alpha^{(i)}\neq 0\}}\\&=\sum_iq_i \left(1- M^{P}_{\underline{\mu}^{(Ji)}}\left[I_{\{\Delta \widebar{X}= 0\}}\big| \widetilde{\cal{P}}\right](\gamma_\alpha^{(i)})\right)I_{\{\gamma_\alpha^{(i)}\neq 0\}},
	\end{split}
	\end{equation}
	\begin{equation}\label{equality2}
	\begin{split}
	\int M^{P}_{\widebar{\mu}}\left[I_{\{\gamma_\alpha^{(i)}\Delta J_i\neq 0\}}\big| \widetilde{\cal{P}}\right](x)\widebar{F}(\D x)=&q_i M^{P}_{\underline{\mu}^{(Ji)}}\left[I_{\{\Delta \widebar{X}\neq 0\}}\big| \widetilde{\cal{P}}\right](\gamma_\alpha^{(i)})I_{\{\gamma_\alpha^{(i)}\neq 0\}}\\=&q_i\left(1- M^{P}_{\underline{\mu}^{(Ji)}}\left[I_{\{\Delta \widebar{X}= 0\}}\big| \widetilde{\cal{P}}\right](\gamma_\alpha^{(i)})\right)I_{\{\gamma_\alpha^{(i)}\neq 0\}}.
	\end{split}
	\end{equation}
	\begin{equation}\label{equality3}
\sum_iq_i M^{P}_{\mu^{(J)}}\left[U\big| \widetilde{\cal{P}}\right](\gamma_\alpha^{(i)})H(\gamma_\alpha^{(i)})=\sum_iq_i M^{P}_{\underline{\mu}^{(Ji)}}\left[U\big| \widetilde{\cal{P}}\right](\gamma_\alpha^{(i)})H(\gamma_\alpha^{(i)}),	\end{equation}
and 
	\begin{equation}\label{equality4}
	\begin{split}
	\sum_i H(\gamma_\alpha^{(i)})\int M^{P}_{\widebar{\mu}}\left[I_{\{\gamma_\alpha^{(i)}\Delta J_i\neq 0\}}\big| \widetilde{\cal{P}}\right](x)\widebar{F}(\D x)+\sum_iH(\gamma_\alpha^{(i)})q_i M^{P}_{\mu^{(J)}}\left[I_{\{\Delta \widebar{X}= 0\}}\big| \widetilde{\cal{P}}\right](\gamma_\alpha^{(i)})=\sum_iH(\gamma_\alpha^{(i)})q_i.
	\end{split}
	\end{equation}
\end{lemma}
\begin{proof}
	Notice that
	\begin{align*}
	I_{\{\gamma_\alpha\Delta J\neq 0\}}\star\widebar{\mu}=&I_{\{\Delta \widebar{X}\neq 0\}}\star\mu^{(J)},
	\end{align*}
	which gives 
	\begin{align*}
	M^{P}_{\widebar{\mu}}\left[I_{\{\gamma_\alpha\Delta J\neq 0\}}\big| \widetilde{\cal{P}}\right]\star\widebar{\nu}=&M^{P}_{\mu^{(J)}}\left[I_{\{\Delta \widebar{X}\neq 0\}}\big| \widetilde{\cal{P}}\right]\star\nu^{(J)},\\
	\int M^{P}_{\widebar{\mu}}\left[I_{\{\gamma_\alpha\Delta J\neq 0\}}\big| \widetilde{\cal{P}}\right](x)\widebar{F}(\D x)=&\sum_iq_i M^{P}_{\mu^{(J)}}\left[I_{\{\Delta \widebar{X}\neq 0\}}\big| \widetilde{\cal{P}}\right](\gamma_\alpha^{(i)})I_{\{\gamma_\alpha^{(i)}\neq 0\}},\\
	\end{align*}
	which is the first equality in (\ref{equality1}). Similarly, we have (\ref{equality2}). 
	
	The second equality in (\ref{equality1}) is the consequence of (\ref{equality3}) or (\ref{equality2}). And to prove (\ref{equality3}), just notice that 
	\begin{align*}
	\mu^{(J)}=\sum_i\underline{\mu}^{(Ji)}.
	\end{align*} 
	
	(\ref{equality4}) is obtained by combining (\ref{equality2}) and (\ref{equality3}) with $U=I_{\{\Delta \widebar{X}=0\}}$ and the fact that \[I_{\{\gamma_\alpha^{(i)}\neq 0\}}H(\gamma_\alpha^{(i)})=H(\gamma_\alpha^{(i)}).\]
\end{proof}

Lemma \ref{Lemma: ConjugateConditionalExpectation} stands for ``conjugation" properties since it relates the two quantities: using the jump measure $\widebar{\mu}$ to predictably filter out the jumps from $\mu^{(J)}$, and the converse one. (\ref{equality4}) is especially useful in the following lemma.
\subsection{Proof of Lemma \ref{Lemma: GRSJDTransitionShock}}\label{Appendix: GRSJDTransitionShock}
	Omit $t$ for notational convenience,
\begin{equation*}
\begin{split}
&\widehat{\kappa}(\alpha_{-},\pi)-\kappa(\alpha_{-},\pi)\\
=&\pi^{tr}(\widehat{b}-\widebar{b}) +
\int \left(\ln (1+\pi^{tr}x)-\pi^{tr}h(x)\right)(\widehat{F}(\D x)-\widebar{F}(\D x)).
\end{split}
\end{equation*}
Next, $\widehat{b}$ is expanded by Proposition \ref{Proposition:GRSJDPredictableCharacteristics}, which yields,
\begin{equation*}
\begin{split}
&\int \pi^{tr}(x-h(x))  (\widebar{F}(\D x)-\widehat{F}(\D x))+
\int \left(\ln (1+\pi^{tr}x)-\pi^{tr}h(x)\right)(\widehat{F}(\D x)-\widebar{F}(\D x))\\
=&\int \left(\ln (1+\pi^{tr}x)-\pi^{tr}x\right)(\widehat{F}(\D x)-\widebar{F}(\D x)).
\end{split}
\end{equation*}
Lemma \ref{Lemma: GRSJDIntegalLevyMeasure} is applied to calculate the integral above, which yields,
\begin{equation*}
\begin{split}
&\int \left(M^{P}_{\widebar{\mu}}\left[I_{\{\gamma_\alpha\Delta J=0\}}\big| \widetilde{\cal{P}}\right](x)-1\right)(\ln (1+\pi^{tr}x)-\pi^{tr}x)\widebar{F}(\D x)\\&+ \sum_iq_iM^{P}_{\mu^{(J)}}\left[I_{\{\Delta \widebar{X}=0\}}\big| \widetilde{\cal{P}}\right](\gamma_\alpha^{(i)})(\ln (1+\pi^{tr}\gamma_\alpha^{(i)})-\pi^{tr}\gamma_\alpha^{(i)})\\ 
&+\sum_i\int\left( M^{P}_{\widebar{\mu}}\left[I_{\{\gamma_\alpha^{(i)}\Delta J_i\neq 0\}}\big| \widetilde{\cal{P}}\right](x)(\ln (1+\pi^{tr}(x+\gamma_\alpha^{(i)}))-\pi^{tr}(x+\gamma_\alpha^{(i)})) \right)\widebar{F}(\D x).
\end{split}
\end{equation*}
We merge the first and third terms above together, which generates,
\begin{equation*}
\begin{split}
&\underbrace{\sum_iq_iM^{P}_{\mu^{(J)}}\left[I_{\{\Delta \widebar{X}=0\}}\big| \widetilde{\cal{P}}\right](\gamma_\alpha^{(i)})(\ln (1+\pi^{tr}\gamma_\alpha^{(i)})-\pi^{tr}\gamma_\alpha^{(i)})}_{\text{extra disjoint jumps}\leq 0}\\ 
&+\underbrace{\sum_i\int M^{P}_{\widebar{\mu}}\left[I_{\{\gamma_\alpha^{(i)}\Delta J_i\neq 0\}}\big| \widetilde{\cal{P}}\right](x)\left(\ln\dfrac{ 1+\pi^{tr}(x+\gamma_\alpha^{(i)})}{1+\pi^{tr}x}-\pi^{tr}\gamma_\alpha^{(i)}\right)\widebar{F}(\D x)}_{\text{distorted joint jumps}}
\end{split}
\end{equation*}
Here we employ (\ref{equality4}) for $H(x)=\pi^{tr}x$ to get the compensation part out, i.e.,
\begin{equation*} 
\begin{split}
&\sum_iq_i M^{P}_{\mu^{(J)}}\left[I_{\{\Delta \widebar{X}=0\}}\big| \widetilde{\cal{P}}\right](\gamma_\alpha^{(i)})\ln (1+\pi^{tr}\gamma_\alpha^{(i)})-\sum_iq_i\pi^{tr}\gamma_\alpha^{(i)}\\ 
&+\sum_i\int\left( M^{P}_{\widebar{\mu}}\left[I_{\{\gamma_\alpha^{(i)}\Delta J_i\neq 0\}}\big| \widetilde{\cal{P}}\right](x)\ln\dfrac{ 1+\pi^{tr}(x+\gamma_\alpha^{(i)})}{1+\pi^{tr}x} \right)\widebar{F}(\D x)
\end{split}
\end{equation*}
Again, we reversely use (\ref{equality4}) for $H(x)=\ln(1+\pi^{tr}x)$, and the above expression equals
\begin{equation*} 
\begin{split}
&\sum_iq_i \ln (1+\pi^{tr}\gamma_\alpha^{(i)})-\pi^{tr}\gamma_\alpha q\\
&+\sum_i\int\left( M^{P}_{\widebar{\mu}}\left[I_{\{\gamma_\alpha^{(i)}\Delta J_i\neq 0\}}\big| \widetilde{\cal{P}}\right](x)\ln\dfrac{ 1+\pi^{tr}(x+\gamma_\alpha^{(i)})}{(1+\pi^{tr}x)(1+\pi^{tr}\gamma_\alpha^{(i)})} \right)\widebar{F}(\D x).
\end{split}
\end{equation*}
Moreover, if for each $i\in\{1,2,...,D\}$, $X^{(i)}$ is special, we can take $h=x$. Thanks to \cite[Equation (5.12)]{Choulli&Yansori2022a}, $\widehat{\kappa}(\alpha_{-},\widetilde{\pi}_\alpha)$ and $\kappa(\alpha_{-},\widetilde{\pi}_\alpha^*)$ take values in $[0,\infty]$. This implies that 
\[\int \left(\ln (1+\widetilde{\pi}_\alpha^{tr}x)-\widetilde{\pi}_\alpha^{tr}x\right)\widehat{F}(\D x)\quad\mbox{and}\quad \int \left(\ln (1+\widetilde{\pi}_\alpha^{*tr}x)-\widetilde{\pi}_\alpha^{*tr}x\right)\widebar{F}(\D x)\]
take values in $(-\infty,0]$. 

Then the term 
\[\int \left(\ln (1+\pi^{tr}x)-\pi^{tr}x\right)(\widehat{F}(\D x)-\widebar{F}(\D x))\]
is always well-defined for $\pi\in\{\widetilde{\pi}_\alpha,\widetilde{\pi}_\alpha^*\}$. Besides, $\mathcal{S}(\widetilde{\pi}_\alpha)$ takes values in $[-\infty,\infty)$ and $ \mathcal{S}(\widetilde{\pi}_\alpha^*)$ takes values in $(-\infty,\infty]$.

\subsection{Proof of Theorem \ref{Theorem: GRSJDSwitchingEffect}}\label{Appendix: GRSJDSwitchingEffect}
Here, we explain more about how the switching feature impacts the optimal utility. Imagine such a dynamic that the non-switching market is altered to be the switching market:\\
{\rm(a)} From (\ref{GRSJDDynamic}), the added switching to make $X^{(1)}\rightarrow\widehat{X}$ has two effects: making the coefficients to be switching, i.e.,
	\[X^{(1)}\rightarrow \widebar{X},\ \kappa(\cdot,1,\cdot)\rightarrow\kappa(\cdot,\alpha(t-),\cdot),\]
	and putting a transition shock as
	\[\widebar{X}\rightarrow\widehat{X},\ \kappa\rightarrow\widehat{\kappa}.\]
{\rm(b)} When the old optimal portfolio has not adapted to be switching, but the market environment (coefficients) has been switching, we call this impact 
	\begin{equation*}
	\kappa(\cdot,\alpha(t-),\cdot)-	\kappa(\cdot,1,\cdot)
	\end{equation*}
	an ``immediate switching impact".\\
{\rm(c)} When the transition shock is added, and the portfolio has not had time to respond, we call this impact a ``transition shock" as $\mathcal{S}(t,\pi)$ in Lemma \ref{Lemma: GRSJDTransitionShock}.\\
{\rm(d)} When the new switching market is fixed, and the optimal portfolio begins adjusting to maximise the current utility under the current market (changing from some old portfolio), we call this adjustment a ``num\'eraire change premium", i.e.
	\begin{equation*}
	\begin{split}
	&\widehat{\kappa}(\cdot,\alpha(t-),\widetilde{\pi}_\alpha(t))-	\widehat{\kappa}(\cdot,\alpha(t-),\pi)\quad\text{or}\quad\kappa(\cdot,\alpha(t-),\widetilde{\pi}_\alpha^*(t))-\kappa(\cdot,\alpha(t-),\pi),\\&\text{for an arbitrary portfolio }\pi.
	\end{split}
	\end{equation*}
	This num\'eraire change premium is nonnegative by Proposition \ref{Proposition: Numeraire}-(a), which actually means that the log-optimal portfolio also maximises the local rate of the utility. This is called a ``myopic" property. One can refer to \cite[Section 3.5]{Karatzas&Kardaras2007} for this property.
	
	We accordingly define these premia as
	\begin{equation*}
	\begin{split}
	\widehat{\mathcal{P}}^{(N)}(\pi):=&\widehat{\kappa}(\cdot,\alpha_{-},\widetilde{\pi}_\alpha)-	\widehat{\kappa}(\cdot,\alpha_{-},\pi)\\
	=&(\widetilde{\pi}_\alpha-\pi) ^{tr}\widehat{b}-\frac{1}{2}\widetilde{\pi}_\alpha^{tr}\widehat{c}\widetilde{\pi}_\alpha + \frac{1}{2}\pi^{tr}\widehat{c}\pi+
	\int \left(\ln \dfrac{1+\widetilde{\pi}_\alpha^{tr}x}{1+\pi^{tr}x}-(\widetilde{\pi}_\alpha-\pi)^{tr}h(x)\right)\widehat{F}(\D x) \geq 0, \\
	\mathcal{P}^{(N)}(\pi):=&\kappa(\cdot,\alpha_{-},\widetilde{\pi}_\alpha^*)-\kappa(\cdot,\alpha_{-},\pi)\\=&(\widetilde{\pi}_\alpha^*-\pi) ^{tr}\widebar{b}-\frac{1}{2}\widetilde{\pi}_\alpha^{*tr}\widebar{c}\widetilde{\pi}_\alpha^* + \frac{1}{2}\pi^{tr}\widebar{c}\pi+
	\int \left(\ln \dfrac{1+\widetilde{\pi}_\alpha^{*tr}x}{1+\pi^{tr}x}-(\widetilde{\pi}_\alpha^*-\pi)^{tr}h(x)\right)\widebar{F}(\D x)  \geq 0.\\
	\end{split}
	\end{equation*}
	In Theorem \ref{Theorem: GRSJDSwitchingEffect}, $\widehat{\mathcal{P}}^{(N)}=\widehat{\mathcal{P}}^{(N)}(\widetilde{\pi}_\alpha^*)$.
	By concavity of $\widehat{\kappa}$ and $\kappa$, \[\widehat{\mathcal{P}}^{(N)}(\pi)=0\quad \text{if and only if}\quad \pi=\widetilde{\pi}_\alpha,\quad \mathcal{P}^{(N)}(\pi)=0\quad \text{if and only if} \quad\pi=\widetilde{\pi}_\alpha^*.\]
{\rm(e)} The difference between the optimal utility with switching coefficients and that with non-switching coefficients is called a ``switching effect". This stands for how the switching coefficient changes the optimal utility. And it is 
	\begin{equation*}
	\kappa(\cdot,\alpha(t-),\widetilde{\pi}_\alpha^*(t))-	\kappa(\cdot,1,\widetilde{\pi}_1^*(t)).
	\end{equation*}
	We can even include the impact from the transition shock, and the ``shock-adjusted switching effect" is 
	\begin{equation*}
	\kappa(\cdot,\alpha(t-),\widetilde{\pi}_\alpha(t))-	\kappa(\cdot,1,\widetilde{\pi}_1^*(t)).
	\end{equation*}
	The rate $\kappa(\cdot,\alpha(t-),\widetilde{\pi}_\alpha(t))$ looks a little strange. While it can be interpreted this way: from the switching and shocked market $\widehat{X}$ with its log-optimal portfolio $\widetilde{\pi}_\alpha$, we truncate the part without transition shock from the rate $\widehat{\kappa}$.

\begin{lemma} \label{Lemma: GRSJDSwitchingEffect2}
	Suppose that the log-optimal portfolios $\widetilde{\pi}_\alpha$ and $\widetilde{\pi}_i^*,i\in\{1,2,...,D\}$ for the regime switching market $\widehat{X}$ and the non-switching markets $X^{(i)},i\in\{1,2,...,D\}$ exist. Then  
	\begin{equation*}
	\begin{split}
	\Delta_T(\widehat{X})=&u_T(\widehat{X})-u_T(X^{(1)})\\
	=& \underbrace{ E\left[\int_0^T \widehat{\mathcal{P}}^{(N)}(t) \D A_t\right] }_{\text{num\'eraire change premium for transition shock}}- \underbrace{ E\left[\int_0^T\mathcal{S}_t(\widetilde{\pi}_\alpha^*) \D A_t\right]}_{\text{transition shock}} \\
	+&\underbrace{E\left[ \int_0^T\kappa(t,\alpha(t-),\widetilde{\pi}_\alpha^*(t))-\kappa(t,1,\widetilde{\pi}_1^*(t)) \D A_t\right] }_{\text{switching effect}} .
	\end{split}
	\end{equation*}
\end{lemma}
\begin{proof}
	This is just \[[u_T(\widehat{X})-u_T(\widebar{X})] +[u_T(\widebar{X})-u_T(X^{(1)})],\]
or in the rate sense by Theorem \ref{Theorem:GRSJDLog-optimal} and Proposition \ref{Proposition: GJDLog-optimal},
\[ [\widehat{\kappa}(t,\alpha(t-),\widetilde{\pi}_\alpha(t))-\kappa(t,\alpha(t-),\widetilde{\pi}_\alpha^*(t))]+[\kappa(t,\alpha(t-),\widetilde{\pi}_\alpha^*(t))-\kappa(t,1,\widetilde{\pi}_1^*(t))].\]
The first term in bracket above is further broken into 
\[[ \widehat{\kappa}(t,\alpha(t-),\widetilde{\pi}_\alpha(t))-\widehat{\kappa}(t,\alpha(t-),\widetilde{\pi}_\alpha^*(t))]+[\widehat{\kappa}(t,\alpha(t-),\widetilde{\pi}_\alpha^*(t))-\kappa(t,\alpha(t-),\widetilde{\pi}_\alpha^*(t))  ]   .\]
By Lemma \ref{Lemma: GRSJDTransitionShock}, we have
\[\widehat{\kappa}(t,\alpha(t-),\widetilde{\pi}_\alpha^*(t))-\kappa(t,\alpha(t-),\widetilde{\pi}_\alpha^*(t)) =-\mathcal{S}_t(\widetilde{\pi}_\alpha^*).\]
This ends the proof of the lemma.\end{proof}

\begin{proof}[Proof of Theorem \ref{Theorem: GRSJDSwitchingEffect}]
The derivation is nearly the same as in the proof of Theorem \ref{Theorem:GRSJDLog-optimal}, with the only difference being that here we cut $\widetilde{L}$ and $\widetilde{V}$ into two parts. 
Since $\widetilde{V}_2$ is continuous, we have 
\begin{align*}
&u_T(\widehat{X})-u_T(X^{(1)})\\
=&  E_P\left[\ln\mathcal{E}_T(\widetilde{\pi}_\alpha\cdot \widehat{X})\right]-E\left[\int_0^T \kappa(t,1,\widetilde{\pi}_1^*(t))\D A_t\right] \\
=&  E\left[-\ln\mathcal{E}_T(\widetilde{L})\right] +E\left[-\ln\mathcal{E}_T(-\widetilde{V})\right] - E\left[\int_0^T \kappa(t,1,\widetilde{\pi}_1^*(t))\D A_t\right] \\
=&E\left[-\ln\mathcal{E}_T(\widetilde{L}_1)-\ln\mathcal{E}_T(\widetilde{L}_2)-\ln\mathcal{E}_T(-\widetilde{V}_1)+\widetilde{V}_{2,T}-\int_0^T \kappa(t,1,\widetilde{\pi}_1^*(t))\D A_t\right]\\
=&E\left[H_T^{(0)}(\widetilde{L}_1)-\ln\mathcal{E}_T(-\widetilde{V}_1)+\widetilde{V}_{2,T}-\int_0^T \kappa(t,1,\widetilde{\pi}_1^*(t))\D A_t\right]+\mathcal{H}(P|\widetilde{\mathbb{P}})-\ln E[\mathcal{E}_T(\widetilde{L}_2)],
\end{align*}
On $\{\Delta A>0\}$,
$$(1+\Delta\widetilde{L}_1)(1-\Delta\widetilde{V}_1)=(1+\widetilde{\pi}_\alpha^{tr}\Delta\widehat{X})^{-1}.$$
The first term above is computed by taking predictable projections, which yields
\begin{align*}
&E\left[\int_0^T \left(\kappa(t,\alpha(t-),\widetilde{\pi}_\alpha(t))-\kappa(t,1,\widetilde{\pi}_1^*(t))\right)\D A_t+\widetilde{V}_{2,T}\right.\\
&\left.\quad+\int_0^T\left(\int (1- M^{P}_{\widebar{\mu}}\left[I_{\{\gamma_\alpha\Delta J= 0\}}\big| \widetilde{\cal{P}}\right](x))\left(\widetilde{\pi}_\alpha^{tr}x-\ln(1+\widetilde{\pi}_\alpha^{tr}x) \right)\widebar{F}(\D x)\right)(t)\D A_t\right]+\mathcal{H}(P|\widetilde{\mathbb{P}})-\ln E[\mathcal{E}_T(\widetilde{L}_2)].
\end{align*}
Finally, notice that by definition the following identity holds.
\begin{equation}\label{PremiumShockIdentity}
\widehat{\mathcal{P}}^{(N)}(\widetilde{\pi}_\alpha^*)-\mathcal{S}(\widetilde{\pi}_\alpha^*)=-\mathcal{P}^{(N)}(\widetilde{\pi}_\alpha)-\mathcal{S}(\widetilde{\pi}_\alpha).
\end{equation}
Then the first equality follows immediately.

The second equality is just combining Lemma \ref{Lemma: GRSJDSwitchingEffect2} and Lemma \ref{Lemma: GRSJDTransitionShock}.
\end{proof}
\bibliography{RSJD_references}

\end{document}